\documentclass[preprint,3p,times,Times,authoryear]{elsarticle}

\usepackage[utf8]{inputenc}
\usepackage[english]{babel}
\usepackage[scaled]{helvet}
\usepackage{amsmath}
\usepackage{color}
\usepackage{natbib}
\usepackage{caption}
\usepackage{float}
\usepackage{sidecap}
\usepackage{subcaption}
\usepackage{tikz}
\usepackage{hyperref}
\usepackage{graphicx}
\usepackage{epsfig}
\usepackage{epstopdf}

\journal{New Astronomy}

\def\jnl@style#1{{\rmfamily#1}}%
\def\jref@jnl#1{{\jnl@style#1}}%

\newcommand\aj{\jref@jnl{AJ}}%
\newcommand\araa{\jref@jnl{ARA\&A}}%
\newcommand\apj{\jref@jnl{ApJ}}%
\newcommand\apjl{\jref@jnl{ApJ}}%
\newcommand\apjs{\jref@jnl{ApJS}}%
\newcommand\ao{\jref@jnl{Appl.~Opt.}}%
\newcommand\apss{\jref@jnl{Ap\&SS}}%
\newcommand\aap{\jref@jnl{A\&A}}%
\newcommand\aapr{\jref@jnl{A\&A~Rev.}}%
\newcommand\aaps{\jref@jnl{A\&AS}}%
\newcommand\azh{\jref@jnl{AZh}}%
\newcommand\baas{\jref@jnl{BAAS}}%
\newcommand\jrasc{\jref@jnl{JRASC}}%
\newcommand\memras{\jref@jnl{MmRAS}}%
\newcommand\mnras{\jref@jnl{MNRAS}}%
\newcommand\na{\jref@jnl{New Astron.}}%
\newcommand\nar{\jref@jnl{New Astron. Rev.}}%
\newcommand\pra{\jref@jnl{Phys.~Rev.~A}}%
\newcommand\prb{\jref@jnl{Phys.~Rev.~B}}%
\newcommand\prc{\jref@jnl{Phys.~Rev.~C}}%
\newcommand\prd{\jref@jnl{Phys.~Rev.~D}}%
\newcommand\pre{\jref@jnl{Phys.~Rev.~E}}%
\newcommand\prl{\jref@jnl{Phys.~Rev.~Lett.}}%
\newcommand\pasp{\jref@jnl{PASP}}%
\newcommand\pasj{\jref@jnl{PASJ}}%
\newcommand\qjras{\jref@jnl{QJRAS}}%
\newcommand\skytel{\jref@jnl{S\&T}}%
\newcommand\solphys{\jref@jnl{Sol.~Phys.}}%
\newcommand\sovast{\jref@jnl{Soviet~Ast.}}%
\newcommand\ssr{\jref@jnl{Space~Sci.~Rev.}}%
\newcommand\zap{\jref@jnl{ZAp}}%
\newcommand\nat{\jref@jnl{Nature}}%
\newcommand\iaucirc{\jref@jnl{IAU~Circ.}}%
\newcommand\aplett{\jref@jnl{Astrophys.~Lett.}}%
\newcommand\apspr{\jref@jnl{Astrophys.~Space~Phys.~Res.}}%
\newcommand\bain{\jref@jnl{Bull.~Astron.~Inst.~Netherlands}}%
\newcommand\fcp{\jref@jnl{Fund.~Cosmic~Phys.}}%
\newcommand\gca{\jref@jnl{Geochim.~Cosmochim.~Acta}}%
\newcommand\grl{\jref@jnl{Geophys.~Res.~Lett.}}%
\newcommand\jcp{\jref@jnl{J.~Chem.~Phys.}}%
\newcommand\jgr{\jref@jnl{J.~Geophys.~Res.}}%
\newcommand\jqsrt{\jref@jnl{J.~Quant.~Spec.~Radiat.~Transf.}}%
\newcommand\memsai{\jref@jnl{Mem.~Soc.~Astron.~Italiana}}%
\newcommand\nphysa{\jref@jnl{Nucl.~Phys.~A}}%
\newcommand\physrep{\jref@jnl{Phys.~Rep.}}%
\newcommand\physscr{\jref@jnl{Phys.~Scr}}%
\newcommand\planss{\jref@jnl{Planet.~Space~Sci.}}%
\newcommand\procspie{\jref@jnl{Proc.~SPIE}}%

\newfont{\Giga}{cmssbx10 scaled 5200}
\newfont{\giga}{cmssbx10 scaled 3600}
\newfont{\Mega}{cmssbx10 scaled 3200}
\newfont{\mega}{cmssbx10 scaled 2500}
\newfont{\Kilo}{cmssbx10 scaled 2000}
\newfont{\kilo}{cmssbx10 scaled 1600}
\newfont{\Deca}{cmssbx10 scaled 1250}
\newfont{\deca}{cmssbx10 scaled 1140}
\newfont{\Normal}{cmssbx10 scaled 1100}
\newfont{\normal}{cmssbx10 scaled 1000}
\newfont{\iGiga}{cmssi10 scaled 6200}
\newfont{\igiga}{cmssi10 scaled 4300}
\newfont{\iMega}{cmssi10 scaled 3200}
\newfont{\imega}{cmssi10 scaled 2500}
\newfont{\iKilo}{cmssi10 scaled 2000}
\newfont{\ikilo}{cmssi10 scaled 1500}
\newfont{\mathGiga}{cmsy10 scaled 6200}
\newfont{\mathgiga}{cmsy10 scaled 4300}
\newfont{\mathMega}{cmsy10 scaled 3200}
\newfont{\mathmega}{cmsy10 scaled 2500}
\newfont{\mathKilo}{cmsy10 scaled 2000}
\newfont{\mathkilo}{cmsy10 scaled 1500}
\newcommand\euro{{\sffamily C%
\makebox[0pt][l]{\kern-.70em\mbox{--}}%
\makebox[0pt][l]{\kern-.68em\raisebox{.25ex}{--}}}}
\newcommand\keuro{k{\sffamily C%
\makebox[0pt][l]{\kern-.70em\mbox{--}}%
\makebox[0pt][l]{\kern-.68em\raisebox{.25ex}{--}}}}
\definecolor{black} {rgb} {0.0,0.0,0.0}
\definecolor{blue}  {rgb} {0,0,1}
\definecolor{blue0} {rgb} {0.0,0.0,1.0}
\definecolor{blue1} {rgb} {0.1,0.1,0.8}
\definecolor{blue2} {rgb} {0.3,0.3,1.0}
\definecolor{blue3} {rgb} {0.50,0.50,0.90}
\definecolor{blue4} {rgb} {0.30,0.30,0.75}
\definecolor{blue5} {rgb} {0.15,0.15,0.90}
\definecolor{blue6} {rgb} {0.05,0.05,0.30}
\definecolor{blue7} {rgb} {0.25,0.25,0.75}
\definecolor{greendk}{rgb}{0.0,0.6,0.4}
\definecolor{green1}{rgb}{0.01,0.65,0.02}
\definecolor{red1}  {rgb} {0.3,0.1,0.1}
\definecolor{verylightgrey}{rgb}{0.95,0.95,0.95}

\newcommand{\sun}{{\odot}}
\newcommand{\cm}{\rm{cm}}
\newcommand{\s}{\rm{s}}
\newcommand{\km}{\rm{km}}
\newcommand{\AU}{\rm{AU}}

\newcommand{\g}{\rm{g}}
\newcommand{\Msol}{\rm{M}_{\sun}}
\newcommand{\Rsol}{\rm{R}_{\sun}}

\newcommand{\au}{\rm{AU}}

\newcommand{\kys}{\rm{kyrs}}

\newcommand{\pd}[2] {
  \frac{\partial#1}{\partial#2} 
}

\newcommand{\bal}{\begin{align}}
\newcommand{\eal}{\end{align}}
\newcommand{\beq}{\begin{equation}}
\newcommand{\eeq}{\end{equation}}
\newcommand{\bef}{\begin{figure}[t]}
\newcommand{\eef}{\end{figure}}
\newcommand{\befSC}{\begin{SCfigure}}
\newcommand{\eefSC}{\end{SCfigure}}

\def\eps@scaling{0.98}

\def\showone#1{
  \centering
  \leavevmode
  \epsfxsize=\eps@scaling\linewidth
  \epsfbox{#1.eps}
}

\def\epssmall@scaling{0.6}

\def\showonesmall#1{
  \centering
  \leavevmode
  \epsfxsize=\epssmall@scaling\linewidth
  \epsfbox{#1.eps}
}

\def\epstwo@scaling{0.48}
\def\epstwolarge@scaling{0.6}
\def\epsthree@scaling{0.32}

\def\showtwo#1#2{
  \centering
  \leavevmode
  \epsfxsize=\epstwo@scaling\linewidth
  \epsfbox{#1.eps} 
  \epsfxsize=\epstwo@scaling\linewidth
  \epsfbox{#2.eps}
}

\def\showtwolarge#1#2{
  \centering
  \leavevmode
  \epsfxsize=\epstwolarge@scaling\linewidth
  \epsfbox{#1.eps} 
  \epsfxsize=\epstwolarge@scaling\linewidth
  \epsfbox{#2.eps}
}
\def\showthree#1#2#3{
  \centering
  \leavevmode
  \epsfxsize=\eps@scaling\linewidth
  \epsfbox{#1.eps} 
  \epsfxsize=\eps@scaling\linewidth
  \epsfbox{#2.eps}
  \epsfxsize=\eps@scaling\linewidth
  \epsfbox{#3.eps}
}

\def\showthreesmall#1#2#3{
  \centering
  \leavevmode
  \epsfxsize=\epssmall@scaling\linewidth
  \epsfbox{#1.eps} 
  \epsfxsize=\epssmall@scaling\linewidth
  \epsfbox{#2.eps}
  \epsfxsize=\epssmall@scaling\linewidth
  \epsfbox{#3.eps}
}

\def\showsix#1#2#3#4#5#6{
  \centering
  \leavevmode
  \epsfxsize=\epstwo@scaling\linewidth
  \epsfbox{#1.eps} \hfil
  \epsfxsize=\epstwo@scaling\linewidth
  \epsfbox{#2.eps} \hfil
  \epsfxsize=\epstwo@scaling\linewidth
  \epsfbox{#3.eps} \hfil
  \epsfxsize=\epstwo@scaling\linewidth
  \epsfbox{#4.eps} \hfil
  \epsfxsize=\epstwo@scaling\linewidth
  \epsfbox{#5.eps} \hfil
  \epsfxsize=\epstwo@scaling\linewidth
  \epsfbox{#6.eps}
}

\def\showsixsmall#1#2#3#4#5#6{
  \centering
  \leavevmode
  \epsfxsize=\epsthree@scaling\linewidth
  \epsfbox{#1.eps} \hfil
  \epsfxsize=\epsthree@scaling\linewidth
  \epsfbox{#2.eps} \hfil
  \epsfxsize=\epsthree@scaling\linewidth
  \epsfbox{#3.eps} \hfil
  \epsfxsize=\epsthree@scaling\linewidth
  \epsfbox{#4.eps} \hfil
  \epsfxsize=\epsthree@scaling\linewidth
  \epsfbox{#5.eps} \hfil
  \epsfxsize=\epsthree@scaling\linewidth
  \epsfbox{#6.eps}
}

\begin{document}

\begin{frontmatter}



\title{Radiation Hydrodynamics using Characteristics on Adaptive Decomposed Domains for Massively Parallel Star Formation Simulations}


\author[addr2]{Lars Buntemeyer}
\author[addr2]{Robi Banerjee}
\author[addr3,addr4]{Thomas Peters}
\author[addr5]{Mikhail Klassen}
\author[addr6]{Ralph E. Pudritz}

\address[addr2]{Hamburger Sternwarte, Universit\"{a}t Hamburg, Gojenbergsweg
112, 21029 Hamburg, Germany, lars.buntemeyer@hzg.de}
\address[addr3]{Institut für Computergest\"{u}tzte Wissenschaften, Universit\"{a}t Z\"{u}rich, 
Winterthurerstrasse 190, CH-8057, Z\"{u}rich, Switzerland}
\address[addr4]{Max-Planck-Institut für Astrophysik, Karl-Schwarzschild-Str. 1, D-85748 Garching, Germany}
\address[addr5]{Department of Physics and Astronomy, McMaster University 1280 Main Street W, 
Hamilton, ON L8S 4M1, Canada}
\address[addr6]{Origins Institute, McMaster University, 1280 Main Street W, 
Hamilton, ON L8S 4M1, Canada}


\begin{abstract}
We present an algorithm for solving the radiative transfer problem on massively parallel computers using adaptive mesh refinement and domain decomposition. The solver is based on the method of characteristics which requires an adaptive raytracer that integrates the equation of radiative transfer. The radiation field is split into local and global components which are handled separately to overcome the non-locality problem. The solver is implemented in the framework of the magneto-hydrodynamics code FLASH and is coupled by an operator splitting step. The goal is the study of radiation in the context of star formation simulations with a focus on early disc formation and evolution. This requires a proper treatment of radiation physics that covers both the optically thin as well as the optically thick regimes and the transition region in particular. We successfully show the accuracy and feasibility of our method in a series of standard radiative transfer problems and two 3D collapse simulations resembling the early stages of protostar and disc formation. 

\end{abstract}


\end{frontmatter}

%
\section{Introduction}
\label{intro}
Radiative feedback plays a crucial role in the process of star and disc formation, the evolution of circumstellar discs and the thermodynamics of the interstellar medium (ISM). 
Massive stars emit a large number of energetic UV photons and strongly determine the structure of giant molecular clouds (GMCs) by creating large
bubbles of ionized gas (HII regions) \citep[e.g.][]{Peters10a,Walch12,Dale13}. 
On smaller scales, low mass and intermediate mass stars also significantly influence their surroundings by
radiative heating. By increasing the fragmentation scale, radiative heating can completely inhibit
further fragmentation in a radius of several AU and prevent, e.g., the formation of a binary system \citep{Price10}. 
\citet{Offner09} investigate the initial mass function (IMF) and the star formation rate (SFR)
by comparing 3D hydrodynamical simulations of low mass star formation with and without the effects of radiative transfer. They find
that the thermal support of a protostar's accretion luminosity suppresses further fragmentation in the cloud core as well
as in the protostellar disc. The SFR in their simulations is about half the value of the simulations without radiative transfer 
and the mass distribution of protostars of very low mass ($M_*<0.1\Msol$) is significantly reduced. \citet{Bate09} finds similar effects.\\
Regarding the formation and evolution of circumstellar discs, radiative feedback is indispensable to understand
their fragmentation behaviour, thermodynamics, and morphology \citep{Chiang97} and to model the infrared excess observed 
in their spectral energy distributions (SEDs) \citep[e.g.][]{Dullemond10}. The initial formation of massive discs during the
Class $0$ phase has been investigated using hydrodynamical and magnetohydrodynamical (MHD) simulations \citep[e.g.][]{Yorke93,Mellon08,Machida10,Peters10a,Seifried11a}, and
\citet{Seifried13} emphasize the importance of turbulence to explain the formation of Keplerian discs even if strong magnetic fields are present.\\
Despite a large number of studies, the actual transition from the early self-gravitating protostellar disc (Class $0$) to the
Keplerian protostellar disc is still poorly understood. Recent observations \citep[e.g.][]{Tobin12} indicate that Keplerian discs might form
very early during the protostellar evolution and the analytic study by \citet{Forgan13} emphasizes the effects of radiative processes.
However, the effects of radiative transfer have usually been neglected in MHD simulations so far or were substantially approximated \citep[e.g.][]{Yorke93,Mellon08,Machida10,Seifried11a}. The self-consistent
modelling of the formation and early evolution of protostars and protostellar discs therefore creates the need for numerical methods to make
3D radiation MHD simulations feasible.\\
In this context, radiative transfer is a rather costly computation compared to solving Euler's equations. 
The reason for this is that the timescale of radiative transfer is usually much shorter than those of hydrodynamics and MHD 
because of the large speed of light compared to the sound speed of the gas in, e.g., a molecular cloud or the characteristic
Alfv\'en wave speeds of the magnetic field. The short timescale on which radiation emerges throughout the complete computational domain makes 
radiative transfer a highly non-local problem compared to MHD which is determined completely by local thermodynamic properties of the gas. 
In this sense, hydrodynamics and radiative transfer are two very different numerical tasks and very challenging to solve consistently.
Modern Eulerian MHD codes like FLASH \citep{FLASH00} mostly solve the Euler equations on a grid with adaptive mesh refinement (AMR)
to resolve fluid features on a wide range of length scales. These codes are parallelized by subdividing the computational domain
into several subdomains each containing a fixed number of cells. Since the Euler equations describe local fluxes of
mass, momentum and energy, all subdomains can be handled in parallel during a hydrodynamical time step. Between the time steps,
boundary values of the subdomains are exchanged using the Message Passing Interface (MPI) for communication.
In contrast, characteristics based radiative transfer codes are usually designed very differently. Instead
of domain decomposition, these codes are parallelized exploiting the formal independence of the radiative transfer equation (RTE) on the solid angle.
Resolving the anisotropy of the radiation field accurately requires a large set of characteristics each covering
a discrete opening angle of the $4\pi$ unit sphere. If all radiative quantities are assumed fixed during one solution step,
characteristics of different directions can be computed independently of each other which makes it ideal for parallelization.
However, the spatial information of the computational domain with all radiative quantities has to be available to the each processor computing 
a certain number of characteristics on the solid angle grid. This can be
a severe drawback in terms of memory requirement if high spatial resolution is required or a large number of frequencies or both (e.g,
synthetic stellar spectra).
Solving both Euler's equations and the RTE consistently requires careful approximations to
the radiative transfer problem to make the coupling of an MHD code with a radiative transfer code feasible. \citet{vanNoort02} present
a radiation solver that is coupled to a hydrodynamical code using AMR and domain decomposition in 2D. The radiation solver uses
short characteristics (SC) for integrating the RTE while boundary values are communicated between Lambda iteration steps. The focus of this approach lies
on modelling the dynamics of scattering dominated stellar atmospheres. The SC approach allows for a fast
converging Gauss-Seidel iteration scheme \citep[e.g.,][]{Trujillo95}, while non-local contributions have to be communicated by a successive exchange
of boundary values between subdomains. This approach was also extended for 3D simulations \citep[e.g.,][]{Hayek10,Davis12}.
However, while the Gauss-Seidel short characteristics approach is well suited for highly scattering dominated regimes, it introduces
a lot of numerical diffusion because a large number of upwind interpolations is necessary.
{\bf \citet{razoumov05} implement a method that is as computationally cheap as the SC method but less diffusive. They create rays on each refinement level separately while their approach is fully threaded but not MPI parallelized. Recently, \citet{Tanaka14} parallelized this approach using the {\it multiple wavefront method} by \citet{Nakamoto01} based on a carefully chosen calculation sequence on a spatially decomposed domain. This method requires successive communication of boundary values.} A similar approach is used with long characteristics (LC) in 3D by \citet{Heinemann06} without AMR.\\
{\bf Another approach for including radiative transfer in hydrodynamical simulations is based on the moment equations (the angular integrated RTE) of the
zeroth, first and second moment of the specific intensity. 
A moment-based scheme does not necessarily require to integrate along large sets of characteristics, however, the anisotropy of the radiation field 
has to approximated reasonably in order to close the set of moment equations for the mean intensity, radiative flux and pressure.
A possible closure relation is the M1-closure used, e.g., in the HERACLES code \citep{Gonzalez07}. The closure relation can also be 
explicitly calculated using, e.g., a characteristics based approach which is known as the Variable Eddington Tensor (VET) method \citep[e.g.][]{Jiang12}.}
A common approach for star formation simulations is the diffusion approximation of the angular moment equations which assumes the radiation field
to be completely isotropic. In regions of high opacities $\chi$, the diffusion approximation is an expansion of the specific intensity
in which all terms $\propto 1/\chi$ are neglected in the RTE. This leads to Eddington's approximation in which the isotropic radiation pressure
is proportional to the radiation energy density. The moment equations of the radiative intensity themselves then form a set of hyperbolic equations, 
like Euler's equations. However, since those two hyperbolic systems would still have to be handled on their
individual timescales, one can even make one further step and neglect the time dependence of the radiation flux by assuming it to be
proportional to the gradient of the radiation energy (Fick's law). The moment equations can then be combined into a single diffusion equation
for the energy of the radiation field. Because the flux in the diffusion approximation lost its finite propagation speed,
one has to introduce a flux-limiter to avoid unphysical propagation speeds depending on the actual opacity. This {\it flux-limited 
diffusion approximation} (FLD) \citep{Levermore81} has been successfully used in radiation hydrodynamical star formation simulations 
coupled within Eulerian grid codes \citep[e.g.][]{Stone92c,Krumholz07,Commercon11,Flock13,Zhang13,Enzo14} 
as well as smoothed particle hydrodynamics (SPH) codes \citep[e.g.][]{Bate13}. However, the diffusion approximation 
is only valid in optically thick regions where the radiation field becomes isotropic. \citet{Kuiper10a} have shown the significant 
drawbacks of exclusively using FLD in the transition regions from optically thick to optically thin regimes where the radiation field becomes highly anisotropic.
Recent efforts have been made to combine raytracing methods with FLD solvers \citep{Kuiper10a, Flock13, Klassen14} to handle, at least, primary stellar or protostellar
radiation separately from the FLD approximation and to avoid the stellar flux from diffusing into shadow regions.\\
Finally, Monte Carlo (MC) methods have become increasingly popular during the last decade, especially in post-processing
MHD simulations. The MC method is a statistical approach and treats individual
photons or photon packages by following its propagation path and computing absorption, emission and scattering probabilities.
Several advances have been introduced, e.g., photon peel-off \citep{Lucy99}, immediate reemission \citep{Bjorkman01} and diffusion approximations \citep{Min09} which
make the MC method a powerful tool to calculate synthetic spectra, SEDs or polarization maps from
the outcome of MHD simulations.
The angular and frequency resolution are, in principle, unlimited since the direction of propagation of a photon package and its frequency are chosen randomly
from a continuous probability function. In that sense, the MC method
always gives a quite reasonable result even in the limit of a small number of photon packages while a low resolution shows mainly up
as statistical noise in the solution. But the statistical approach also has a severe drawback since we do not
know in advance the exact path a photon package will travel, and how and when it is emitted or absorbed. Therefore, it is
extremely difficult to implement on a decomposed domain. MC methods are extremely successful in post-processing
the outcome of MHD simulations but are rarely used in combination with hydrodynamical simulations. Those approaches
which does include MC methods \citep[e.g.][]{Acreman10} are fairly restricted in their spatial resolution of the AMR grid, since each
processor has to get a copy of the complete computational domain to be able to follow the path of an arbitrary photon package.
For our approach, we therefore choose a discrete ordinate method using characteristics to integrate the RTE
which requires a raytracer that works on an AMR grid with domain decomposition.\\ 
The paper is organized as follows: In Section \ref{sec:theory_numerics}, we give a brief introduction into the theory of radiative transfer as far as it concerns our method and we describe in detail the method of hybrid characteristics. 
We also describe the coupling between our radiation solver and the FLASH code. In Section \ref{sec:tests}, we show results from test calculations we performed to investigate the accuracy of our radiation code as well as the coupling to the FLASH code. 
In Section \ref{sec:collapse}, we present results from 3D radiation hydrodynamical collapse simulations and the parallel scaling performance of our 
code is described in Section \ref{sec:performance}. In Section \ref{sec:summary}, we discuss our results and put it into context with other state-of-the-art radiation transfer methods. 


%
\section{Theory and Numerics}
\label{sec:theory_numerics}
In this Section, we describe the theory of radiative transfer (RT) that forms the basis of our solution method as well as the numerics. We describe the hydrodynamics only as it becomes important in the coupling with the radiation solver. For a more detailed description of the FLASH code and its capabilities we refer to \citet{FLASH00}.
\subsection{The Equation of Radiative Transfer}
The theory of radiative transfer in this section is based on the work by \citet{Mihalas84} in the limit of geometrical optics. The energy of the radiation field is described by a scalar field of specific intensities $I(\mathbf{x},\mathbf{n}(\theta,\phi),\nu)$, where $\mathbf{x}$ is the position in space, $\theta$ and $\phi$ define the direction of propagation $\mathbf{n}$, and $\nu$ is the frequency. 
The radiative transfer equation (RTE) describes the change of the specific intensity during its propagation in a medium which is determined by an energy balance between emission and absorption processes. It reads
\beq
\frac{1}{c} \pd{I_{\nu}}{t} + \mathbf{n} \cdot  \nabla I_{\nu} = \eta_{\nu} - \chi_{\nu} I_{\nu},
\label{equ:rte}
\eeq
where $\eta_{\nu}$ denotes the emissivity (energy volume density per unit time and solig angle), $\chi_{\nu}$ is the extinction coefficient (1/length) and $c$ is the speed of light. 
The specific intensity denotes the radiative energy flux {\it per solid angle} $d\Omega=sin\theta\,d\theta\,d\phi$, thus in vacuum, it is constant along a line of sight. Interaction processes between radiation and matter determine the extinction coefficient $\chi_{\nu}$.
However, for this work, we solve the time independent RTE since we assume the radiation field to emerge instantaneously throughout the entire computational domain during a hydrodynamical time step. Furthermore, we use the definition of the {\it source function} $S_{\nu}=\eta_{\nu}/\chi_{\nu}$ and rewrite the RTE in terms of the optical depth without explicit frequency dependence:
\begin{equation}
\frac{dI(\mathbf{n})}{d\tau(\mathbf{n})} = S - I(\mathbf{n}). 
\label{equ:time_independent_rte}
\end{equation}
This form of the RTE describes the propgation of the specific intensity along a specific line element $ds$ in the direction $\mathbf{n}$ and the optical depth element $d\tau=\chi\,ds$ respectively. This requires a proper paramerization depending on the coordinate system and, hence, the definition of $\mathbf{n}\cdot\nabla$. Note that the RTE in the form of Equation (\ref{equ:time_independent_rte}) becomes a 1D ordinary differential equation. However, for the numerical solution in 3D, the optical depth element $d\tau$ is discretised and parameterised in Cartesian coordinates and the solution is obtained by integrating the RTE on a solid angle grid. The source function $S$ is a more general form of Kirchoff's law. It describes the ratio of emission and extinction of radiative energy and allows arbitrary contributions from thermal emission as well as scattering contributions. In fact, the complexity of the model and the solution of the RTE depend strongly on the source function we choose to accurately describe the current radiation transfer problem {\bf (e.g. local thermodynamic equilibrium (LTE), non-LTE (NLTE)}, grey or non-grey, anisotropy, dust continuum radiation, line transfer, etc.).
Describing the complete radiation field would require 6 dimensions, which makes it an extremely challenging task to compute and store a 3D solution. In order to handle the radiation field numerically, the intensity is only computed on the fly and accumulated in form of the solid angle averaged mean intensity
\begin{equation}
  J = \frac{1}{4\pi}\oint_{4\pi} I \; d\Omega . 
  \label{equ:mean}
\end{equation}
The mean intensity is the zeroth moment of the specific intensity and closely related to the radiative energy density $E_r = \frac{4\pi J}{c}$. Depending on the model setup, the source function usually depends on the radiation field itself and the mean intensity becomes a part of the source function, which makes the RTE an integro-differential equation. In order to find a self-consistent solution, one has to invoke an iteration scheme of some form.
Formally however, the RTE from Equation (\ref{equ:time_independent_rte}) can be solved by the {\it formal solution}:
\begin{equation}
I(\tau_2) = I(\tau_1)\; e^{(\tau_2-\tau_1)} + 
     \int_{\tau_1}^{\tau_2} S(\tau)\; e^{(\tau_2-\tau)} d\tau. 
\label{equ:fs}
\end{equation}
The formal solution describes the intensity propagation along a line element with the optical depth $\Delta \tau = \tau_2 - \tau_1$. It contains the incoming intensity $I(\tau_1)$, which is partially extinct and additional energy from emission processes.
The RTE and the formal solution are linear in the intensity and allow us to split the radiation field in as many components as the solution method requires. This is a crucial part in our approach of solving the RTE on a decomposed computational domain such as the adaptive grid embedded in the FLASH code.

\subsection{Numerical Radiative Transfer}
\label{numerical_rt}
\begin{figure}
	\begin{minipage}[H]{0.49\textwidth}
	\includegraphics[width=\textwidth]{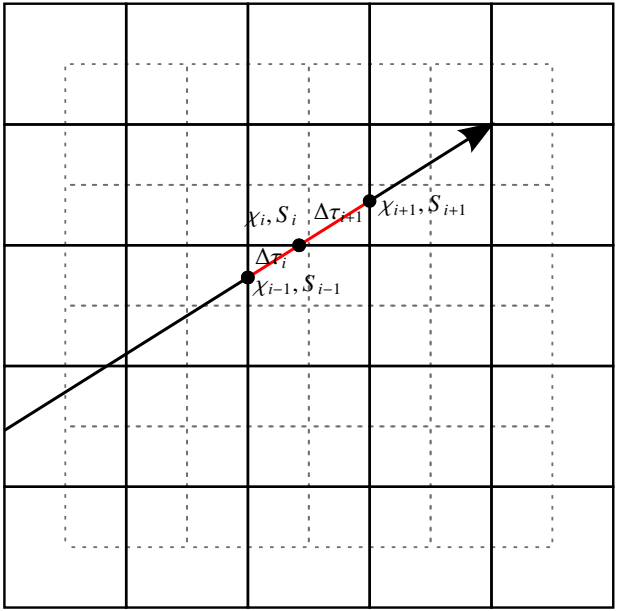}
        \put(-135,120){\small $\chi_{i-1},S_{i-1}$} 
        \put(-135,130){\small $\Delta \tau_{i}$} 
        \put(-113,147){\small $\Delta \tau_{i+1}$} 
        \put(-138,146){\small $\chi_i,S_i$} 
        \put(-88,150){\small $\chi_{i+1},S_{i+1}$} 
        \captionsetup{font=small}
	\caption{The staggered RT-grid (solid lines) defined by the cell centers of the underlying finite-volume hydro-grid (dashed lines); a long characteristic at an arbitrary direction is shown, which integrates the RTE for the hydro cell-center at the upper right corner of the domain.}
	\label{rt_grid} 
	\end{minipage}
\end{figure}

Integrating the RTE along a set of rays of different directions $\mathbf{n}$ using Equation (\ref{equ:fs}) is based on the {\it method of characteristics}. It was first introduced into the radiation transfer community by \citet{Olson86}. The RTE is integrated for each cell in the computational domain and each direction by computing a stepwise formal solution along a ray, or {\it long characteristic} (LC), according to the discretised formal solution
\begin{align}
\label{discr_formal_solution}
I_i &= I_{i-1}\; \exp{(-\Delta \tau_{i-1})} + \Delta I_i,
\end{align}
where $\Delta \tau_i$ is the finite optical depth element given by a piecewise linear interpolation
\begin{equation*}
\Delta \tau_i = \frac{1}{2} (\chi_{i-1} + \chi_i) \Delta s.
\end{equation*}
$\chi_i$ is the opacity at the discretization point $s_i$ on the characteristic. $\Delta I_i$ is the discretised counterpart to the integral in the formal solution (Equation (\ref{equ:fs})) and is solved by either a linear or parabolic interpolation according to
\begin{equation}
\Delta I_i = \alpha_i S_{i-1} + \beta_i S_{i} + \gamma_i S_{i+1}.
\label{delta_I}
\end{equation}
The coefficients $\alpha_i$, $\beta_i$ and $\gamma_i$ depend on the optical depths between $s_{i-1}$, $s_i$ and $s_{i+1}$. They are given in \citet{Olson86}. Figure \ref{rt_grid} shows the geometrical situation of a characteristic passing through a homogenous grid at an arbitrary direction $\mathbf{n}_j(\theta,\phi)$.
{\bf Since the opacity and source function are stored in the cell centers of the finite volume FLASH grid (dashed lines), they are assumed to be constant inside the cell. However, since we use a parabolic interpolation \footnote[1]{A simple cell based linear interpolation leads to a loss of radiative energy in optically thick regimes, especially if iteration is required (e.g. Dullemond 2013, \url{http://www.ita.uni-heidelberg.de/~dullemond/lectures/radtrans\_2013/index.shtml}, Chapter 3.8.4 and 4.4.8} of the source function integral, we introduce a point-based RT grid which is based on the cell centers of the FLASH grid. These cell centers define the vertices from which we interpolate the values at the intersection points of the ray bilinearly. Consequently, the point-based RT grid is staggered by half a grid cell since the ray does not intersect with the grid faces of the finite volume grid but with the faces of the point based grid defined by the FLASH cell centers.}
The characteristic is traced on the RT-grid using the fast voxel traversal algorithm introduced by \citet{Amanatides87}. The opacity $\chi_i$ and the source function $S_i$ at the intersection points of the characteristic with the RT-grid are interpolated bilinearly from the adjacent vertices.\\
While the RTE is integrated along each direction $\mathbf{n}$, the mean intensity is computed by accumulating all intensities:
\begin{equation}
J = \frac{1}{4\pi} \; \sum_{\mathbf{n}} I(\mathbf{n}) \Delta \Omega,
\label{equ:mean_discr}
\end{equation}
which requires a discretization of the solid angle $\Omega$. If no information about the anisotropy of the
radiation field is available, one should choose a homogeneous discretization which is a non-trivial problem if one considers spherical coordinates on the unit sphere. 
For this purpose, we use the HEALPix (Hierarchical Equal Area isoLatitude Pixelization) 
scheme introduced by \citet{Gorski05}. HEALPix ensures an optimal discretization of the unit sphere 
into a number of finite solid angles $\Delta\Omega$. The discretization is based on 12 base pixels which
are subdivided depending on the required resolution level. Consequently, typical numbers of directions for
the integration of the specific intensity are $N_{\rm{pix}}=12, 48, 192$ or $768$ (\ref{sec:healpix})\\
Characteristics based radiative transfer is the attempt to approximate the radiative interaction of each cell with each other cell in the computational domain. Although the method of long characteristics is very accurate in doing this, it is rather inefficient as it requires to shoot a large number of rays for each cell to sample the radiation field accurately in 3D. An alternative is to use a short characteristics (SC) approach, in which only neighbouring cells are used to interpolate incoming intensities from different directions. This requires to sweep the cells in an ordered fashion to make sure that all intensities which are required for interpolation are available. The SC approach introduces numerical diffusion because of the large number of interpolations involved but reduces the cost of the RT calculations by a factor of $n_c$ (the total number of cells involved). Either way, the method of characteristics must invoke a raytracer, which samples radiative interactions between arbitrary regions in the computational domain. 
\subsection{Raytracing on the decomposed AMR Grid}
The parallel design of the FLASH framework, in which our solver is currently implemented, forbids to trace rays over the entire domain as it is necessary for the method of characteristics. FLASH invokes PARAMESH (\citet{PARAMESH99}), and lately also the CHOMBO library\footnote[2]{\url{https://seesar.lbl.gov/anag/chombo/}}, for implementing an adaptive mesh refinement (AMR) grid. Paramesh uses a {\it block} structured AMR mesh, in which the fundamental data structure is a block containing cells which are logically indexed by a coordinate triple (i,j,k). The entire computational domain consists of a number of blocks of different physical sizes ordered hierarchically in an octree data structure. Blocks at the bottom of the tree structure, called {\it leaf blocks}, contain valid data and they cover the entire physical size of the computational domain. FLASH allows for massively parallel computation by invoking the Message Passing Interface (MPI) for the communication of ghost cell information between the blocks. Optimal load balancing is guaranteed by splitting the AMR tree equally between all available MPI tasks to ensure that each task receives more or less the same number of leaf blocks. E.g., the AMR tree of a star formation simulation typically requires more than 10 levels of spatial resolution with up to several $10^5$ blocks each containing $8^3$ cells, which is only made possible (in terms of cpu-time and memory requirements) by the parallelization described above.\\ 
The method of characteristics stays in direct contrast to the spatial parallelization of the AMR grid (Figure \ref{fig:two_cpus}). However, in order to account for non-local coupling of the radiation field, we adapt a raytracing technique originally developed by \citet{Rijkhorst06} and improved by \citet{Peters10a}, which uses a combination of local long characteristics and a global "short-characteristics-like" interpolation of outgoing intensities from the decomposed domains of the AMR grid. The basic idea is to split the radiation field in two components:
\begin{itemize}
\item 1. A local component which uses long characteristics to compute only radiative contributions to both the cells inside a block as well as contributions that leave the block ({\it face values}). The computation is done in parallel and in accordance with the design of the block structured AMR grid.
\item 2. A global component which is computed by communicating and accumulating face values (see Figure \ref{fig:grid_step}). This step invokes raytracing over the block structure of the AMR tree and a linear interpolation of face values very similar to the SC method (but on the level of subdomains). After the communication of face values and the tree hierarchy, this step is also done in parallel. 
\end{itemize}
This approach, called {\it hybrid characteristics}, only needs to communicate the face values of the blocks and information about the AMR tree hierarchy but no 3D data. By this, the amount of communicated data is reduced significantly. {Originally, this method was developed by \citet{Rijkhorst06} to compute column densities only with respect to point sources for UV ionization. The original method requires to communicate the whole AMR tree structure at the highest level of spatial resolution during the raytracing step on the AMR block structure. This stands in contrast to the parallel design of the FLASH code and restricts the available range of refinement levels of the AMR tree substantially because of the large memory overhead. \citet{Peters10a} add some major improvements to the algorithm by introducing a walk through the AMR tree, which only requires the communication of basic AMR information and conserves the idea of shared memory parallelization.\\
{\bf However, the method was, originally, restricted to compute column densities along rays which originate at a certain point in the grid and used to represent, e.g., a stellar source. For this work, we removed this restriction and implemented a radiative transfer framework which is able to compute the radiation field independently from any point source by solving the RTE for large sets of characteristics along parallel rays. By combining our improvements with the original method, the solver can not only account for the primary emission by point sources (as in \citet{Rijkhorst06} and \citet{Peters10a}) but also for the reemitted, diffuse component of the radiation field.} Figure \ref{fig:testTrans1} shows a 2D example of a simple test setup with an irradiated central density clump using AMR. From the figure we can see the ability of the method to create sharp shadows and to transport incoming radiation over the entire domain. 
\begin{figure*}
\begin{minipage}{0.49\textwidth} 
   \centering
   \begin{tikzpicture}
   \draw[step=.5cm,black,very thin] (-2,-1.0) grid (-0,0);
   \draw[step=1.0cm,black,very thin] (-2,-2) grid (-0,0);
   \draw[step=1.0cm,black,very thin] (0,-2) grid (1,0);
   \draw[step=.5cm,black,very thin] (-2,0) grid (-1,2);
   \draw[step=.25cm,black,very thin] (-1.0,0.0) grid (0.0,1.0);
   \draw[black,very thick] (-2,-2) rectangle (2,2);
   \usetikzlibrary{arrows}
   \draw[dashed,red,->] (0.18,0.0) -- (2.0,1.5);
   \draw[blue] (-2.0,-1.8) -- (0.18,0.0);
   \end{tikzpicture}
   \caption*{processor 0}
\end{minipage}
\begin{minipage}{0.49\textwidth} 
   \centering
   \begin{tikzpicture}
   \draw[step=1.0cm,black,very thin] (1,-2) grid (2,0);
   \draw[step=.5cm,black,very thin] (0,0) grid (2,2);
   \draw[step=.25cm,black,very thin] (-1.0,1.0) grid (1.0,2.0);
   \draw[step=1.0cm,black,very thin] (-1.0,1.0) grid (1.0,2.0);
   \draw[black,very thick] (-2,-2) rectangle (2,2);
   \usetikzlibrary{arrows}
   \draw[blue,->] (0.18,0.0) -- (2.0,1.5);
   \draw[dashed,red] (-2.0,-1.8) -- (0.18,0.0);
   \end{tikzpicture}
   \caption*{processor 1}
\end{minipage}
\captionsetup{font=small}
\caption{\bf Example for a 2D AMR grid distributed over two processors without shared memory. The bold rectangle shows the boundary of the
whole computational domain. Thin lines show the leaf blocks at different refinment levels that make up the whole subdomain a processor
is assigned to. Raytracing through the domain is obviously restricted to the subdomain each processor is assigned to.}
\label{fig:two_cpus}
\end{figure*}
\begin{figure*}
\centering
  \begin{subfigure}{0.48\textwidth}
    \caption{local contributions}
    \showone{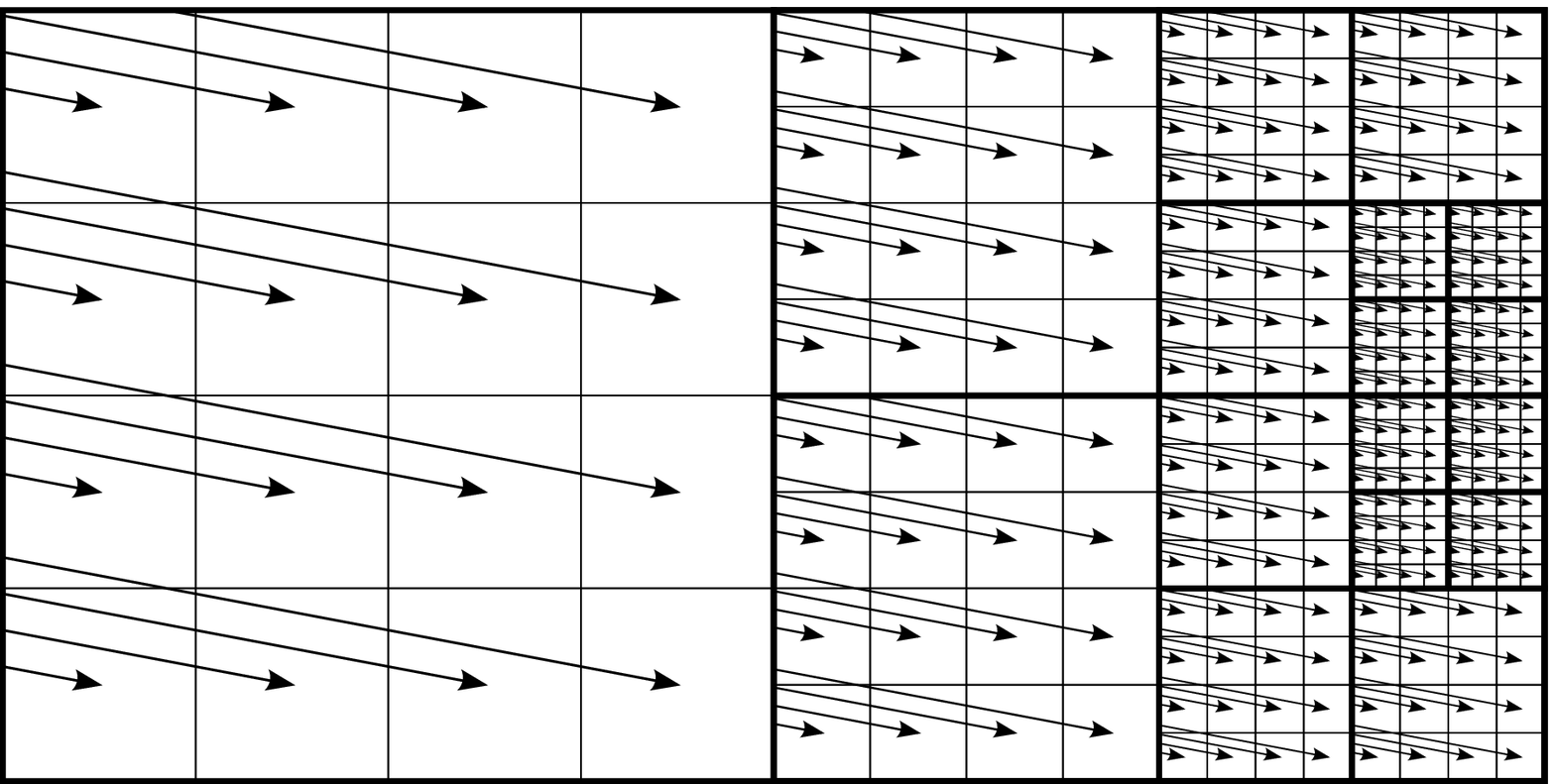}  
    \label{local_lc}
  \end{subfigure}
  \hspace{3mm}
  \begin{subfigure}{0.48\textwidth}
    \caption{face values}
    \showone{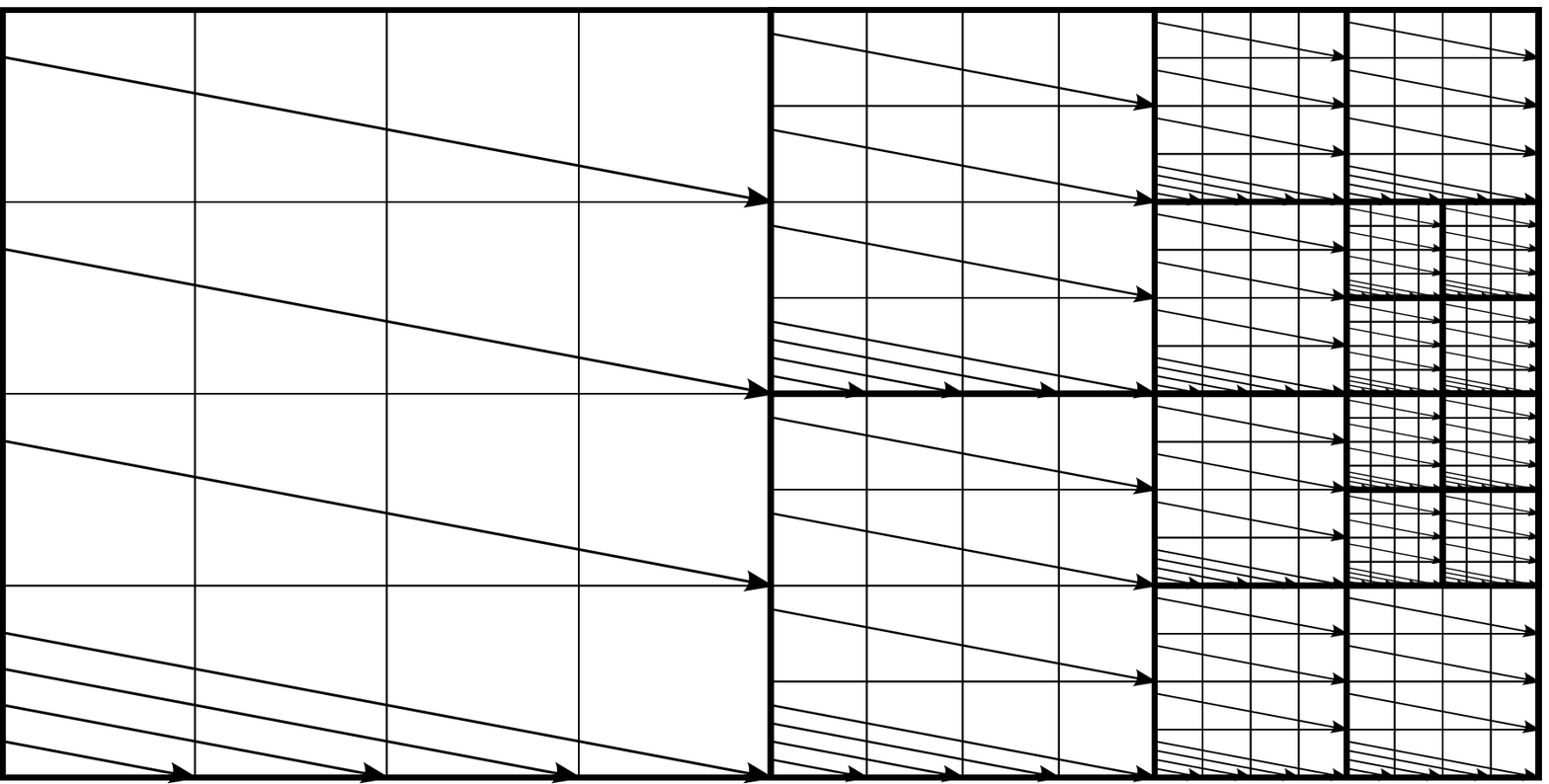}
    \label{face} 
  \end{subfigure}
  \begin{subfigure}{0.48\textwidth}
    \caption{interpolation}
    \showone{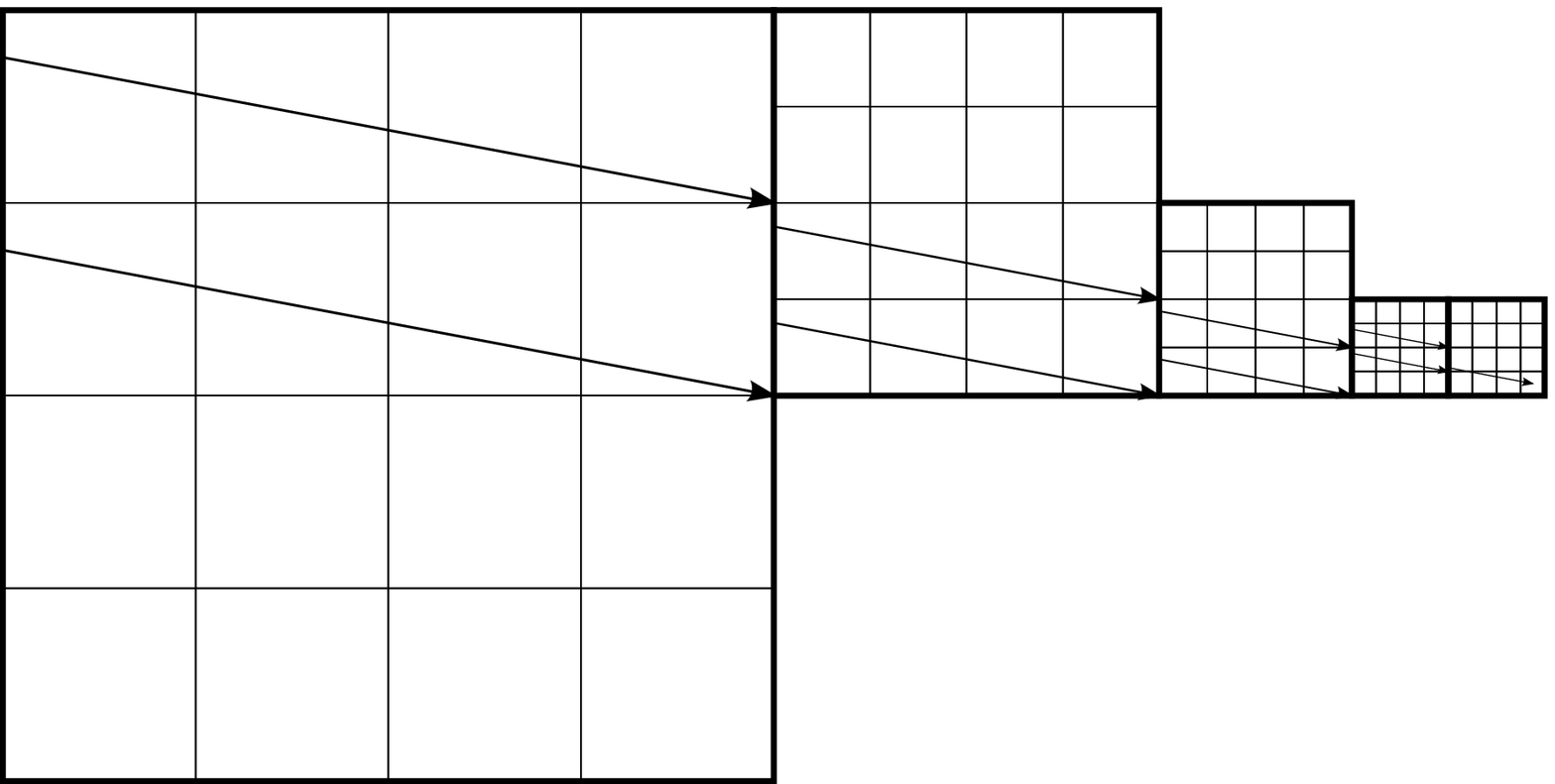}
    \label{interpol} 
  \end{subfigure}
  \linethickness{0.5mm}
  \thicklines
  \color{red}
  \put(0,0){\qbezier(10,5)(90,5)(90,70)}
  \put(10,5){\vector(-1,0){10}}
  \put(5,40){\text{communication}}
\captionsetup{font=small}
 \caption{\bf The basic steps of the hybrid characteristics method for parallel rays (compare to \citet{Rijkhorst06}) in a 2D AMR domain that is refined from left to right. 
Bold lines show the boundaries of the patches at different AMR levels (in FLASH, these patches are called {\it blocks} and are distribted equally over the available number of MPI tasks). 
In this example, each block contains 4x4 cells (indicated by thin lines).
a) Local contributions as calculated with long characteristics.
b) The outgoing face values which are communicated. In fact, we communicate all face values even though they might be part of the same subdomain of a certain MPI process since we need them
in the following interpolation step.
c) Example for the linear interpolation of face values for a particular target cell after the communication step. The linear interpolation requires to weight the face values from two rays.
The weights depend on where a certain ray segment starts at the boundary of a block (which is a 4x4 cell patch at a certain refinement level) or subdomain respectively.}
\label{fig:grid_step}
\end{figure*}
\begin{figure*}
\centering
\includegraphics[width=0.49\textwidth]{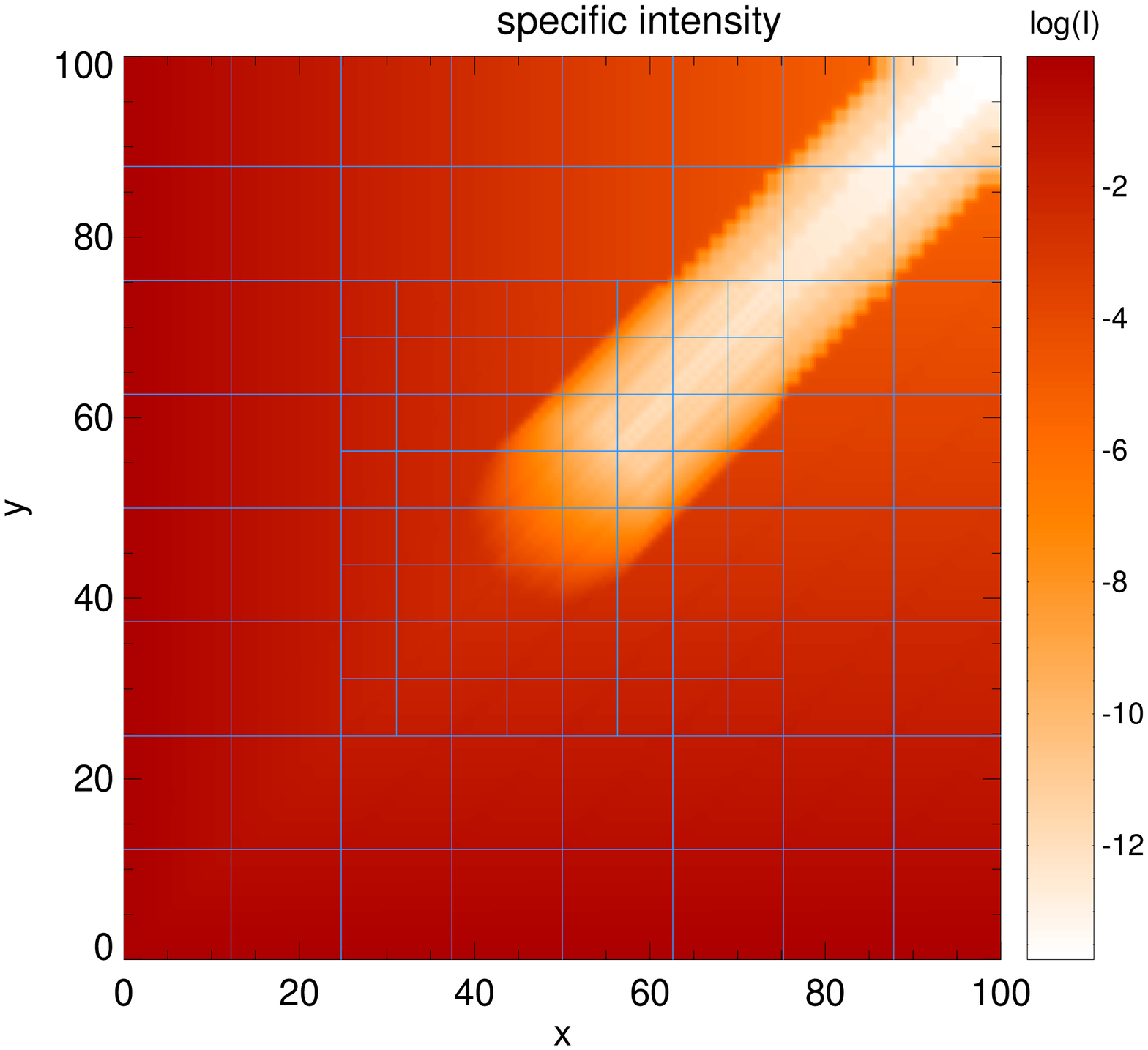}  
\includegraphics[width=0.49\textwidth]{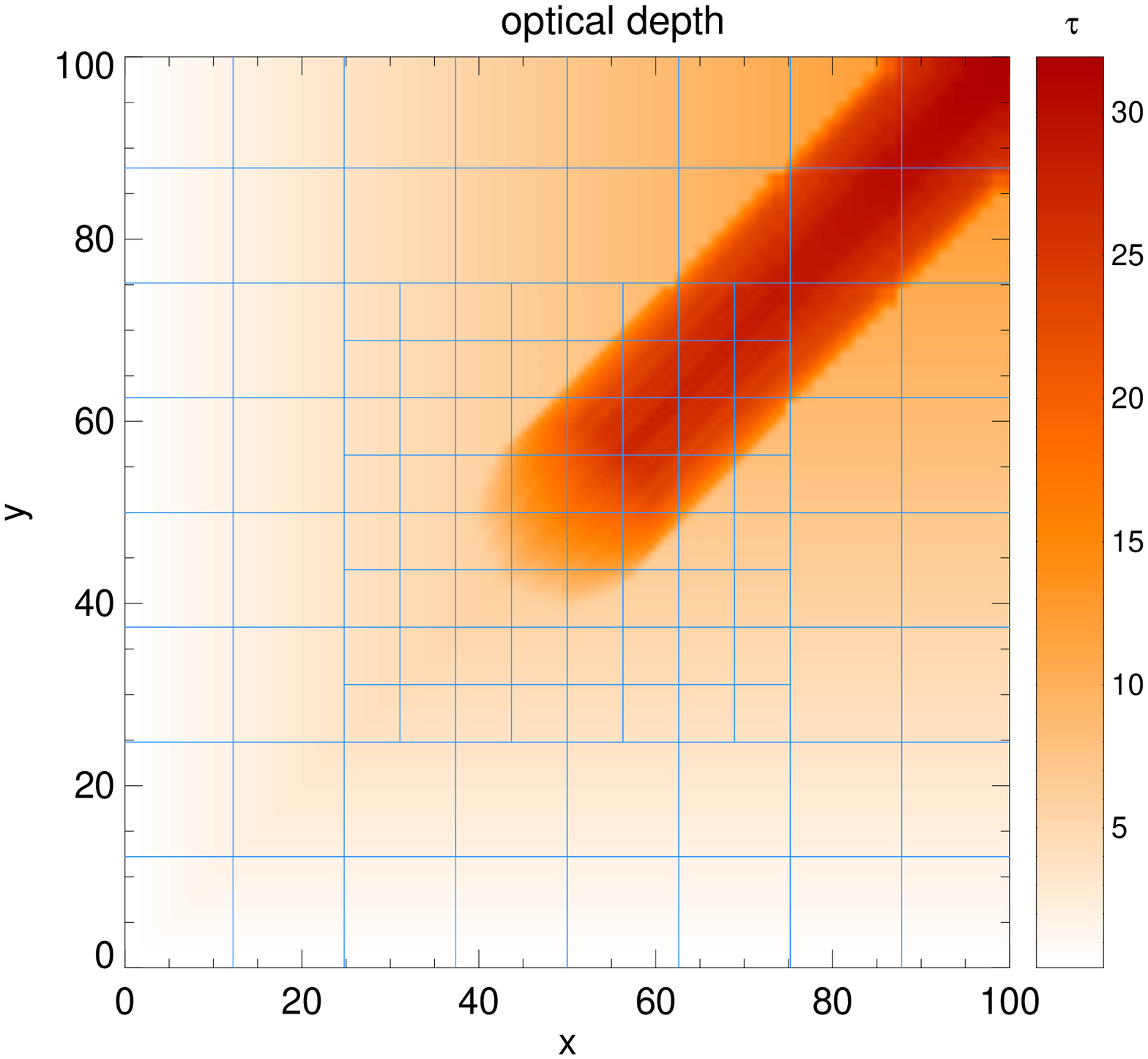}  
\captionsetup{font=small}
\caption{The specific intensity and the optical depth computed diagonally in the xy-plane with a central density clump. The source function is set to unity at the left and bottom outermost boundaries and zero everywhere else. The opacity of the central clump is one order of magnitude higher than the ambient opacity. The grid indicates the block structure of the AMR grid, units are arbitrary.}
\label{fig:testTrans1}
\end{figure*}
\subsection{Coupling to the FLASH Code}
Since our method is implemented in the FLASH framework, it is straightforward to couple the radiative transfer module to the 
hydrodynamical and MHD modules of the FLASH code. The coupling is done by accounting for radiative emission 
and absorption processes, which are determined by the thermal emission opacity $\chi_e=\kappa_e\rho$ and the thermal absorption 
opacity $\chi_a=\kappa_a\rho$. The opacities are calculated from mass specific cross sections $\kappa_e$ and $\kappa_a$. 
Note that the total extinction coefficient $\chi$, which is used 
for the solution of the RTE, may include an additional scattering opacity $\chi_s$ and therefore $\chi=\chi_a+\chi_s$. 
The coupling of both the radiation and the MHD solver is achieved by computing a source term according to \citet{Mihalas84} which 
describes the total net gain or loss of energy due to radiative heating and cooling. It reads
\begin{equation}
Q_{rad} = 4\pi \chi (J-S) = 4\pi \chi_a (J-B).
\label{equ:qrad}
\end{equation}
This source term is computed from the time-independent solution of the radiation field as described in the previous sections and
it is coupled to the MHD integrator in an operator splitting step. Hence, the set of 
compressible MHD equations in dimensionless form including gravitation and radiative energy exchange are those of 
\begin{align}
&\text{continuity} \nonumber \\
&\frac{\partial \rho}{\partial t} + \nabla \cdot (\rho \mathbf{v}) = 0, \\
\label{equ:MHD_momentum} 
&\text{momentum conservation} \nonumber \\
&\frac{\partial (\rho\mathbf{v})}{\partial t} + \nabla \cdot (\rho \mathbf{v}\otimes\mathbf{v}+p_*\mathsf{1}-\mathbf{B}\otimes\mathbf{B}) = \rho\mathbf{g} \\
\label{equ:MHD_energy} 
&\text{energy conservation} \nonumber \\
&\frac{\partial E}{\partial t} + \nabla \cdot (\mathbf{v}(E + p_*)-\mathbf{B}(\mathbf{v} \cdot \mathbf{B})) = \rho\mathbf{v}\cdot\mathbf{g} + Q_{\rm rad}, \\
&\text{and the induction equation} \nonumber \\
&\frac{\partial \mathbf{B}}{\partial t} - \nabla \times (\mathbf{v} \times \mathbf{B}) = 0,
\end{align}
with the gas velocity field $\mathbf{v}$, the magnetic field vector $\mathbf{B}$ and the gravitational acceleration $\mathbf{g}$. $p_*$ 
is the total pressure and $E$ the total energy density of a fluid element containing magnetic contributions according to
\begin{align}
\centering
p_* &= p + \frac{B^2}{2}, \\
E &= \frac{1}{2} \rho u^2 + e_{\rm int} \rho + \frac{B^2}{2},
\end{align}
with the gas density $\rho$, the thermal pressure $p$ and the internal specific energy $e_{\rm int}$. $\mathsf{1}$ denotes the unity matrix.
The MHD equations are closed by an ideal gas equation of state (EOS) which relates the internal energy
to the thermal gas pressure according to
\beq
p = (\gamma-1) \rho e_{\rm int}.
\eeq 
We assume $\gamma=5/3$ which corresponds to a mono atomic (hydrogen) gas. The temperature is also related to
the internal energy by:
\beq
e_{\rm int} = (\gamma-1)^{-1} \frac{k_b}{\mu m_{\rm p}} T,
\eeq
where $k_b$ is the Boltzmann constant, $m_{\rm p}$ the proton mass and $\mu$ the mean molecular weight of the gas.\\
Note that we solve the equations of MHD and RT successively by an operator splitting step and not simultaneously. Furthermore, for the following test cases, the thermal pressure dominates the hydrodynamics and it is several orders of magnitude larger than the radiation pressure, which we therefore neglect in the momentum equation (\ref{equ:MHD_momentum}). However, since our method explicitly computes the angular dependency of the radiation field, it is straightforward to couple it into the MHD equations.
\subsubsection{Choosing the Time Step}
\label{sec:timestep}
The current coupling is done by an update of the internal gas energy $e_{\rm int}$ and temperature $T$ respectively. 
Since we solve the time independent RTE, there is no update of the radiative energy or the source function during the solution of Euler's equations. Instead
this is done in the following time step when the gas quantities have been updated. The update of the internal energy is done explicitly by
\beq
\Delta e_{\rm int} = \Delta t \; Q_{\rm rad}.
\label{equ:qrad_dt}
\eeq
Due to the explicit update, we have to make some restrictions on the time step. The radiation field does not have an explicit influence
on the CFL time step since the energy update is done after the solution of the MHD equations.
Instead, we compute a cooling time step which is chosen if it is shorter than any other time step from a FLASH module. The cooling time step 
is chosen so that the energy contribution $\Delta e_{\rm int}$ does not exceed a fixed percentage of the internal energy.
Otherwise, if the time step $\Delta t$ is chosen too large, the total radiative energy could become negative (e.g., $\Delta e_r > e$). 
This leads to the following time step restriction:
\beq
\Delta t_{\rm c} = \min\left(\frac{e_{\rm int}}{|\Delta e_{\rm int}|}\right)_i \, k_{\rm c} \, \Delta t_{\rm CFL},
\eeq
{\bf where $k_{c}$ determines how much change in the internal energy is allowed, $\Delta t_{\rm CFL}$ is the CFL time step, and $i$
denotes the indices of all grid cells in the computational domain. Because of the explicit energy update, the cooling time step is usually 
shorter than the CFL time step. So far, there is no subcycling involved and the FLASH code chooses a global minimum time step from all physics modules involved (including, e.g., self-gravity).}
The cooling time step highly depends on the absorption coefficient $\chi$ since it determines the optical depth of the medium and 
how much radiation is absorbed and emitted during a single time step. 
Typically the choice of $0.2 > k_{c} > 0.01$ is convenient as it produces accurate results (Section \ref{sec:diffusion_tests}) and time steps about one oder of 
magnitude lower than the CFL time step.
\subsection{The Lambda Formalism}
\label{sec:Lambda}
Computing the radiation field in the form of the mean intensity in Equation (\ref{equ:mean_discr}) requires a formal solution of the
RTE in the way described above. Usually, this task is described in a rather compact form by using the {\it Lambda Operator}: 
\begin{equation}
J = \Lambda[S].
\label{equ:lambda_step}
\end{equation}
Formally, the Lambda operator for {\it one cell} in the computational domain contains the radiative contributions from
each other cell. The construction of the operator would require to explicitly calculate the radiative coupling
between a cell and each other cell. But this is far too costly concerning computation time and memory requirements.
Instead, we do not construct the operator but we approximate the Lambda step from Equation (\ref{equ:lambda_step}) by using
the formal solution from Equation (\ref{equ:fs}) to compute the radiation field $J$ from the source function $S$
in the way described above. The accuracy of this approximation in a 2D or 3D computation depends crucially on the angular 
resolution, since it determines whether we actually "hit" each other cell during the angular integration of the mean intensity or not.
{\bf We avoid this problem partly by calculating the radiation from point sources (e.g. a stellar source) explicitly for each cell by combining our method with 
the original hybrid characteristics method by \citet{Rijkhorst06} and \citet{Peters10a}.}
However, the Lambda step from Equation \ref{equ:lambda_step} requires that we know the source function in advance.
If we take the temperature from FLASH's hydro solver, we can compute the source function simply as being $S=B(T)$ then solve for the radiation field, 
couple it back to the hydro solver and we are done. This approach assumes the gas to be in a state of thermodynamical equilibrium but this is, of course, not always the case.
If the radiation field is decoupled from the gas temperature, we do not know the source function in advance. The solution then requires some kind of iterative procedure to account for 
the non-local coupling of the radiation field with the gas. In the theory of radiative transfer, this iterative method is called {\it Lambda iteration}, 
which requires iterating over Equation (\ref{equ:lambda_step}) until a self-consistent solution for $J(S)$ is found. Strictly speaking, even in the LTE case with $S=B(T)$,
we have to iterate to find a temperature that is consistent with the internal energy of the gas since this determines the thermal emission. 
However, the Lambda iteration may need several hundreds of iteration steps, 
which is too costly and ineffective to be employed in a hydrodynamical simulation. One way of resolving this problem, is to partly solve Equation 
(\ref{equ:lambda_step}) analytically by splitting the Lambda operator. These approaches, called {\it Accelerated Lambda iteration} (ALI), 
have been investigated and used extensively in the stellar atmosphere community \citep[e.g.][]{Trujillo95}. 
We have implemented the most simple form of ALI, the local lambda operator, to solve radiative transfer problems even in regions of 
high optical depths and strong decoupling where the classical Lambda iteration usually fails (\ref{sec:ALI}).

\section{Tests}
\label{sec:tests}

In this section, we show test results from the implementation of our radiation solver. The tests include time independent
(Sections \ref{sec:1D_atmo} and \ref{sec:pascucci_benchmark}) as well as dynamical tests (Section \ref{sec:diffusion_tests}) in 1D and 3D.
We also show results from the combined FLASH/RT code in a series of 1D radiative shock calculations in Section \ref{sec:shocks}. 

\subsection{1D Atmosphere}
\label{sec:1D_atmo}

In the first test, we compute the radiation field in a grey, isothermal, scattering dominated 1D atmosphere. This test
is typically used to verify a radiation solver's iterative performance and accuracy in a non local thermodynamic equilibrium (NLTE) situation 
on a wide range of optical depths. It is also particularly useful to ensure that the solver accurately reproduces the diffusion limit in an 
optically thick regime, e.g., in the lower parts of the atmosphere. This test also requires the ALI method
since the classical Lambda iteration fails to reproduce the solution in the case of strong scattering contributions.
The amount of scattered radiation is quantified by the ratio of the thermal absorption coefficient to the total extinction coefficient
according to
\beq
\epsilon = \frac{\chi_a}{\chi_a + \chi_s},
\eeq
where we neglect the frequency dependence and $\epsilon$ is the {\it photon destruction probability}. The grey source function
in the atmosphere contains a thermal part and a scattering contribution, and it reads
\begin{align}
S &= \frac{\eta}{\chi} = \frac{\eta_s}{\chi_a + \chi_s} + \frac{\eta_e}{\chi_a + \chi_s},\\[8pt]
  &= (1-\epsilon) J + \epsilon B,
\label{equ:isotropic_source}
\end{align}
{\bf where we defined $J=\eta_s/\chi_s$, the thermal emission is $B=\eta_e/\chi_a$, and $\eta_s$ and $\eta_e$ denote the scattering and the thermal
emissivity respectively.} Since the atmosphere is isothermal, we
assume that we know the temperature and normalize it so that $B=1$. The crucial part in this test is to find the source function
which has to be consistent with the mean intensity $J$ which is
\beq
J = \frac{1}{2} (I_-+I_+), 
\eeq
where $I_-$ and $I_+$ are the down and upward (2 stream solution) integrated specific intensity respectively.
Since we assume a uniform mass specific opacity $\kappa$ and constant temperature $T$, the intensity is only a function of optical depth $d\tau=\chi dz$,
thermal emission $B$ and the photon destruction probability $\epsilon$. The mean intensity is then given by the analytic solution
\beq
J = B \left(1-\frac{\exp(-\sqrt{\epsilon}\tau)}{1+\sqrt{\epsilon}}\right).
\eeq
The density $\rho$ of the model increases exponentially with distance from the upper boundary and we assume that $\chi\propto\rho$ but with $\epsilon$ being constant.
There is no incoming radiation at the upper boundary of the atmosphere at $\tau=0$ while at the lower boundary the incoming radiation is $I=B$. 
The resulting model atmosphere provides an exponentially varying optical depth $\tau$ which resolves the transition region from the optically thick 
inner LTE-regions to the optically thin NLTE-regions at the outer boundary. We test the solver for a wide range of photon destruction probabilities 
from $\epsilon=10^{-1}$ to $10^{-8}$. The domain consists of 8 subdomains each containing 8 cells which results in a total spatial resolution of 64 cells.
Figure $\ref{fig:pp_atmo_1D}$ shows the results. In the outer optically thin parts of the atmosphere, the scattering contribution in the source function becomes dominant 
since radiation leaves the atmosphere. The numerical solution is in excellent agreement with the analytic solution.\\
\bef
\showone{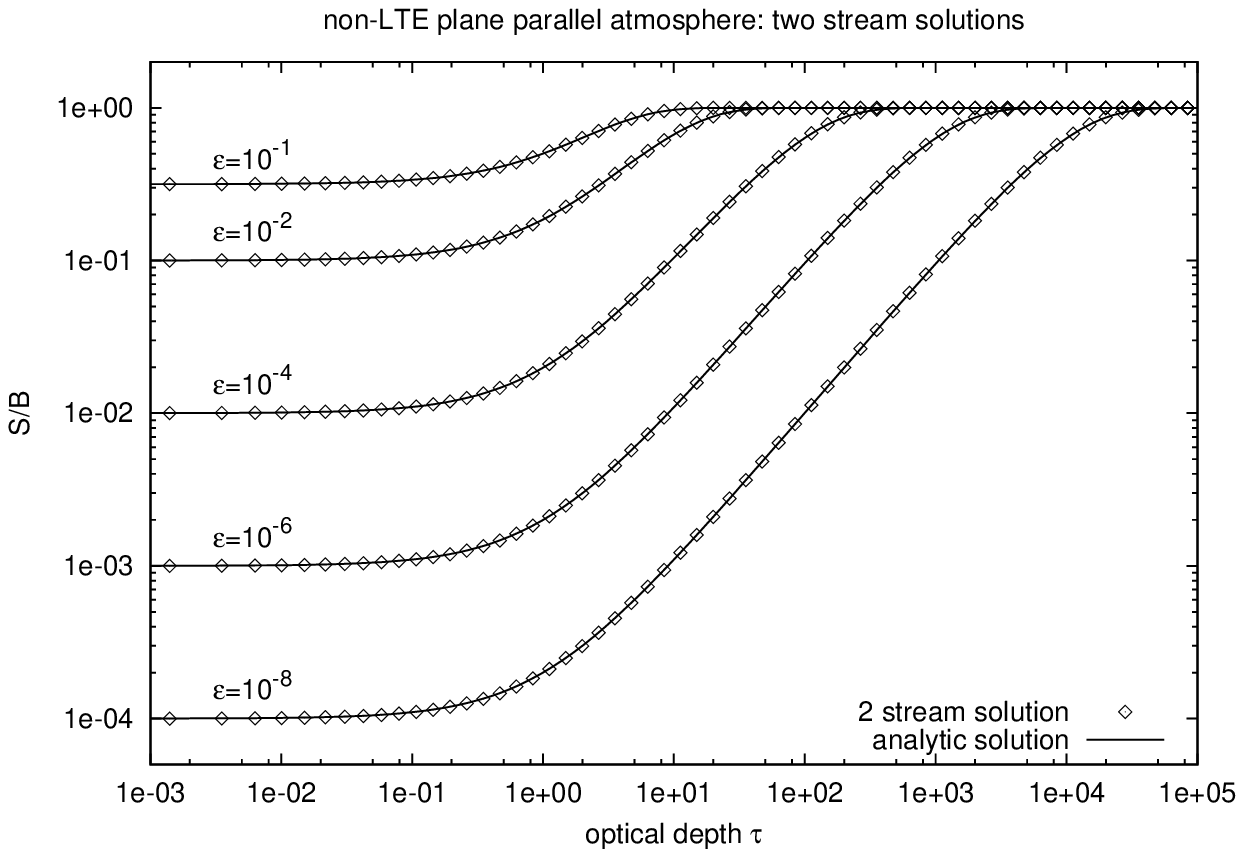}
\captionsetup{font=small}
\caption{Scattering dominated 1D atmosphere problem. The solutions from the radiation solver (symbols) are compared to the analytic
solutions (lines) for five different photon destruction probabilities.}
\label{fig:pp_atmo_1D}
\eef
\subsection{Hydrostatic Protostellar Disc}
\label{sec:pascucci_benchmark}
Cosmic dust is one of the most important constituents of the ISM. By mass, it makes up only a small fraction of typically about 1\%, 
but dust has important radiative and chemical properties. Dust particles have strong continuum opacities which are highly frequency-dependent. 
Especially in the optical regime, dust absorbs light much more efficiently than in the infrared regime. That is why young 
protostars, which are surrounded by gaseous and dusty envelopes, are difficult to observe in the visible wavelengths but require
infrared observations.
Thermal absorption and reemission of radiation by dust (a process called {\it reprocession}) strongly determines the thermodynamical properties of a 
protostellar disc, especially in those regions where the disc is opaque to direct stellar radiation and dominated by thermal reemission of dust molecules. This is mainly the case near 
the equator of the disc because radial optical depths with respect to the central star are typically much larger than unity ($\tau_*\gg 1$). 
Therefore, modelling the temperature structure requires diffuse radiative transfer to be taken into account.\\
In this test setup, we combine emission from a point source with the solution of the RTE. The goal is to determine
the self-consistent temperature structure of the gas in a protostellar disc. The setup is based on the benchmark by \citet{Pascucci04}, which is based
on the theoretical work by \citet{Chiang97}. We compare our solutions from a 3D calculation with the results from the Monte Carlo
radiative transfer code RADMC-3D \citep{RADMC3D}.

\subsubsection{Thermal Radiative Transfer}
A protostellar disc combines optically thick and thin regimes, which requires the computation of primary stellar radiation and the thermal reemission 
from dust molecules in the disc. Our approach follows the idea of splitting the radiation field in two components handling each separately. 
Following the work of \citet{Dullemond02b}, the first component we compute is the extinct stellar flux. This can be handled by using the original 
hybrid characteristics method, which computes the optical depth with respect to a central stellar source ($\tau_*$). 
The extinct stellar flux $F_*$ at a distance $r$ from a star of luminosity $L_*$ is given by
\begin{equation}
  F_*(\mathbf{x}) = \frac{L_*}{4\pi r^2} \exp(-\tau_*(\mathbf{x})),
  \label{equ:disc_stellar_flux}
\end{equation}
assuming that the star can be approximated as a point source. The amount of energy per unit time that is absorbed this way is determined 
by the absorption coefficient $\chi$ and given by
\begin{equation}
  Q(\mathbf{x}) = \chi F_{*}(\mathbf{x}).
  \label{equ:disc_absorbed_energy}
\end{equation}
The reemitted radiation of the dust grains in the disc is treated as a secondary component of the radiation field.
This component is computed with the general transfer algorithm using parallel rays. Assuming LTE, 
the dust grains will acquire an equilibrium temperature such that they emit exactly the same amount of energy which they absorb
\begin{equation}
  \frac{\sigma}{\pi}\chi T^4 = \frac{Q}{4\pi} + \chi \frac{1}{4\pi}\oint_{4\pi} I \; d\Omega . 
  \label{equ:disc_temperature}
\end{equation}
where $I$ is the specific intensity of the reprocessed radiation field. The first term in Equation (\ref{equ:disc_temperature}) accounts for the direct stellar radiation 
while the second term describes the energy of the reprocessed radiation field. The transfer equation for reemitted radiation by dust grains is
\begin{equation}
\frac{\partial I}{\partial \tau} = \frac{\sigma}{\pi} T^4 - I.
\label{equ:rte_dust}
\end{equation}
Hence, the source function in this setup is the frequency integrated thermal emission from dust grains $S=\frac{\sigma_{\rm SB}}{\pi}T^4$.
The task at hand is to find a temperature that is consistent with the coupled set of Equations (\ref{equ:disc_temperature}) and (\ref{equ:rte_dust}). 
This is done by iterating the equations until convergence is reached (Lambda-iteration).

\subsubsection{The Disc Model}
For the simulation setup we are following the benchmark test of \citet{Pascucci04} which resembles a {\it flared disc} \citep{Chiang97}. 
The idea is to define a radial gas surface density distribution and to assume that the vertical density structure is only determined by the 
hydrostatic equilibrium in the vertical direction. The gas density distribution is given by
\begin{equation}
\begin{aligned}
  \rho(r,z) &= \rho_0 \; f_1(r) \; f_2(r), \\
  f_1(r) &= \left(\frac{r}{r_d}\right)^{-1.0},\\
  f_2(r) &= \exp\left(-\frac{\pi}{4} \; \left(\frac{z}{h(r)}\right)^{2}\right),\\
  h(r)   &= z_d \left(\frac{r}{r_d}\right)^{1.125},
  \label{equ:pascucci_setup}
\end{aligned}
\end{equation}
where $r$ is the radial distance in the disc midplane, $z$ is the height above the disc, and $\rho_0$ is the gas density in the midplane at $r=r_d=500\,\au$ and $z=0$. 
The outer disc radius is defined by $r_{\rm out}=1000\,\AU=2\,r_d$ and we crop the disc at an inner radius $r=r_{\rm in}$.
$z_d$ determines the height of the disc which we choose to be $0.25\,r_d$ consistent with \citet{Pascucci04}. We choose the central source to have
solar properties with $M_*=1\,\Msol, R_*=1\,\Rsol$ and $T_*=5800\,\rm{K}$. We use a grey opacity at the visible wavelength of $\lambda=550$ nm from the opacity
tables used in \citet{Pascucci04} ($\kappa = 8736\,\rm{cm}^2\,\g^{-1}$).\\
In contrast to the Pascucci benchmark, we perform our calculations in 3D instead of 2D. Therefore, we can not directly compare our results to the Pascucci results 
but instead use the results from RADMC-3D as a reference. 
We perform calculations for three cases of $\rho_0$ so that the total radial optical depth of the disc in the midplane varies from $\tau_{\rm disc}=1$, $\tau_{\rm disc}=10$ 
and $\tau_{\rm disc}=100$. We do not explicitly distinguish between a gas and a dust temperature and 
assume both to be tightly coupled and the dust density is defined as a fixed fraction of the gas density ($1\%$). The 
dust density distribution through the xz-midplane of the disc setup for the optically thin case ($\tau_{\rm disc}=1$) is shown in Figure \ref{fig:pascucci_setup}.\\
The linear spatial resolution varies over 4 refinement levels from $\Delta x = 31.25\,\AU$ in the outer regions to $\Delta x = 1.953\,\AU$ in the center of the disc.
The solid angle integration is performed using 768 directions (nSide=8).
\bef
\includegraphics[width=\linewidth]{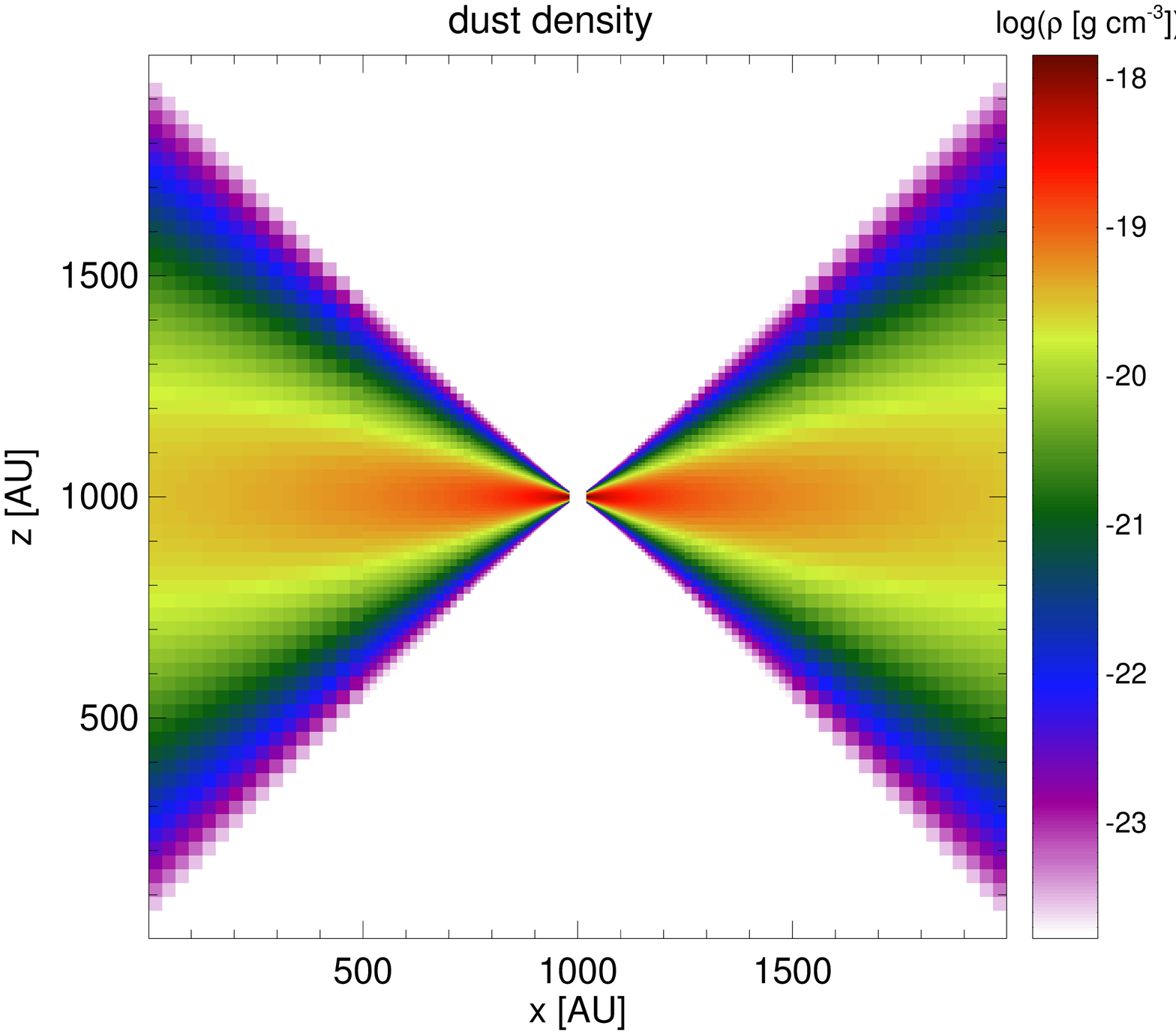}
\captionsetup{font=small}
\caption{The dust density in the xz-midplane for the Pascucci benchmark for a total optical depth of $\tau_{\rm disc}=1.$}
\label{fig:pascucci_setup}
\eef
\subsubsection{Results}
The resulting temperature structures and averaged midplane profiles are shown in Figure \ref{fig:pascucci_results}. As it turns out, the accuracy of the
solution is very sensible to the spatial resolution of the inner edge of the disc at $r=r_{\rm in}$ which is a result of discretizing the inner circular
rim on a Cartesian grid. Therefore, we increase the inner radius from $r_{\rm in}=10\,\AU, 20\,\AU$ to $40\,\AU$ for the three different setups
to guarantee sufficient resolution at the point where the disc becomes optically thick.\\
In the optically thin case ($\tau_{\rm disc}=1$), the midplane temperature is almost entirely dominated by the direct illumination of 
the central source. In the optically thick cases, the midplane temperature is dominated by the reprocessed radiation from dust in the photosphere of the disc, 
which is directly illuminated by the central source. At the point where the disc becomes optically thick, a bump emerges in the temperature profile since
the dust distribution becomes dense enough to absorb a considerable amount of radiation from the central source.
Our results are in excellent agreement with the reference computed by RADMC-3D and within the $10$\% range of the results from the different codes 
used for the \citet{Pascucci04} benchmark.\\
{\bf However, the temperature structure in the left panel of Figure \ref{fig:pascucci_results} is sensitive to the angular resolution. 
Although the raytracer takes care of the known primary stellar radiation, the solution in the outer regions depend on whether
the reprocessed radiation from the hot inner rim is accounted for correctly. Especially in the optically thick case ($\tau_{\rm disc}=100$), single rays become
visible in the temperature strucutre even for a large number of directions ($N_{\rm{pix}}=768$) since not each cell is correctly connected to the hot inner rim in
terms of radiative exchange. A larger number of directions is very costly, but an alternative is to also model the emission from such "hot spots" as a part of the primary emission. The problem
is to identify these hot spots in the domain since they can not be represented by point sources or sink particles. But once these regions are identified, their
emission can be handled by an inverse raytracer similar to the approach for point sources \citep{Peters10a}. However, this approach requires an adaptive angular grid while
our approach is only capable of using a homogenous angular resolution at the moment (\ref{sec:healpix}).}
\begin{figure*}
\begin{minipage}{0.46\textwidth}
\centering
\includegraphics[width=\linewidth]{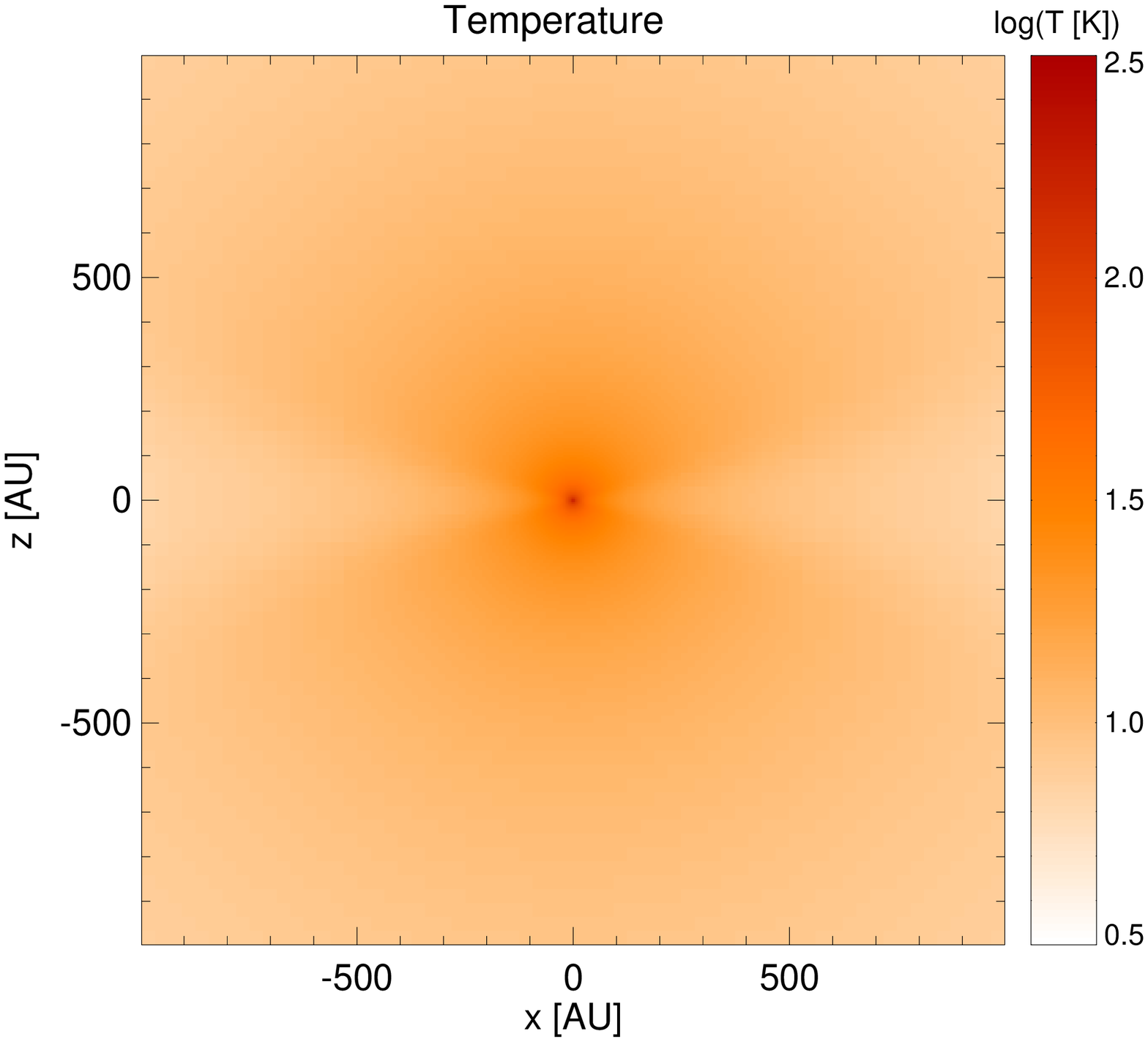}
\includegraphics[width=\linewidth]{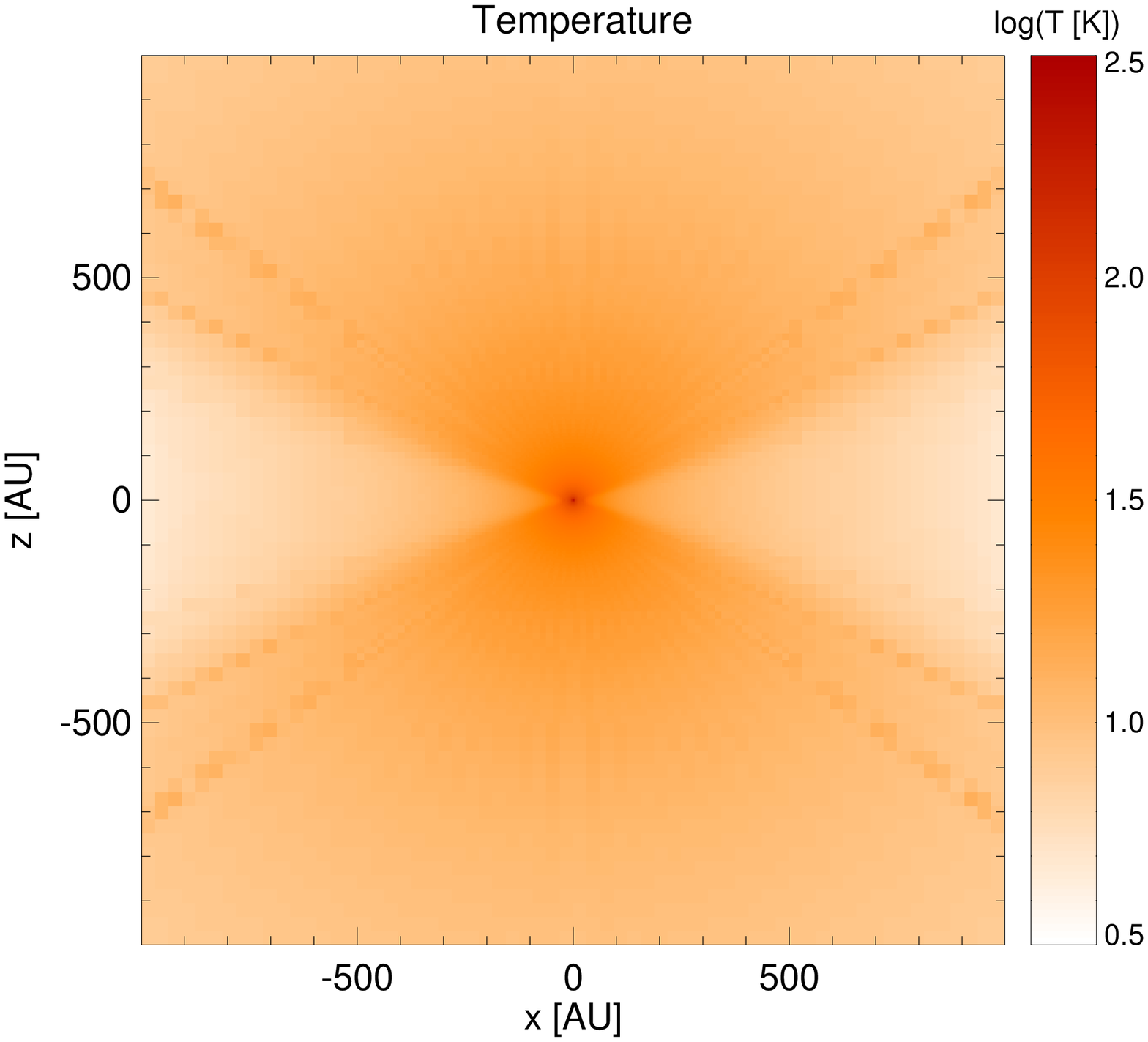}
\includegraphics[width=\linewidth]{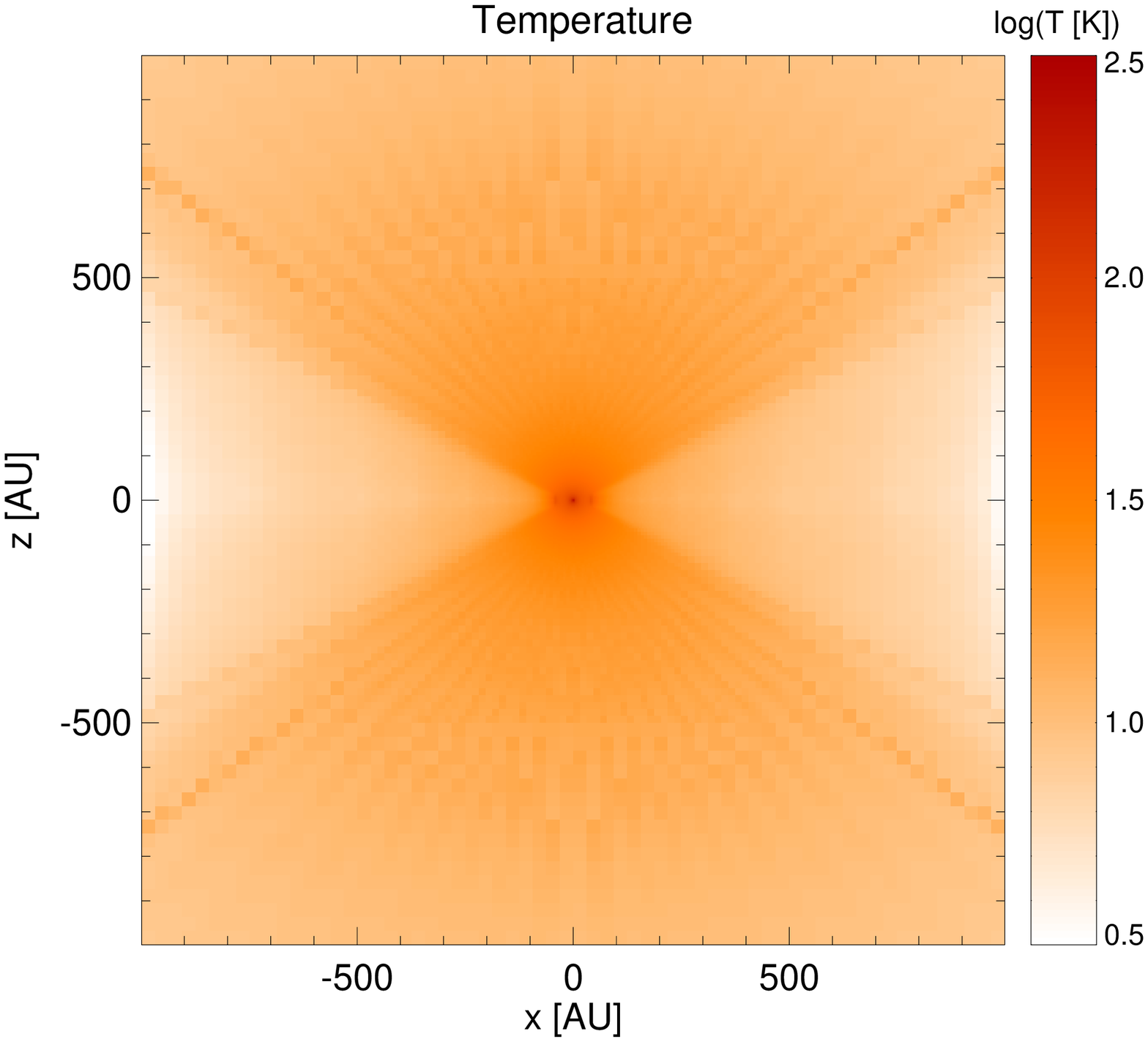}
\end{minipage}
\begin{minipage}{0.46\textwidth}
\centering
\includegraphics[width=\linewidth]{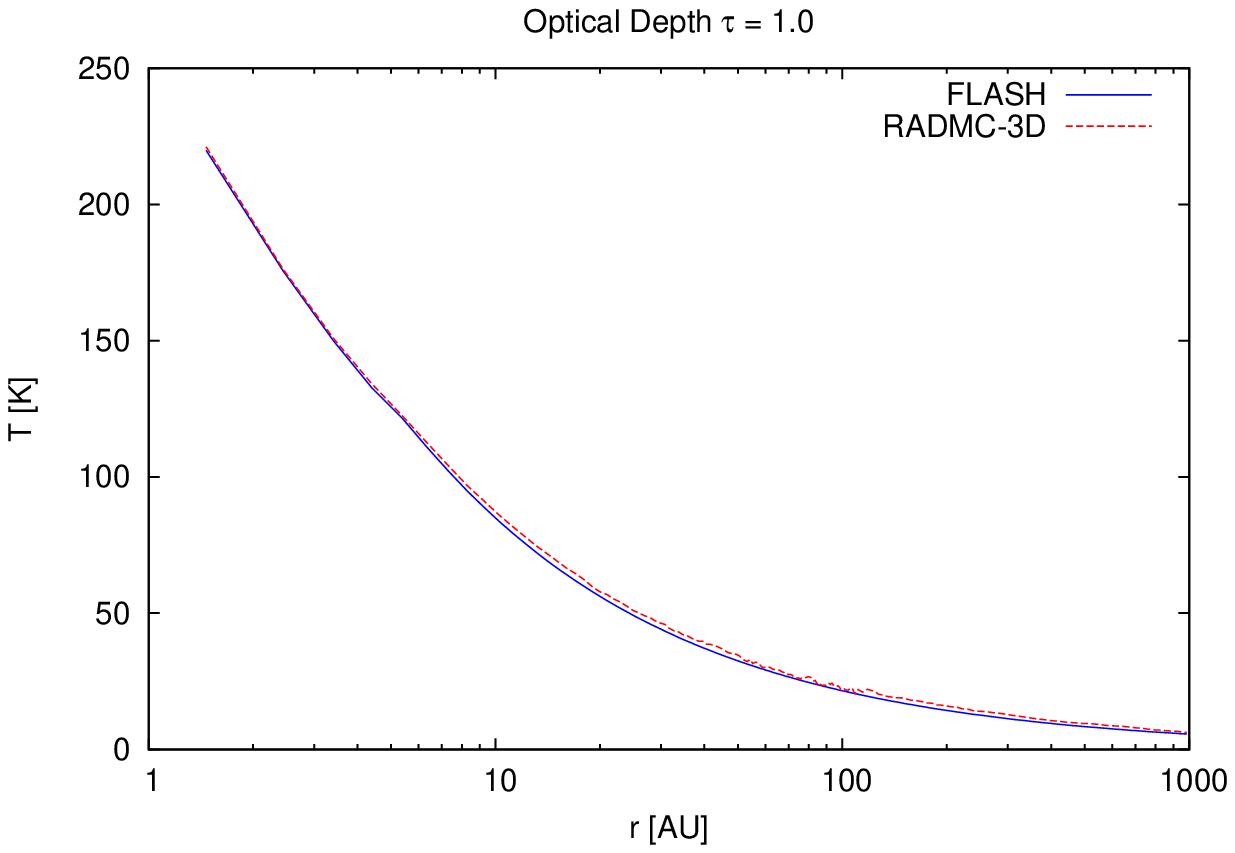}

 \vspace{12mm}

\includegraphics[width=\linewidth]{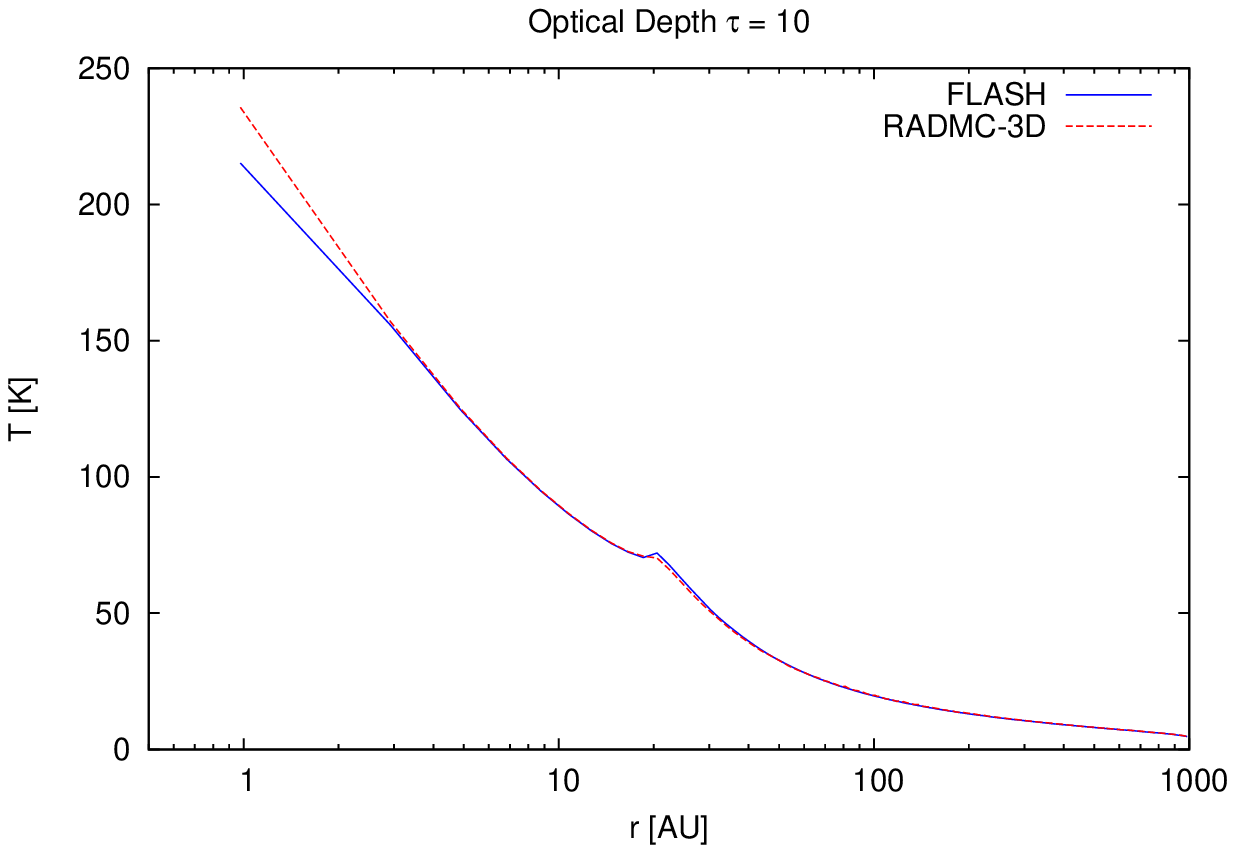}

 \vspace{12mm}

\includegraphics[width=\linewidth]{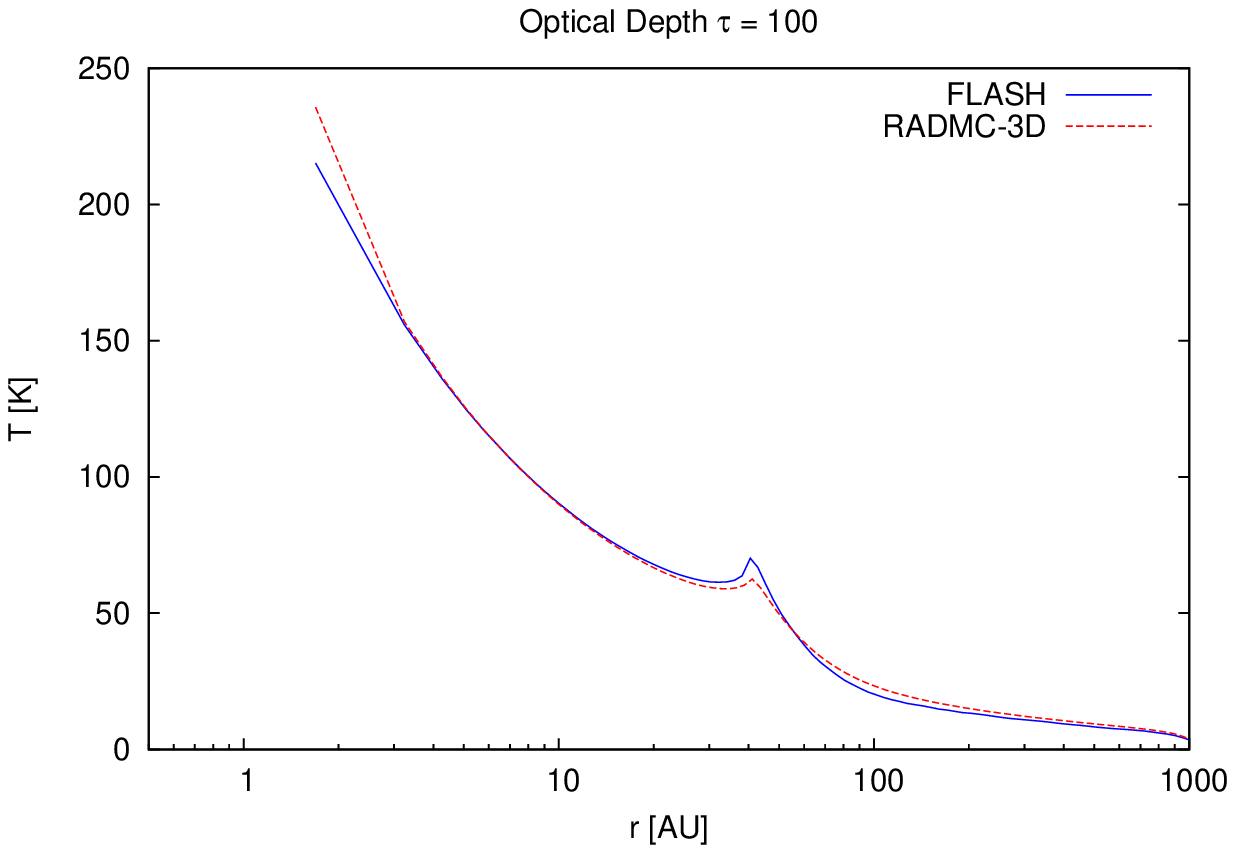}
\end{minipage}
\captionsetup{font=small}
\caption{The solutions of the \citet{Pascucci04} benchmark problem. Left column: the temperature structure through the xz-midplane of the disc for total 
radial optical depths of $\tau=1$ (top), $\tau=10$ (mid), and $\tau=100$ (bottom). Right column: averaged temperature profiles in the xy-midplane 
in comparison with the solutions of RADMC-3D. Solutions obtained with FLASH/RT use 768 directions for the angular discretization. 
Monte Carlo computations with RADMC-3D were performed using $10^8$ photon packages.}
\label{fig:pascucci_results}
\end{figure*}

\subsection{Diffusion Test}
\label{sec:diffusion_tests}
In this section, we show results from a time dependent radiative transfer calculation. 
Solving the time dependent RTE on the timescale of the speed of light would lead to time steps far too small for the use in a hydrodynamical simulation on astrophysical scales. 
However, since we are not interested in the dynamics of the propagation of the radiation itself but in its contribution to the energy budget of the gas, we assume the hydrodynamical
timescale to be much larger than the timescale on which radiation is transported. This means that the radiation field emerges instantaneously everywhere, and we assume 
the solution of the time independent RTE as being convenient. Consequently, the time dependence of the radiation field originates exclusively from the
coupling to the FLASH code using the energy source term from Equation (\ref{equ:qrad}).\\
In this section, we show results from testing the evolution of the source term by following the propagation of the radiation field in a highly opaque medium.
The source function is updated using a simple forward Euler time integration of the energy source term. Since the radiation field shows a diffusion like evolution in the limit of high optical depths,
we compare our numerical solution to the analytic solution of the diffusion equation.
%
\subsubsection{Setup}
In this test, we investigate the ability of our solver to follow the flux of radiative energy into a highly opaque medium.
In this case, the propagation of the radiation field can be described by the diffusion approximation, and we show that our approach reproduces the diffusion limit accurately. 
The diffusion approximation is derived from the moment equations of the RTE by invoking a closure relation between the radiative energy and the radiative pressure (e.g., the Eddington approximation). 
The radiation equations themselves then form a hyperbolic system. By neglecting the explicit time dependence of the radiative flux ${\bf F}$ and assuming that ${\bf F} \propto \nabla E_r$, 
the flux can be eliminated from the equations. The dynamics of the radiation field $J=cE_r/(4\pi)$ can then be described in a single equation, the diffusion equation \citep{Mihalas84}:
\begin{equation}
\frac{\partial J}{\partial t} - \nabla \left( \frac{c}{n\,\chi} \nabla J \right) = c \, \chi \, (S-J).
\label{equ:diffusion}
\end{equation}
where $n$ denotes the number of dimension. We do not allow any interaction of the radiation field with the hydrodynamics and only follow the propagation of the radiation field. 
Hence, the diffusion equation becomes homogeneous since $S=J$. In this case, the solution to the diffusion equation is described by the Gaussian function 
\begin{equation}
\label{equ:gaussian}
J_D(\mathbf{x},t) = \frac{J_0}{(4\pi D t)^{n/2}} \exp \left({-\frac{(\mathbf{x}-\mathbf{x}_0)^2}{4 D t}}\right),
\end{equation}
{\bf where $J_0$ denotes the initial mean intensity at $t=t_0$, $\mathbf{x}_0$ is its initial position, and $D=c/(\eta\chi)$ is the diffusion coefficient.} We use Equation (\ref{equ:gaussian}) to compute the initial conditions $J(\mathbf{x},t_0)$ for our test setup. 
We perform 1D and 3D computations with the initial conditions $J_0=J(\mathbf{x}_0,t_0)=10^5\,\rm{erg}\,\rm{s}^{-1}\,\rm{cm}^{-2}\,\rm{sr}^{-1}$ with $t_0=10^{-11}\,\s$ in 3D and $t_0=10^{-10}\,\s$ in 1D respectively, 
The center of the Gaussian is at $\mathbf{x}_0=0$, and we evolve the radiation field until $t=20\times t_0$ is reached. The length of the computational domain is $1\,\cm$ with a homogeneous density distribution 
of $\rho=1\,\g\,\cm^{-1}$ and a constant absorption coefficient $\kappa=1000\,\cm^{2}\,\g^{-1}$, which results in a highly optically thick medium. 
The temperature is constant and arbitrarily set to $T=1K$. Since no heating or cooling is allowed, there is no hydrodynamical response from the medium and all hydrodynamical quantities are constant in space and time.\\
Since we solve the time-independent RTE, there is a problem in reproducing the time-dependent term in Equation (\ref{equ:diffusion}). Strictly speaking, the static source function vanishes
since we do not couple the radiation field to the medium through which it propagates. Consequently, the mean intensity would also vanish in the time independent solution. However, the time dependence 
causes an effective contribution in the source function \citep[e.g.][]{Jack12} if the time discretization is carried out implicitly in the RTE (\ref{equ:rte}). This contribution depends 
on the specific intensities of the previous time step, is evolved through time, and describes the evolution of the radiation field. Since we do not account for this implicit contribution 
(which would require to store the complete scalar field of angle dependent specific intensities), we solve the problem by
operator splitting using the right-hand side of Equation (\ref{equ:diffusion}) to calculate the new source function at the following time step.
The evolution is done using a simple forward Euler time integration scheme of the form
\begin{equation}
\label{equ:source_euler}
S_n = S_{n-1} + \Delta t_n \, \chi \, c \, (J_{n-1}-S_{n-1}),
\end{equation}
where $\Delta t_n$ is the length of the current time step $n$. Therefore, the time step is restricted to be (Section \ref{sec:timestep})
\begin{equation}
\Delta t_n = \min\left(\frac{S_{n-1}}{(|S_{n-1}-S_{n-2}|)}\right)_i\,k_{\rm{rad}} \Delta t_{n-1},
\end{equation}
where the min function denotes the minimum change in the source function with time from all cells $i$ in the computational domain (FLASH does not support adaptive 
time stepping on a block level but rather uses a uniform global time step). $k_{\rm rad}$ limits the maximum change in the
source function, and we found a value of $k_{\rm rad}\approx0.1$ to give stable and accurate results in 3D.
\subsubsection{Results}
The results of the 1D solutions are shown in Figure \ref{fig:diffusion_1D}. We compare the numerical results with the analytic solution given by Equation (\ref{equ:gaussian}) 
and found our results to be within $1\%$ accuracy at a resolution larger than 32 cells. At the edge of the domain, the numerical solution deviates from the diffusion solution 
as radiative energy can leave the domain and we allow no irradiation from the outside. The results from the 3D computation are shown in Figure \ref{fig:diffusion_3D} and compared 
to the diffusion solution along the three main axes of the domain. In the 3D case, the domain is subdivided by the AMR grid into 4 blocks in each dimension. Each block contains $8^3$ 
cells giving a total linear resolution of 32 cells. In the 3D case, the setup consists of a Gaussian kernel around the origin which diffuses outwards. 
The solutions along each coordinate axis are obviously indistinguishable, emphasizing the accuracy and importance of the homogeneous angular HEALPix tessellation (\ref{sec:healpix}). 
The 3D computations were performed using 192 directions.
\begin{figure*}
\includegraphics[width=0.32\textwidth]{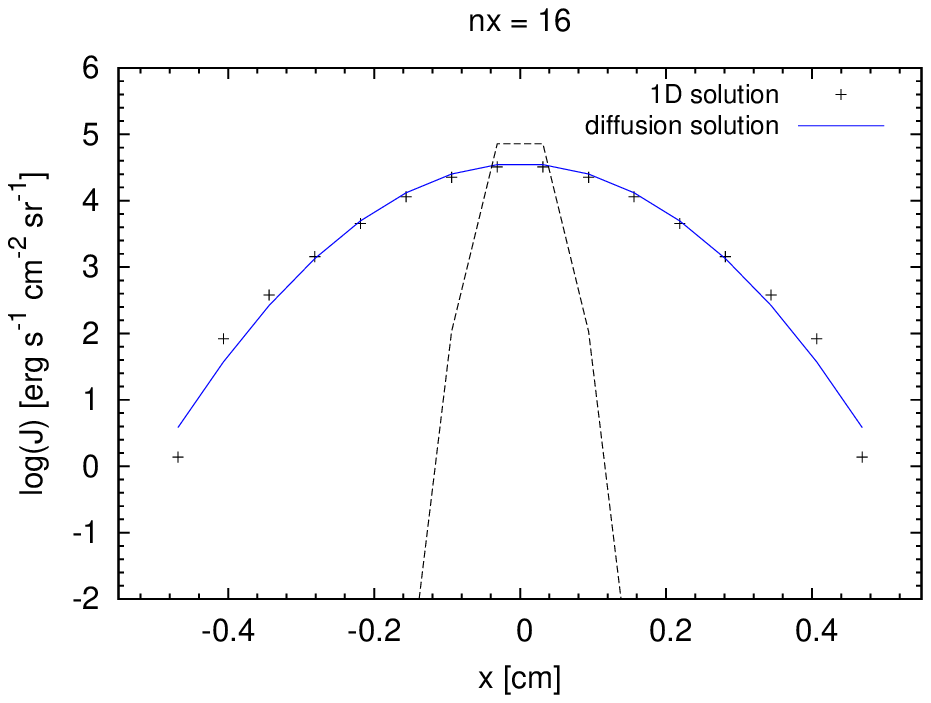}
\includegraphics[width=0.32\textwidth]{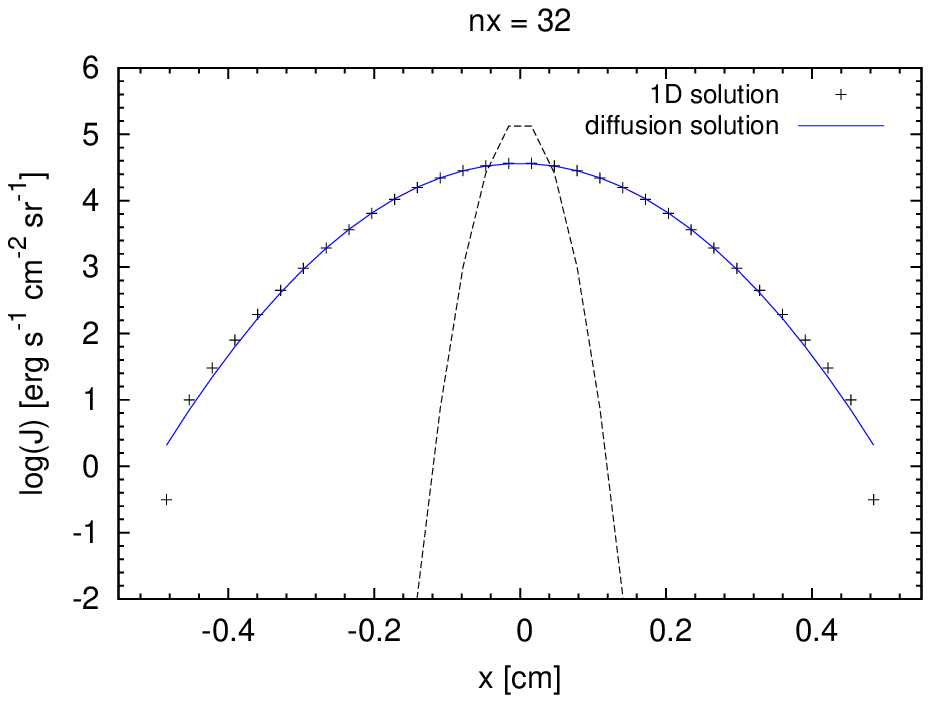}
\includegraphics[width=0.32\textwidth]{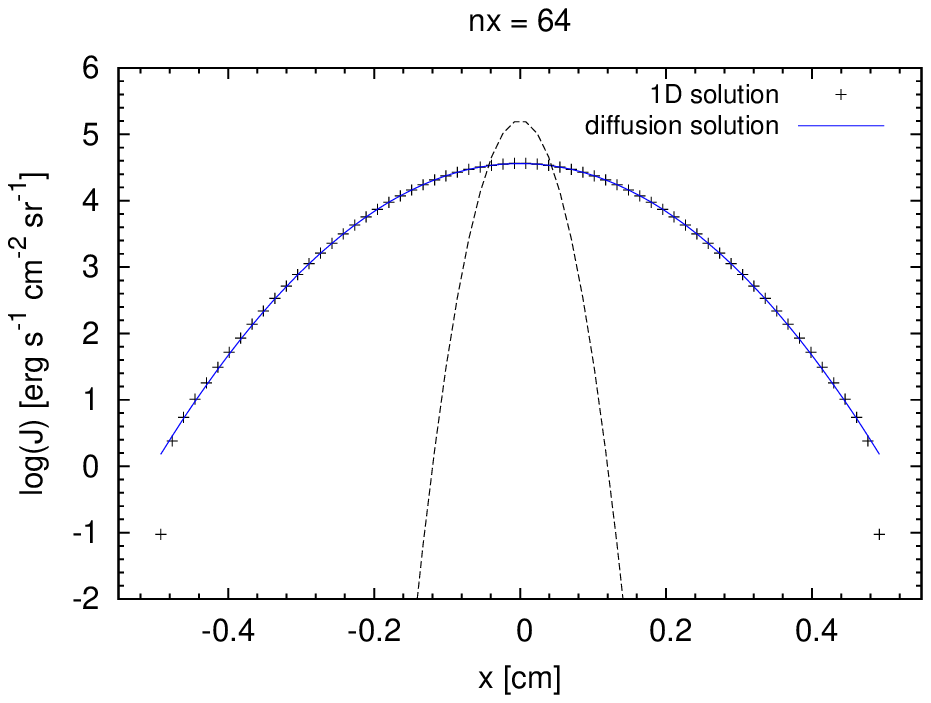}
\includegraphics[width=0.32\textwidth]{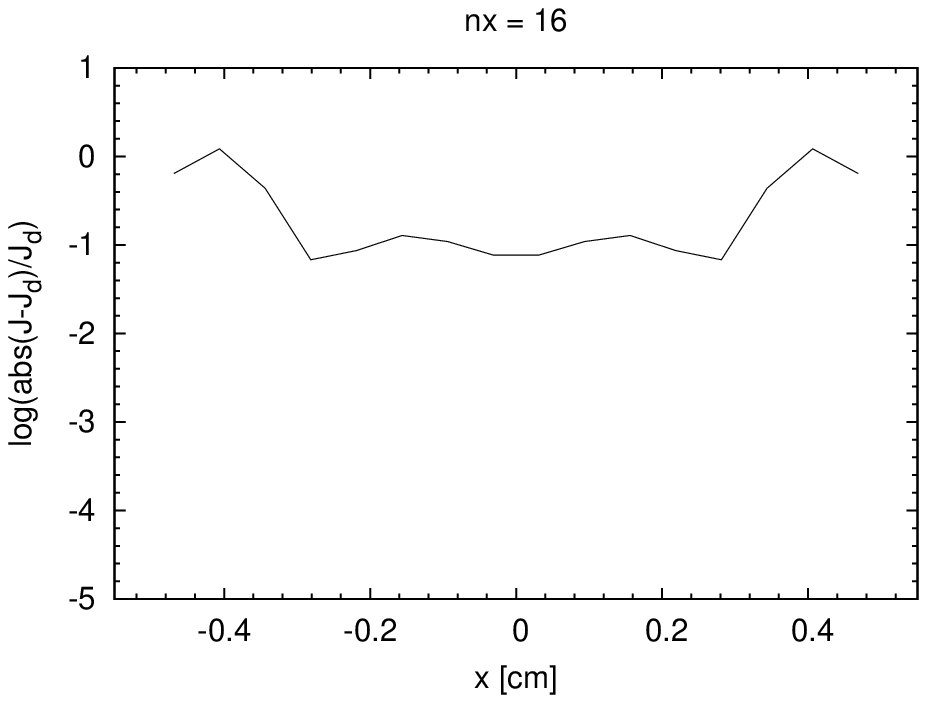}
\includegraphics[width=0.32\textwidth]{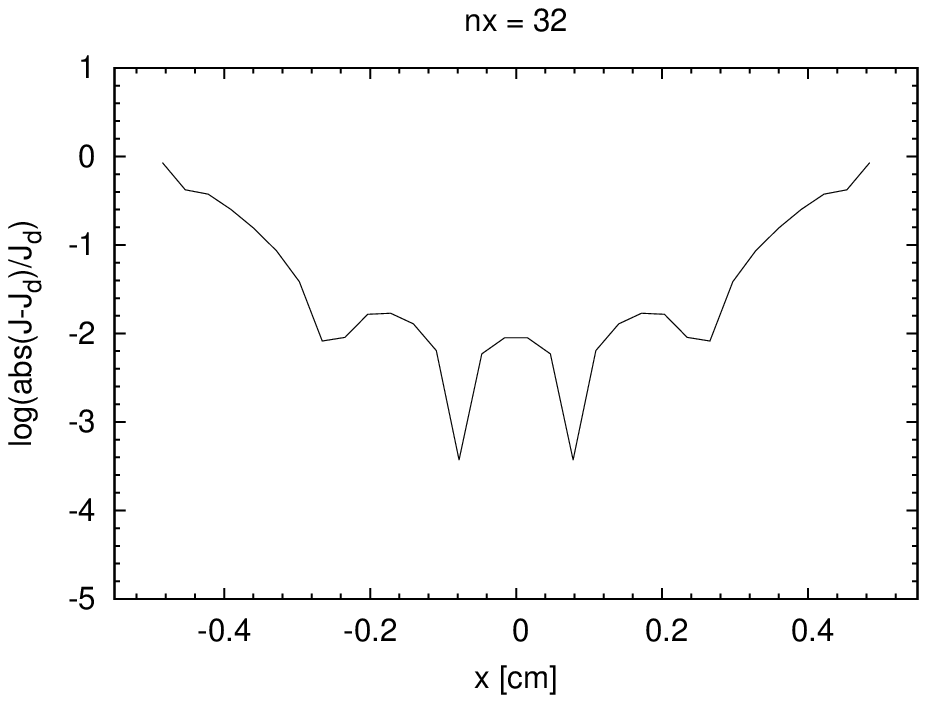}
\includegraphics[width=0.32\textwidth]{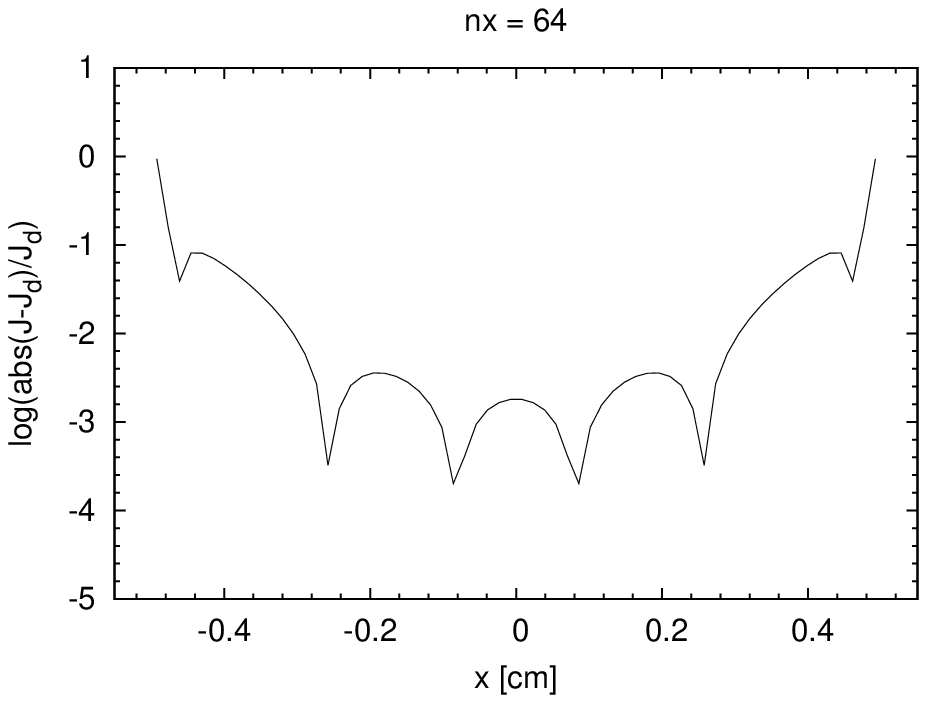}
\captionsetup{font=small}
\caption{Results of the 1D diffusion test for different homogeneous spatial resolutions, {\it top:} nx=16, {\it mid:} nx=32, {\it right:} nx=64. The dashed lines show the initial conditions at $t=t_0$ determined by the Gaussian solution of the diffusion equation. The initial radiative energy (symbols) is evolved and diffuses outwards until $t=20\times t_0$ is reached and compared to the analytical solution (solid lines) of the homogeneous diffusion equation. For a sufficient spatial resolution, the numerical solution stays within $1\%$ accuracy. At the edge of the domain, the radiation solver deviates from the diffusion solution as radiation leaves the domain while the diffusion solution is valid for an infinite domain.}
\label{fig:diffusion_1D}
\end{figure*}
\begin{figure*}
\includegraphics[width=0.32\textwidth]{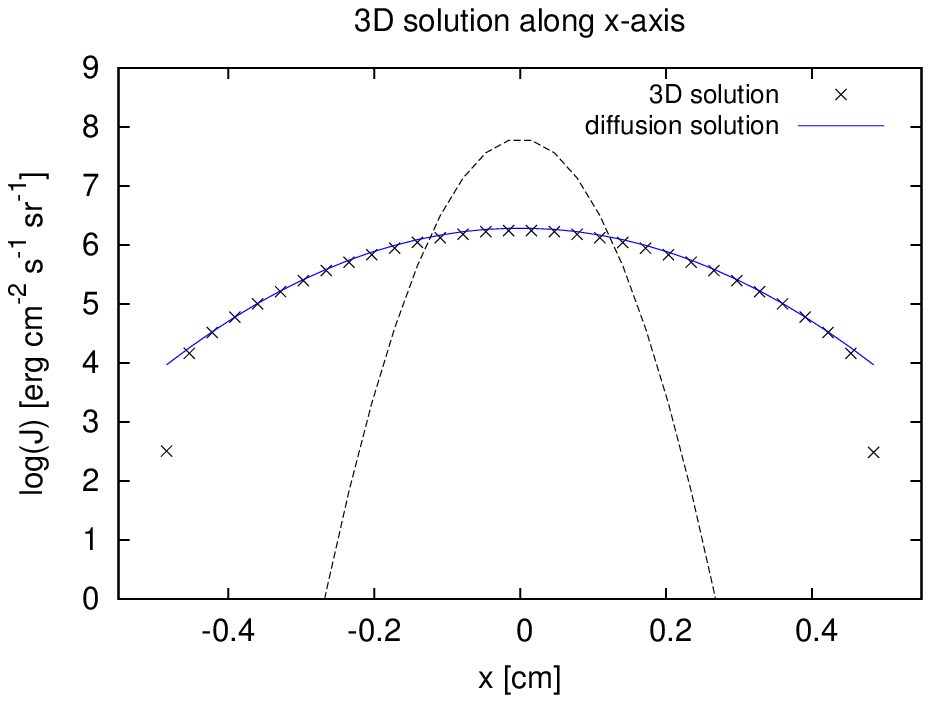}
\includegraphics[width=0.32\textwidth]{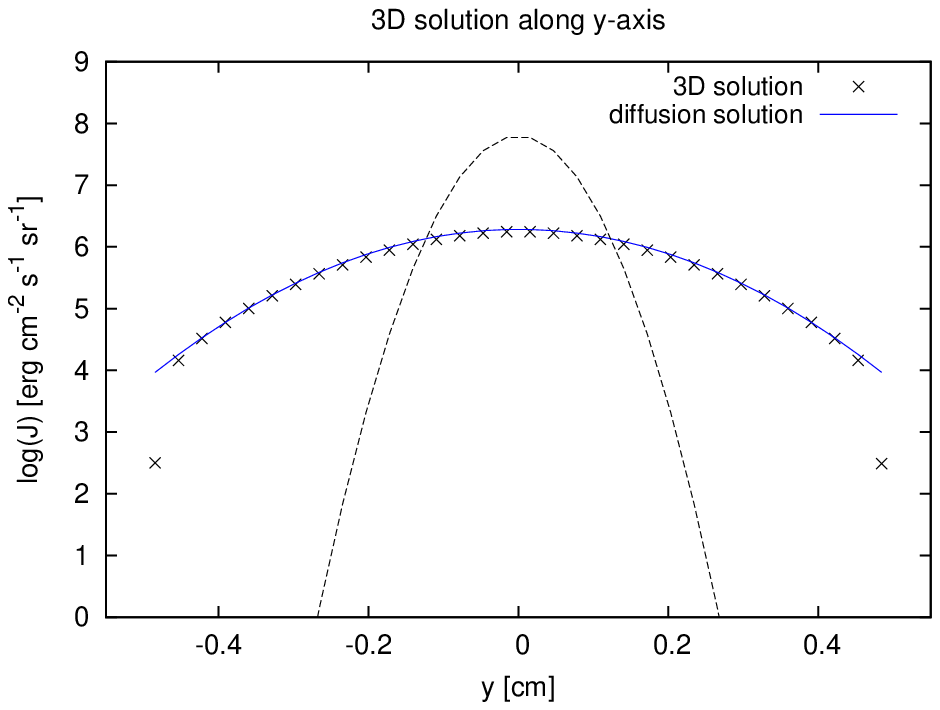}
\includegraphics[width=0.32\textwidth]{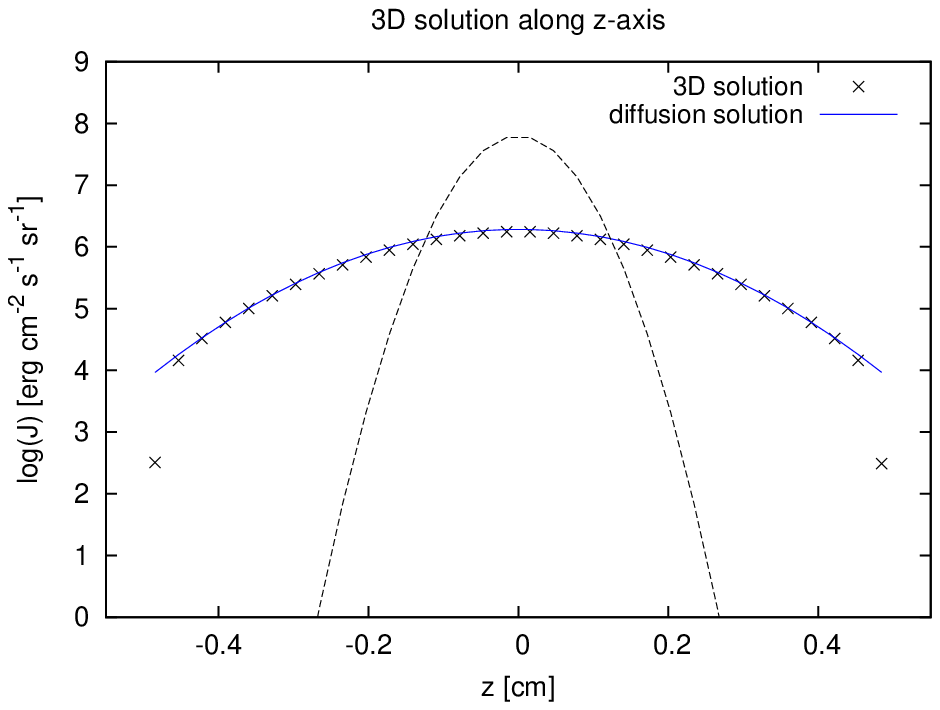}
\includegraphics[width=0.32\textwidth]{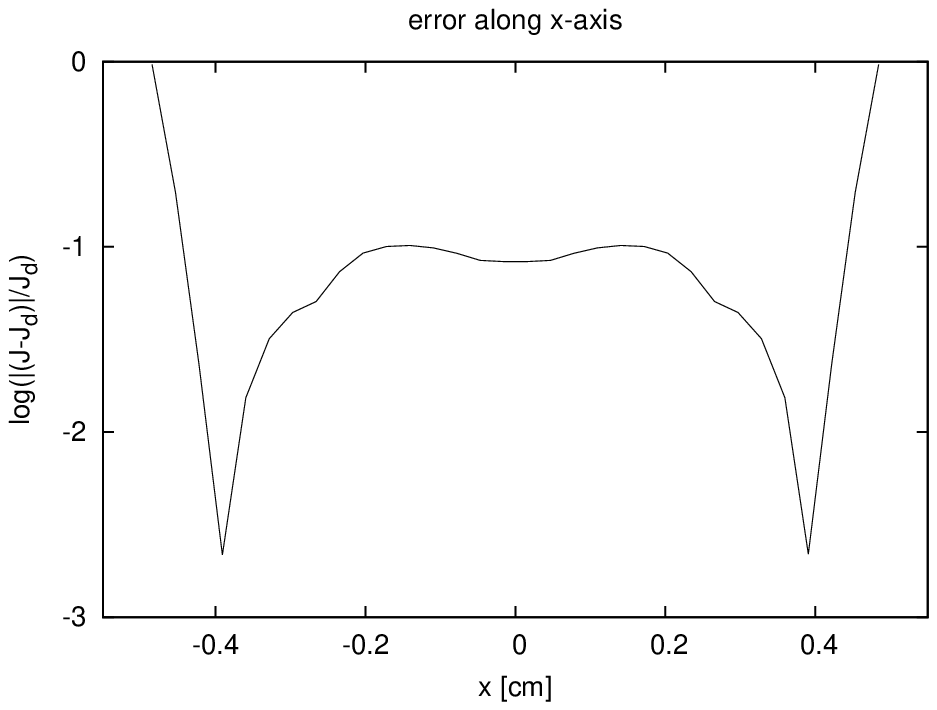}
\includegraphics[width=0.32\textwidth]{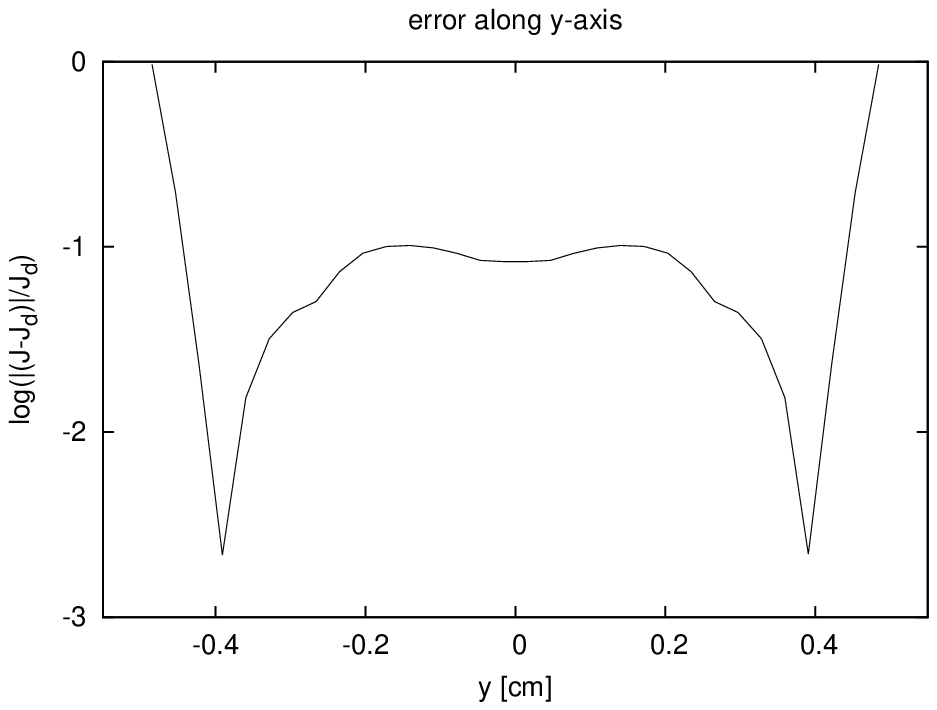}
\includegraphics[width=0.32\textwidth]{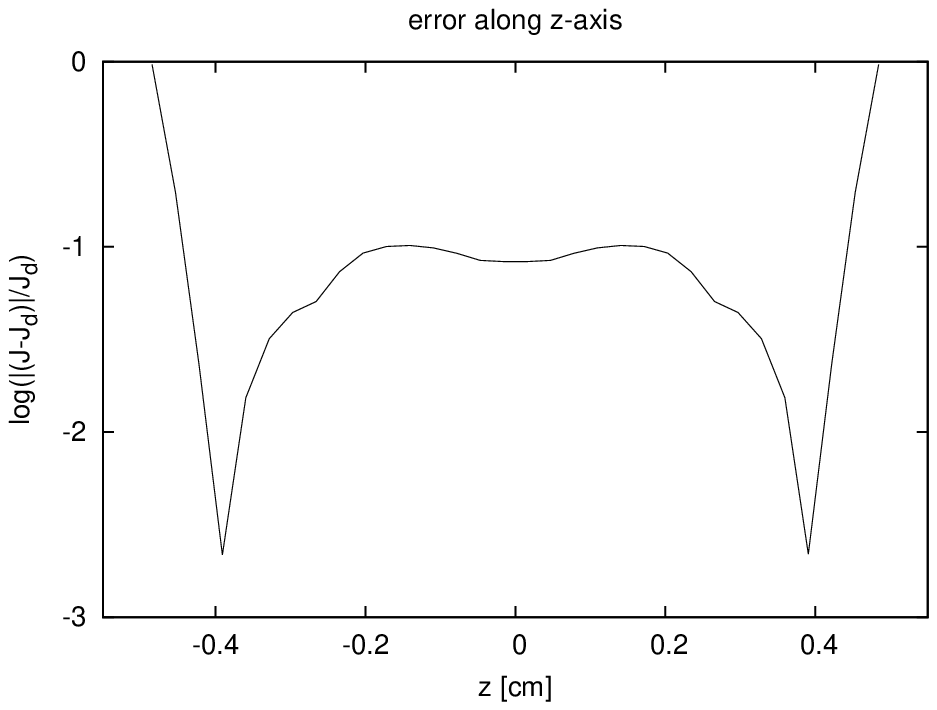}
\captionsetup{font=small}
\caption{Results of the 3D diffusion test along the x-({\it left}), y-({\it mid}) and z-axis ({\it right}) of the simulation box with a homogeneous spatial resolution of nx=ny=nz=32 (symbols). {\it Top row:} The dashed lines show the initial conditions at $t=t_0$ determined by the Gaussian solution of the diffusion equation. Blue solid lines show the analytic solution according to Equation \ref{equ:gaussian}. Symbols show the numerical results from our radiation solver. {\it Bottom row:} The relative error of the numerical solution. The 3D solution is not as accurate as the 1D results but still within $10\%$ of the analytical solution. The obvious independence of the solution on the direction axis results from the homogeneous angular HEALPix tessellation. The calculations were performed using 192 directions.}
\label{fig:diffusion_3D}
\end{figure*}

\subsection{1D Non-equilibrium Radiative Shocks}
\label{sec:shocks}
Testing the radiative transfer solver for radiative shock computations is the next crucial step and requires the combination
of our radiation solver with the FLASH code. The source term is determined by the energy budget of absorption and emission processes. 
We recall the frequency integrated source term from Equation (\ref{equ:qrad}) here:
\begin{equation}
Q_{\rm rad} = 4\pi \chi_a (J-B),
\label{equ:source_term_2}
\end{equation}
which is coupled to the hydrodynamical solver by adding it to the right-hand side of the Euler equation for the internal gas energy. 
For this test case, the emission and absorption opacities are equal. Since the shock setup is used for test purposes, we neglect the magnetic field.
\subsubsection{Initial Conditions}
The initial conditions are consistent with the theoretical work of \citet{Lowrie08}. In their work, the jump conditions and the equations of radiation hydrodynamics 
are given in a dimensionless form. The equations are normalized using reference material quantities and a constant $\rm{P}_0$ which arises 
from the normalization process and is given by
\begin{equation}
\rm{P}_0 = \frac{\tilde{\alpha}_r \tilde{T}_0^4}{\tilde{\rho}_0 \tilde{a}_0^2}.
\label{norm_constant}
\end{equation}
The quantities denoted with a tilde are the dimensional reference material attributes (temperature $\tilde{T}_0$, density $\tilde{\rho}_0$, sound speed $\tilde{a}_0$) 
and $\tilde{\alpha}_r$ is the radiation constant. The "$0$"-subscript indicates pre-shock state initial values. $\rm{P}_0$ gives a measure for the relative 
importance of gas and radiation pressure or alternatively, the radiative energy to the material energy \citep{Mihalas84}. 
For our test setups, we choose a grey non-equilibrium shock setup with Mach numbers of $M_0=1.2$, $M_0=2$ (subcritical), and $M_0=5$ (supercritical), 
which we compute in the reference frame of the shock with $\rm{P}_0=10^{-4}$ and $\gamma=5/3$. \citet{Lowrie08} give a dimensionless absorption and transmission 
cross section, which determine the radiative energy exchange and diffusivity of the radiating materials. Evaluating the dimensionless values
gives an absorption coefficient of $\kappa_a\approx 423.0$ $\cm^2/g$ and a total extinction coefficient of $\chi \approx 788.0$ $\cm^2/g$, which results in an effective photon destruction probability 
of $\epsilon=\kappa_a/\chi \approx 0.5377$. The initial dimensionless pre-shock gas temperature $T_0$ and density $\rho_0$ are set to unity, the post-shock 
initial values ($T_1, \rho_1 $) are computed using the Rankine-Hugoniot jump conditions. The actual dimensional initial conditions can then be calculated using 
their dimensional reference material values (for more details we refer to \citet{Lowrie08}). Finally, the radiation temperature
\beq
T_{\rm r} = \left(\frac{\pi}{\sigma_{\rm SB}}\;J \right)^{1/4}
\eeq
is initially in equilibrium with the gas temperature. For the radiation shock test problem, the source function is determined by a thermal emission and a diffusive part. 
This is equivalent to using the isotropic scattering source function  
\begin{equation}
S = (1-\epsilon) J + \epsilon B
\label{eq:source_shock}
\end{equation}
with the appropriate photon destruction probability and a thermal energy contribution given by the frequency integrated Planck emission $B=\frac{\sigma_{\rm SB}}{\pi} \tilde{T}^4$.
Since the radiation field will not be not be in thermal equilibrium with the material throughout the simulation, we need to iterate until a consistent solution of the mean intensity $J$ is found. 
However, since $\epsilon \approx 0.5377$ gives only a moderate scattering contribution and using the solution from the previous time step, the accelerated lambda iteration usually 
converges after 2 or 3 iteration steps.
\subsubsection{Results}
The shocks need a few nanoseconds to relax into a static equilibrium state. Figure \ref{fig:shock_M1.2}, \ref{fig:shock_M2} and
\ref{fig:shock_M5} show the resulting temperature and density profiles after 10 nanoseconds. Sufficiently far upstream (left) and downstream (right) 
of the hydrodynamical shock (at $x=0$), gas and radiation are in a thermodynamical equilibrium and the radiation temperature coincides with the gas 
temperature computed from the initial conditions. Since the total extinction coefficient $\chi$ is about twice the thermal absorption and emission coefficient, 
the temperature of the radiation field and the gas are out of equilibrium near the shock front.\\
The subcritical shock with $M_0=1.2$ (Figure \ref{fig:shock_M1.2}) shows a hydro shock but no spike in the
radiation temperature. For $M_0=2$ the so called {\it Zel'Dovich spike} in the gas temperature appears for the first time as seen in Figure \ref{fig:shock_M2}.
The spike appears since radiation is transported through the hydrodynamical shock and preheats the inflowing gas, which is initially in a thermal equilibrium 
with the radiation field in the upstream region. After the gas has passed the hydrodynamical shock, it cools down until the radiation field and the gas are again in 
thermal equilibrium on the downstream side of the shock. Since the upstream temperature at the shock front is still less than the downstream temperature the
shock is subcritical. For $M_0=5$, the shock becomes supercritical, since the upstream gas is preheated until it reaches the downstream gas temperature even before passing
the hydrodynamical shock front. The discontinuity in the gas temperature is then restricted to the narrow range of the Zel'dovich spike (Figure \ref{fig:shock_M5}). 
Our solutions resemble the semi-analytical results from \citet{Lowrie08} and show the correct spike evolution. However, a closer look at the results show a slight deviation
of the shock front from its initial position (at $x=0$). Especially in the supercritical case, the shock front drifts very slowly into the downstream direction.
This drift is due to the absence of the radiation pressure in our approach, which becomes important for high Mach numbers (with a high downstream gas temperature).
While the shock front drifts very slowly, the temperature and density profiles do not change since the radiation source term is still very well approximated in our approach.
\begin{figure*}
\centering
\showtwo{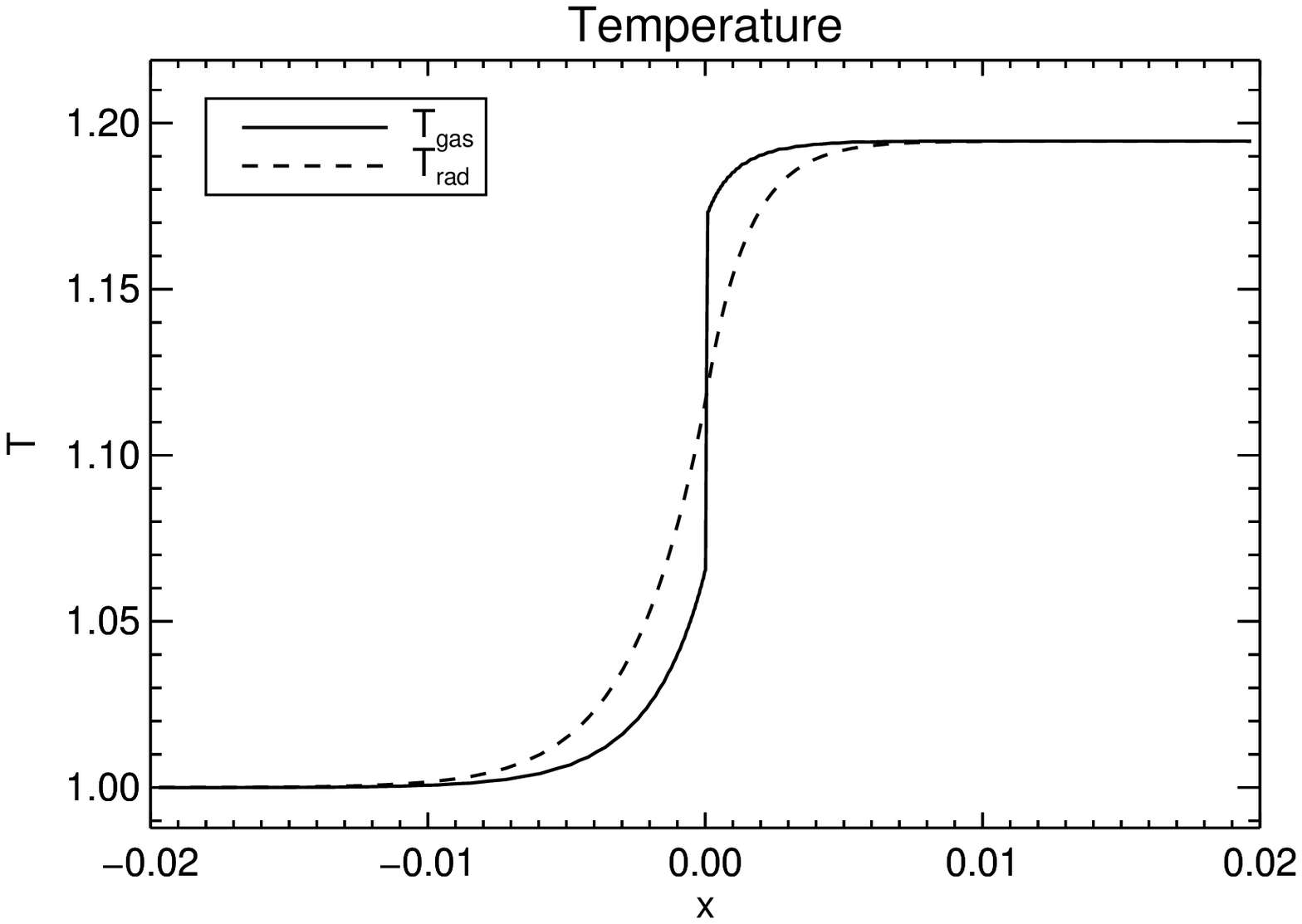}
        {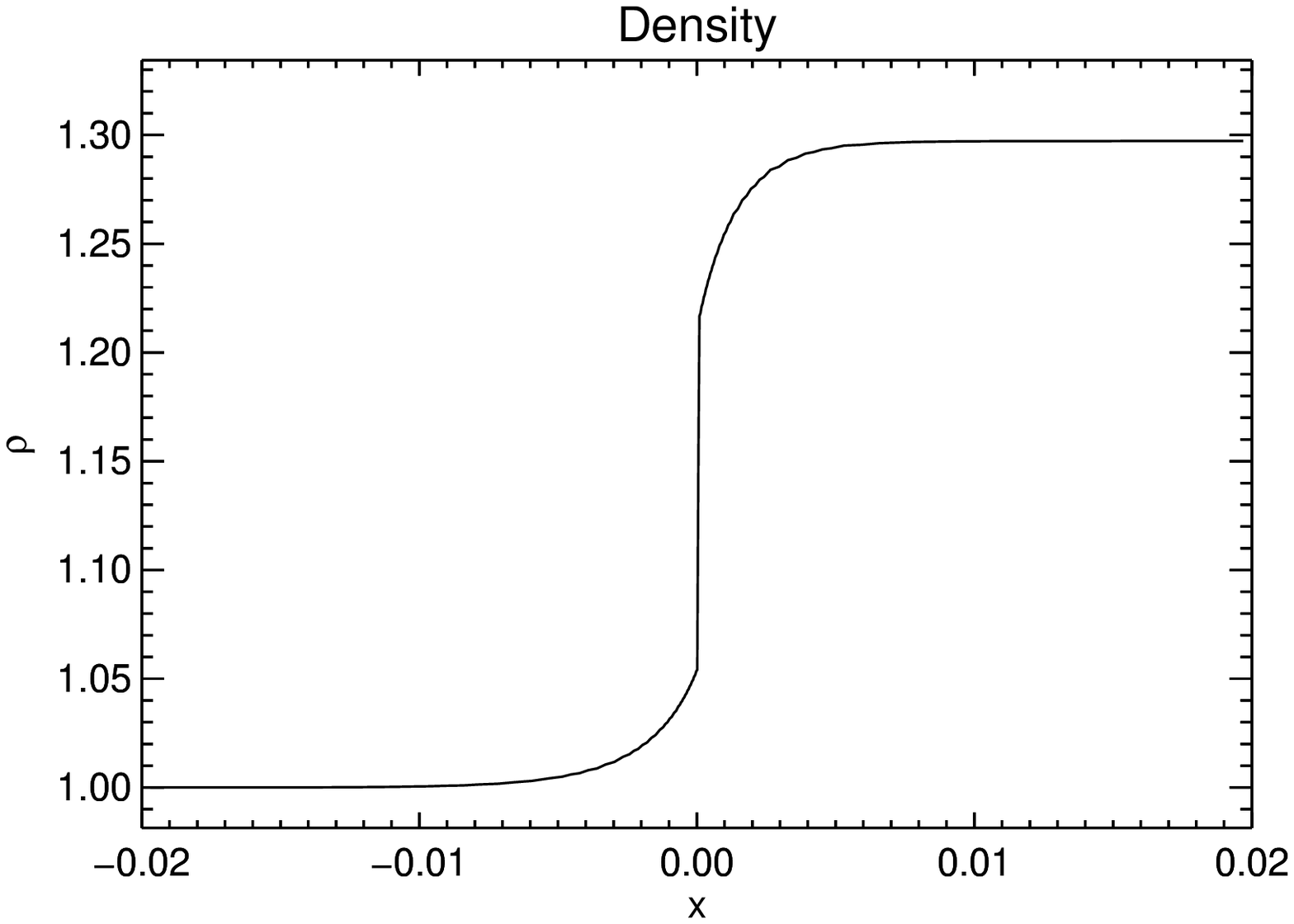}
\captionsetup{font=small}
\caption{Normalized temperature and density profiles for the subcritical shock with $M_0=1.2$ in the equilibrium state after 10 nanoseconds. The gas is preheated on the upstream side and cools downe on the downstream side of the hydro shock front.}
\label{fig:shock_M1.2}
\end{figure*}
\begin{figure*}
\centering
\showtwo{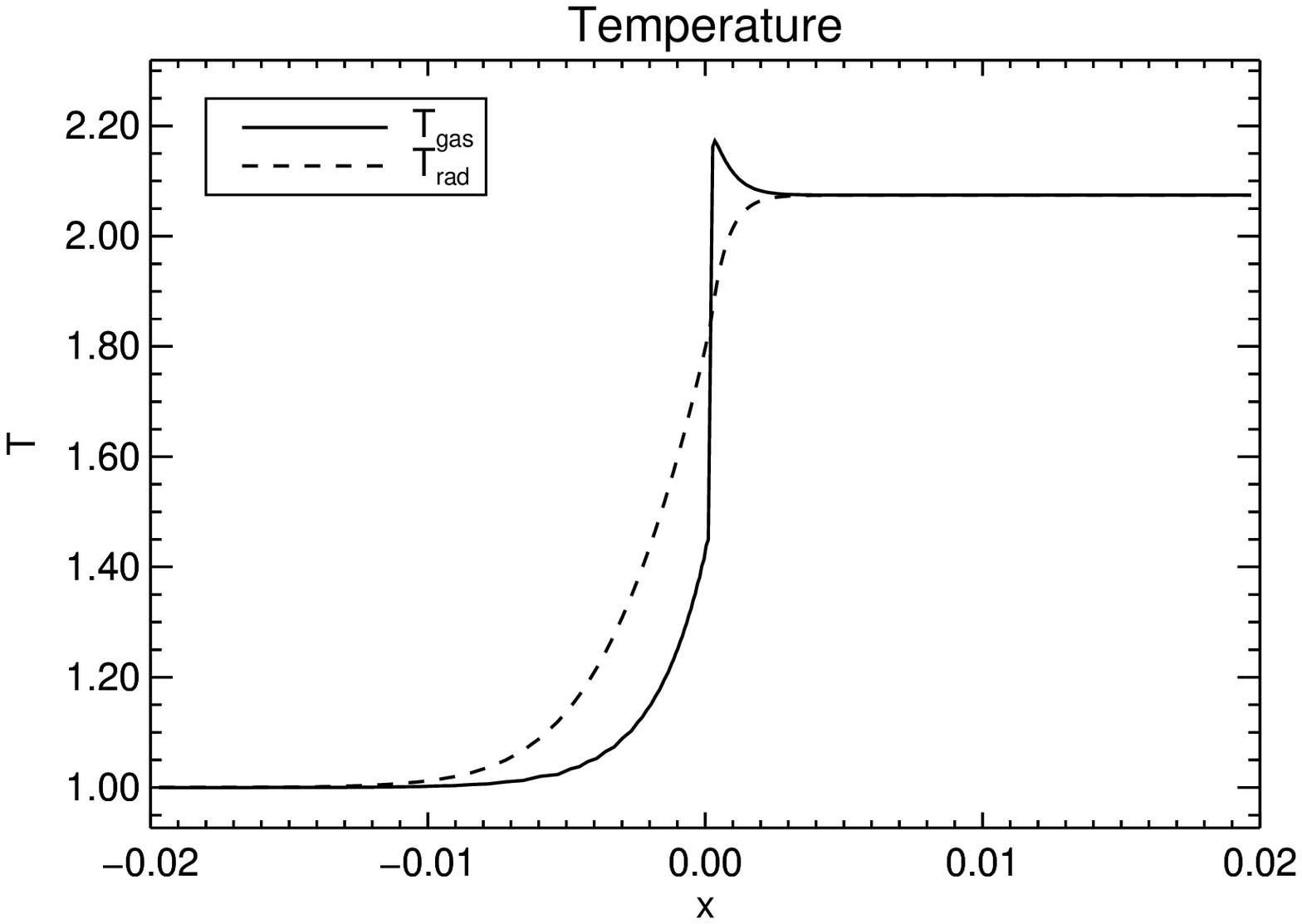}
        {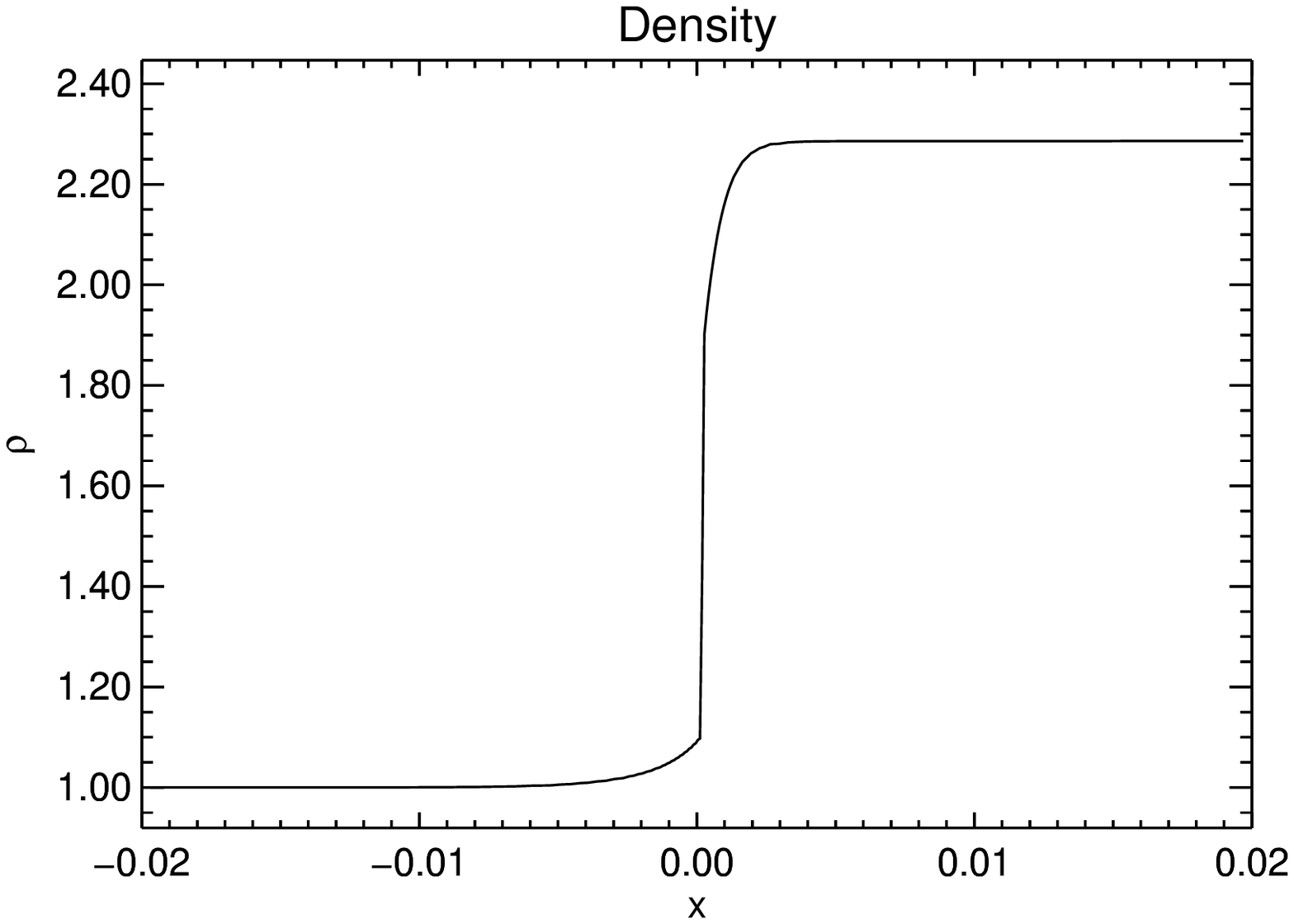}
\captionsetup{font=small}
\caption{Same conditions as in Figure \ref{fig:shock_M1.2} but with $M_0=2$. The maximum temperature at the shock begins to exceed the downstream equilibrium temperature which
results in the Zel'dovich spike. Since the temperature at the upstream side of the shock is still well below the downstream temperature, the shock is subcritical.}
\label{fig:shock_M2}
\end{figure*}
\begin{figure*}
\centering
\showtwo{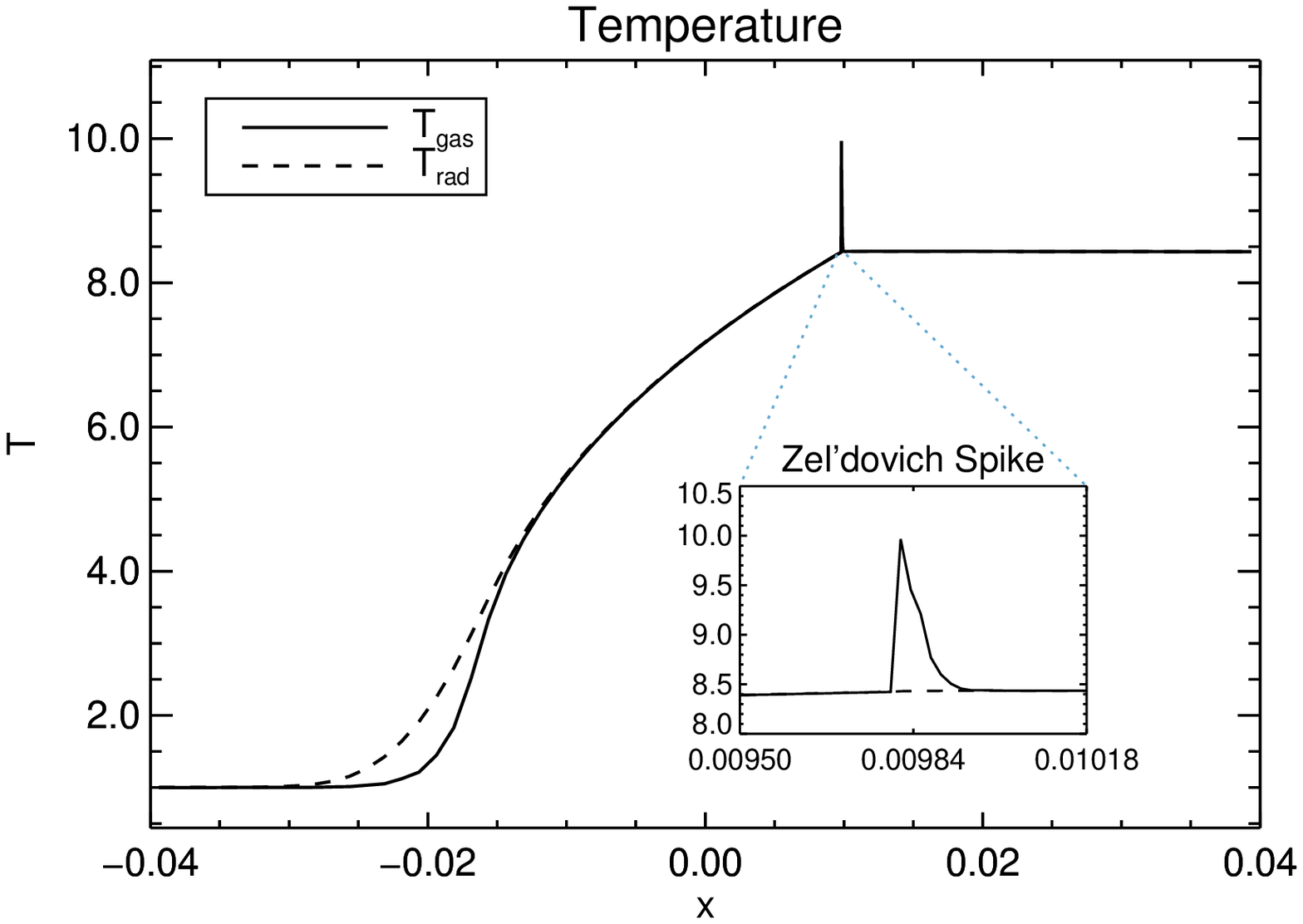}
        {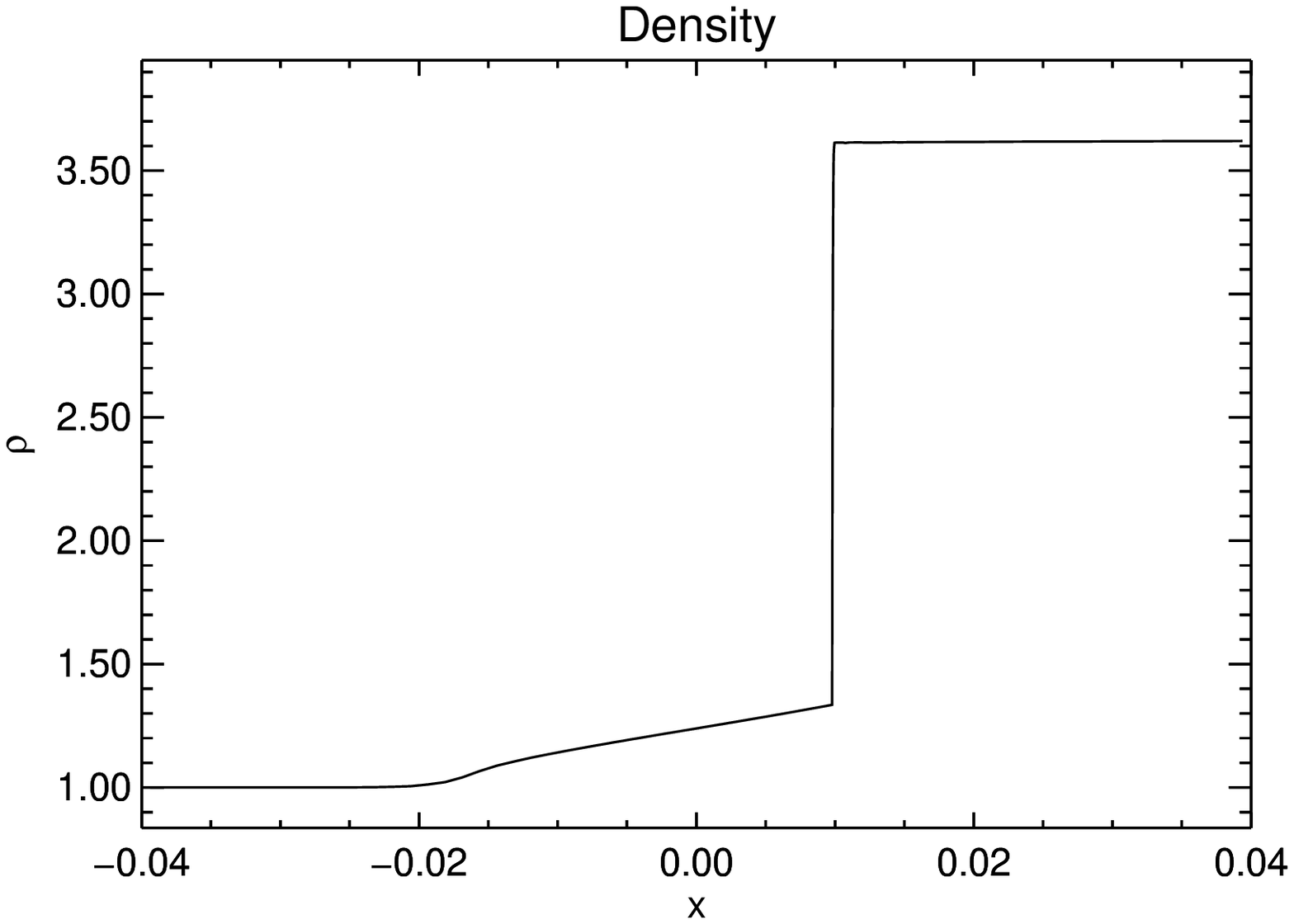}
\captionsetup{font=small}
\caption{Same conditions as in Figure \ref{fig:shock_M1.2} but with $M_0=5$. The temperature on the upstream side of the hydro shock front reaches
the downstream equilibrium value. The Zel'dovich spike gets very narrow and the shock becomes supercritical.} 
\label{fig:shock_M5}
\end{figure*}

\section{3D Collapse Simulations}
\label{sec:collapse}

In this section, we show results from full 3D radiation hydrodynamical simulations performed with FLASH/RT.
Since we aim to use our framework for the modelling of radiative feedback in star formation simulations, 
we show the capabilities of our method in two self-gravitating collapsing cloud simulations. 
We follow the collapse until the first hydrostatic core is formed and before the dissociation 
of hydrogen molecules start (the first collapse). In Section \ref{sec:non_rotating_collapse},
we show results from a basic collapse simulation without rotation and compare the resulting profiles
to other similar works. Afterwards, we show results from a more complex simulation including rotation and
turbulence (Section \ref{sec:rotating_collapse}) and compare the results to a simulation without modelling
radiative transfer. The angular resolution of the radiative transfer calculations are the same for both collapse simulations,
and we use 768 directions to compute the radiation field (nSide=8 for the HEALPix tessellation).

\subsection{Opacities}
Since our solver does not yet support any frequency dependence, the source function $S$ is only determined by the frequency-integrated thermal emission of the gas ($S = B = \frac{\sigma_{\text{SB}}T^4}{\pi}$), 
and we neglect any scattering processes. Consequently, we have to use frequency-integrated mean dust opacities. For this purpose, we choose the Planck mean opacities by \citet{Semenov03}. 
In their work, the dust composition model takes into account the evaporation temperatures of ice, silicates, iron as well as their density dependencies. 
We coupled their subroutines\footnote[3]{http://www.mpia-hd.mpg.de/homes/henning/\\Dust\_opacities/Opacities/opacities.html} for computing temperature and density dependent dust opacities 
into FLASH, and we choose the input parameters for spherical homogeneous dust grains with a normal relative iron content in the silicates of Fe/(Fe+Mg) = 0.3.

\subsection{Collapse without Rotation}
\label{sec:non_rotating_collapse}
In this section, we study the collapse of a spherical, homogeneous, and gravitationally unstable density distribution. 
The initial conditions do not contain any turbulence or density perturbations and hence, the results are spherically symmetric. This setup represents
a common benchmark for the capabilities of a radiation hydrodynamical astrophysical computer code, and we compare 
our results to similar work done by \citet{Commercon11}, \citet{Masunaga98}, and the pioneering simulations of \citet{Larson69}.  

\subsubsection{Initial Conditions}
We start with highly gravitationally unstable initial conditions. The cloud core of one solar mass consists of a homogeneous sphere with radius $R_0=7.07\times10^{16}\,\cm\,(\approx4725\,\AU)$ and 
and density $\rho_0=1.38\times10^{-18}\,\g\,\cm^{-3}$, which results in an initial free fall time of $t_{\rm{ff}}\approx56.67\,\kys$. The linear size of the 3D computational domain is 
four times the initial cloud radius $R_0$ in each dimension. The surrounding gas density is a hundred times less than the initial cloud density $\rho_0$, and the cloud is initially in thermal equilibrium 
with the ambient gas at a temperature of $T_0=10\,\rm{K}$ resulting in an initial isothermal sound speed of $c_{\rm s}\approx0.195\,\km\,\s^{-1}$.
Since the cloud is initially not in pressure equilibrium with its surroundings, FLASH's hydrodynamical solver drives a weak shock wave 
into the ambient gas which is soon dissipated. To prevent our radiation solver from resolving this shock in terms of radiative energy exchange 
(which would result in rather small time steps), we do not couple the radiation field to the hydrodynamics outside of $R_0$ but rather keep the ambient gas and radiation temperature fixed. \\
The initial conditions result in a gravitationally unstable cloud core which contains nearly two Jeans masses. To ensure a proper resolution and avoid artificial fragmentation during the collapse, 
we use the Jeans condition by \citet{Truelove97} as the refinement criterion of the AMR grid. In our case, we use at least $\rm{N}_\text{j}=9$ grid cells per Jeans length.
To resolve the first hydrostatic core properly, we allow a maximum linear resolution of $\Delta x \approx 0.07\AU$ which requires the AMR grid to cover 11 levels of resolution.\\
The summarized initial conditions are:
\begin{align*}
\text{Mass}\quad M&=1.0\,\Msol,\\
\text{Density}\quad \rho_0 &= 1.38\times10^{-18}\,\g\,\cm^{-3},\\
\text{Temperature}\quad T_0 &= 10\,\rm{K},\\
\text{Angular Velocity}\quad \Omega &= 0.0\,\rm{rad}\,\s^{-1},\\
\text{Radius}\quad R_0 &= 7.07\times10^{16}\,\cm,\\
\text{Free Fall Time}\quad t_{\rm ff} &= 56.67\,\kys.
\end{align*}
\subsubsection{Results}
The cloud core starts to collapse, and as soon as the maximum density in the cloud exceeds about $10^{-13}\,\g\,\cm^{-3}$, the central regions of the cloud core become optically thick. 
At this point, the central temperature starts to rise rapidly and the following evolution proceeds almost adiabatically with more gas falling onto the central quasi-hydrostatic core. 
Since the simulation does not contain any rotation or turbulence, the 3D solution is spherically symmetric, and we present the results in the form of averaged radial profiles. 
The profiles for density, radial velocity, temperature, optical depth, and central mass after $1.036\times t_{\rm ff}$ are shown 
in Figure \ref{fig:bb_profiles}. The resulting protostellar core has a mass of $M_{\rm fc}\approx1\times10^{-2}\,\Msol$, a radius of $R_{\rm fc}\approx4\,\AU$, and a central
temperature of $T_{\rm c}\approx186\,{\rm K}$. The boundary of the core can be identified easily in the velocity profile, where there is a sudden decrease in the infall velocity (the accretion shock). Inside the core,
the infall does not stop completely indicating that the core is only quasi-hydrostatic.\\
Our results are quantitatively very similar to those of \citet{Larson69} and qualitatively very similar to the more recent works by \citet{Masunaga98} and \citet{Commercon11}.
Table \ref{bb_table} shows an overview of the characteristic temperature, mass and radius of the first core in comparison to these works (the common reference point is when the maximum central density of the
first core reaches $\rho_{\rm fc}\approx 2\times10^{-10}\,\g\,\cm^{-3}$). Apparently, our computations produce qualitatively similar results, 
although the methods invoked in the other works are quite different and use different initial conditions and opacity models.
\begin{figure*}
\centering
\showsix{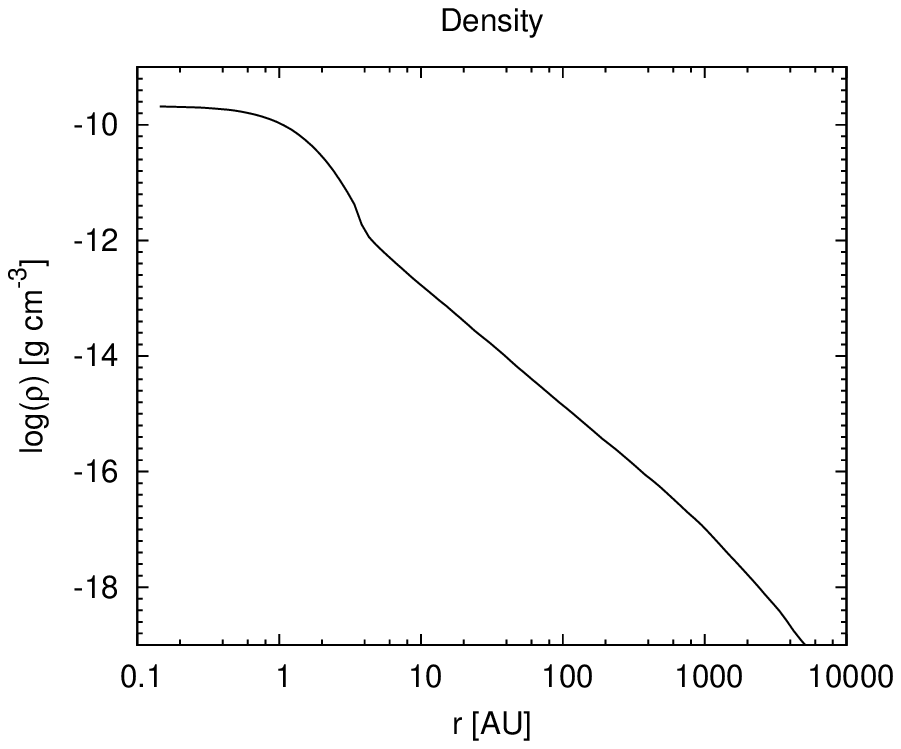}  
{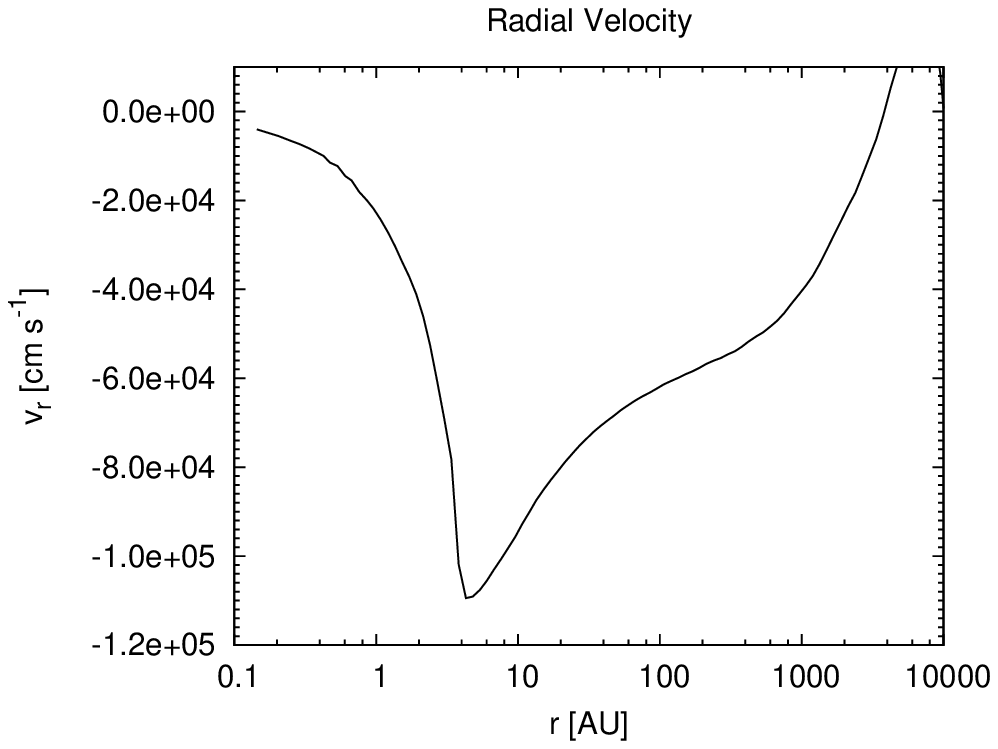}  
{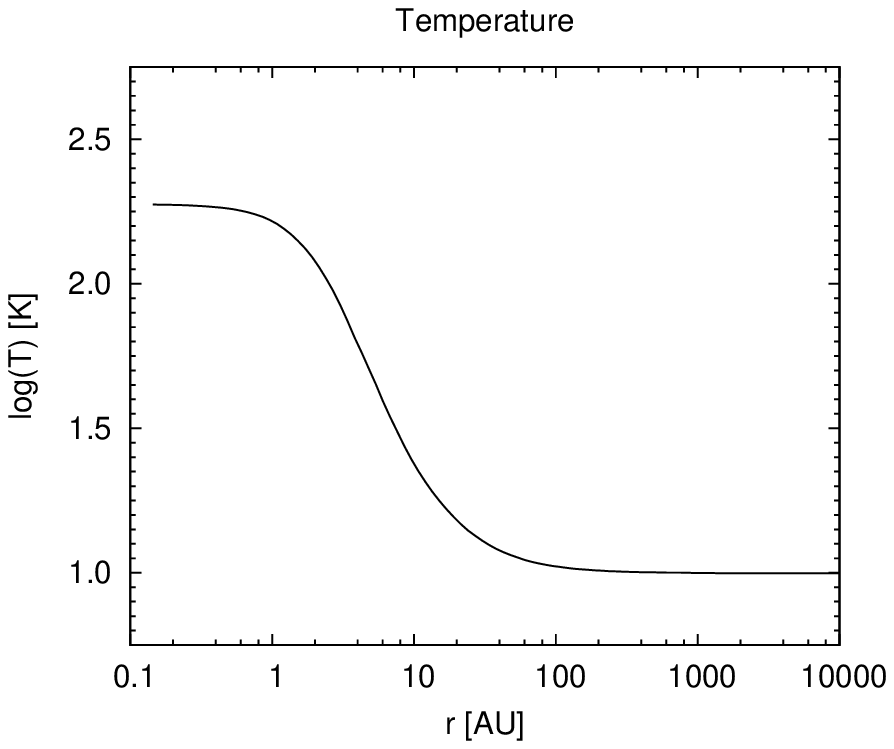}  
{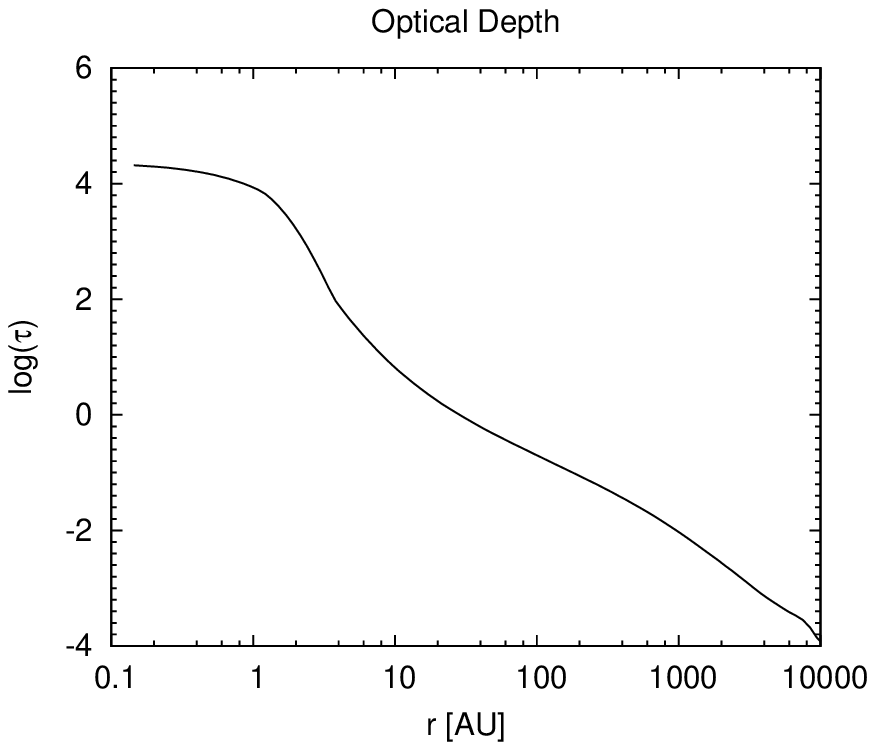}  
{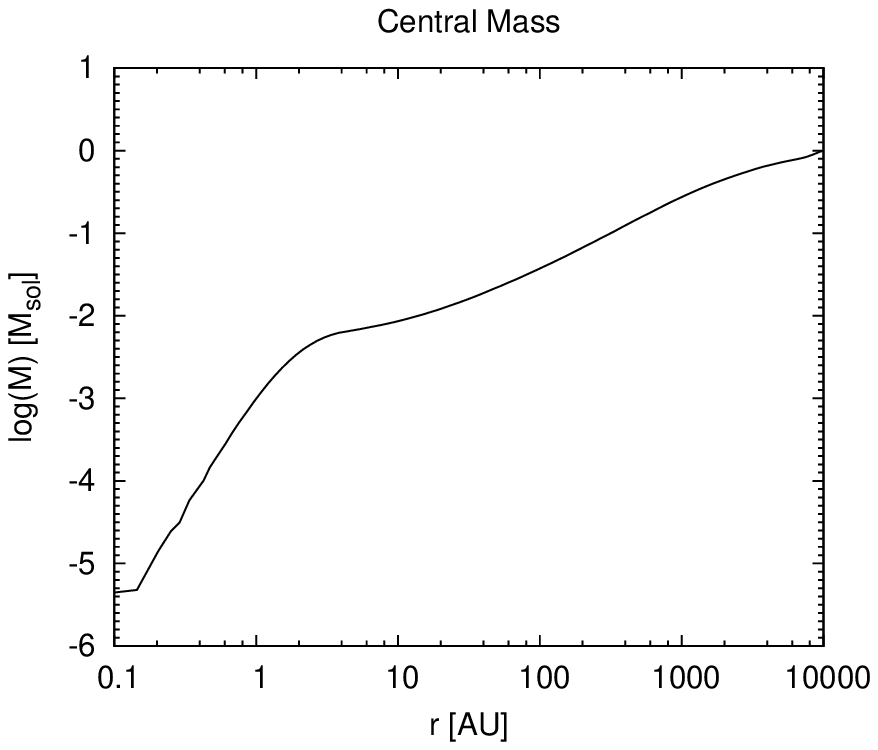}  
{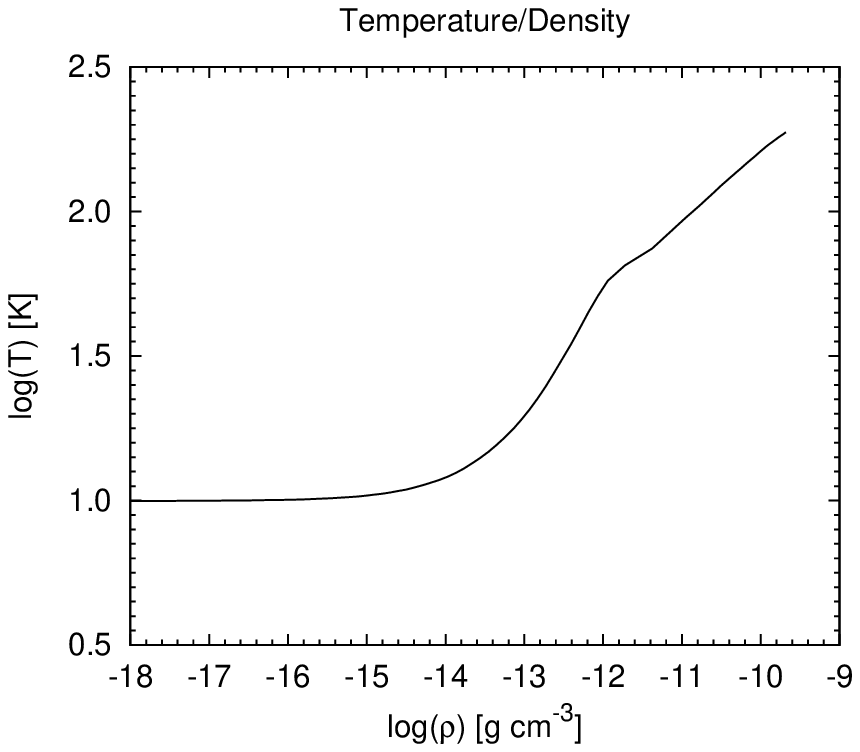}  
\captionsetup{font=small}
\caption{Profiles of the collapse simulation after $t = 1.036\,t_\text{ff}$; the maximum density at the core center is $\rho_c\approx 2.0 \times 10^{-10}\,\text{g}\,\text{cm}^{-3}$ with a temperature of $T_\text{c}\approx\,186\,\text{K}$, a radius of $R_\text{fc}\approx 4\,\text{AU}$ and a mass of $M_\text{c}\approx10^{-2}\,\Msol$}
\label{fig:bb_profiles}
\end{figure*}
\begin{table*}
    \centering
    \begin{tabular}{  l | l | l | l | l }
    \hline 
    \hline
    Reference & $\text{R}_{\text{fc}}$ [AU] & $\text{M}_{\text{fc}} [\text{M}_{\odot}]$ & $\text{T}_{\text{fc}}$[K] & $\text{T}_{\text{c}}$ [K]\\ \hline
    This work & 4 & $1\times10^{-2}$ & 50  & 186 \\ 
    {\citet{Commercon11}} & 8 & $2.1\times10^{-2}$ & 81 & 396 \\
    {\citet{Masunaga98}}& 8 & $\approx$ $10^{-2}$ & 60 & 200 \\
    {\citet{Larson69}}& 4 & $1\times10^{-2}$ & - & 170 \\ \hline
    \end{tabular}
    \caption{Comparison of simulation results; R$_\text{fc}$ is the radius of the first core, M$_\text{fc}$ is the core mass, T$_\text{fc}$ the central temperature and T$_\text{fc}$ is the temperature at R$_\text{fc}$.}
\label{bb_table}
\end{table*}
\subsection{Collapse with Rotation and Turbulence}
\label{sec:rotating_collapse}
This simulation run has very similar initial conditions as described in the previous section except that we add rotational
and turbulent energy. The cloud is initially in a rigid body rotation around the z-axis at the center of the simulation box.
The ratio of rotational and gravitational energy is given by 
\beq
\beta = \frac{1}{3}\,\frac{R_0^3\,\Omega_0^2}{\rm{G}\,\rm{M}_0}.
\eeq 
We choose $\beta=0.03$ which gives an initial angular velocity of $\Omega_0=1.886\times10^{-13}\,\rm{rad}\,\s^{-1}$ and agrees with typically
observed values of molecular cloud cores \citep{Goodman93}. In addition, we superimpose a turbulent velocity perturbation
on the initial uniform angular velocity field. The construction of the velocity perturbation is based on the theory for
incompressible turbulence by \citet{Kolmogorov41}, in which the kinetic energy $E$ of the velocity fluctuation with wave number $k$
is described by a power spectrum
\beq
E(k) \propto k^{p}.
\eeq
The wave number $k=2\pi/l$ is the inverse of the length scale $l$ of a turbulent fluctuation (sometimes called {\it eddy}).
In our case, the spectrum has a power law index of $p=-2$ resembling a Burgers type model of turbulent energy decay.
The geometries and density distribution of the initial cloud core are the same as for the simulation 
without rotation and turbulence.\\
In addition to the simulation run with FLASH/RT, we also run the simulation without modelling radiative transfer. Instead,
we use a barotropic EOS with a density-dependent effective adiabatic exponent $\gamma$ that mimics radiative cooling. The internal
energy/temperature is fixed at $T_0=10\,\rm{K}$ as long as the gas density is less than $\rho\approx10^{-15}\,\g\,\cm^{-3}$ (isothermal). Above
this threshold density, the temperature rises slowly with $\gamma=1.1$ until the adiabatic exponent becomes $\gamma=4/3$ above $\rho\approx10^{-13}\,\g\,\cm^{-3}$ (adiabatic).
We ran the simulation including radiative transfer as well as the reference run with the barotropic EOS until
the formation of the first hydrostatic core with a central density of $\rho_{\rm fc}\approx10^{-11}\,\g\,\cm^{-3}$. At this point, 
both simulations cover 9 different levels of resolution in the AMR grid with a maximum linear resolution of $\Delta x\approx0.57\,\AU$
while the whole simulation box has a linear extent of $18903\,\AU$.\\
The summarized initial conditions are: 
\begin{align*}
\text{Mass}\quad M&=1.0\,\Msol,\\
\text{Density}\quad \rho_0 &= 1.38\times10^{-18}\,\g\,\cm^{-3},\\
\text{Temperature}\quad T_0 &= 10\,\rm{K},\\
\text{Angular Velocity}\quad \Omega &= 1.886\times10^{-13}\,\rm{rad}\,\s^{-1},\\
\frac{\text{\small Rotational Energy}}{\text{\small Gravitational Energy}}\quad \beta &= 0.03,\\
\text{Radius}\quad R &= 7.07\times10^{16}\,\cm,\\
\text{Free Fall Time}\quad t_{\rm ff} &= 56.67\,\kys.
\end{align*}
\subsubsection{Results}
The rotational energy and the superimposed turbulent velocity perturbations break the symmetry of the simulation.
Figure \ref{fig:col_dens_run06} shows the column densities along the main axes of the inner region where the dense first core has formed
after about $60\,\kys$ ($\approx1.07\,t_{\rm ff}$) with a maximum gas density of $\rho_{\rm fc}\approx10^{-11}\,\g\,\cm^{-3}$.
Because of the additional rotational and turbulent energy, the formation of the first core is deferred and forms later
than in the previous simulation (Section \ref{sec:non_rotating_collapse}). The conservation of angular momentum
causes the first core to be flattened roughly along the z-axis and the density distribution shows a flat disc-like
structure revolving around the central compact hydrostatic core. The resulting density distribution is roughly the same
as in the reference run without radiative transfer. The initial collapse which seeds the formation of the central core
does mostly occur in the isothermal phase, hence, modelling radiative feedback does not influence the initial formation
of the core significantly. However, Figure \ref{fig:dens_temp_run06} shows the resulting density weighted temperature averages along
the main axes in the central regions around the first core (e.g $\int \rho\,T\,dz / \int \rho\,dz$). The left column shows
the results including radiative transfer (FLASH/RT) while the right column shows results from the reference run.
The FLASH/RT model clearly shows how the central core heats the surrounding gas to a temperature roughly 30\% higher
than in the reference run (like in \citet{Price10}). The resulting temperature density distribution in comparison
to the barotropic EOS is shown in Figure \ref{fig:temp_dens_histo_run06}.\\
Unfortunately, our FLASH/RT simulations are very costly (see Section \ref{sec:performance} for more details)
and currently, it is not feasible to continue these simulations without coupling the radiative transfer solver to
a subgrid model for the formation of the central core, e.g., sink particles \citep{Federrath10}.
However, our current test simulations show the first stages of disc formation and the importance of modelling radiative
transfer accurately. Since the thermodynamics of the gas significantly influence the fragmentation behaviour,
modelling radiative transfer is indispensable to study the further evolution of the protostar, the circumstellar disc, and the surrounding gas envelope.
\begin{figure*}
\centering
\includegraphics[height=0.36\linewidth]{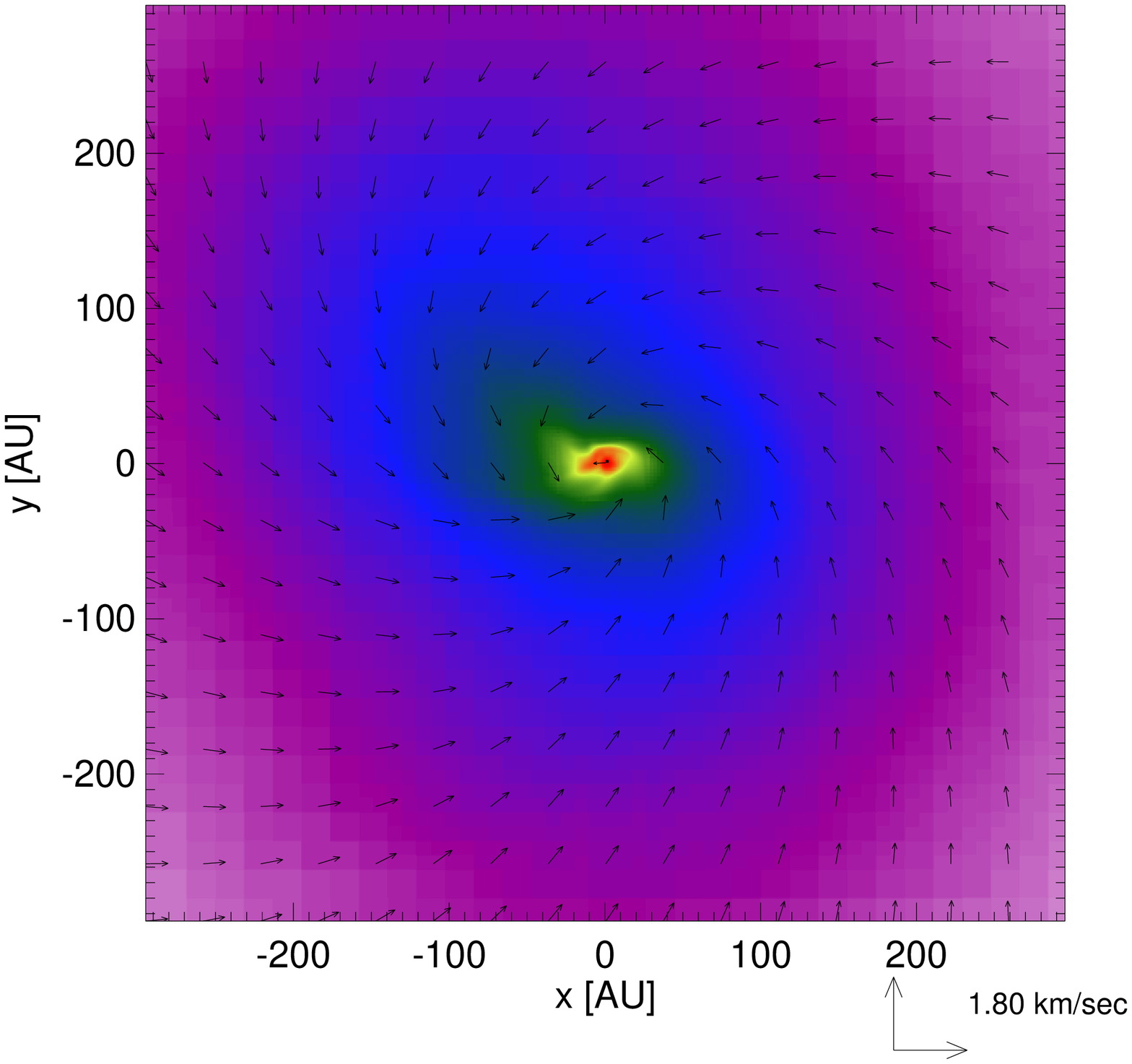}
\includegraphics[height=0.36\linewidth]{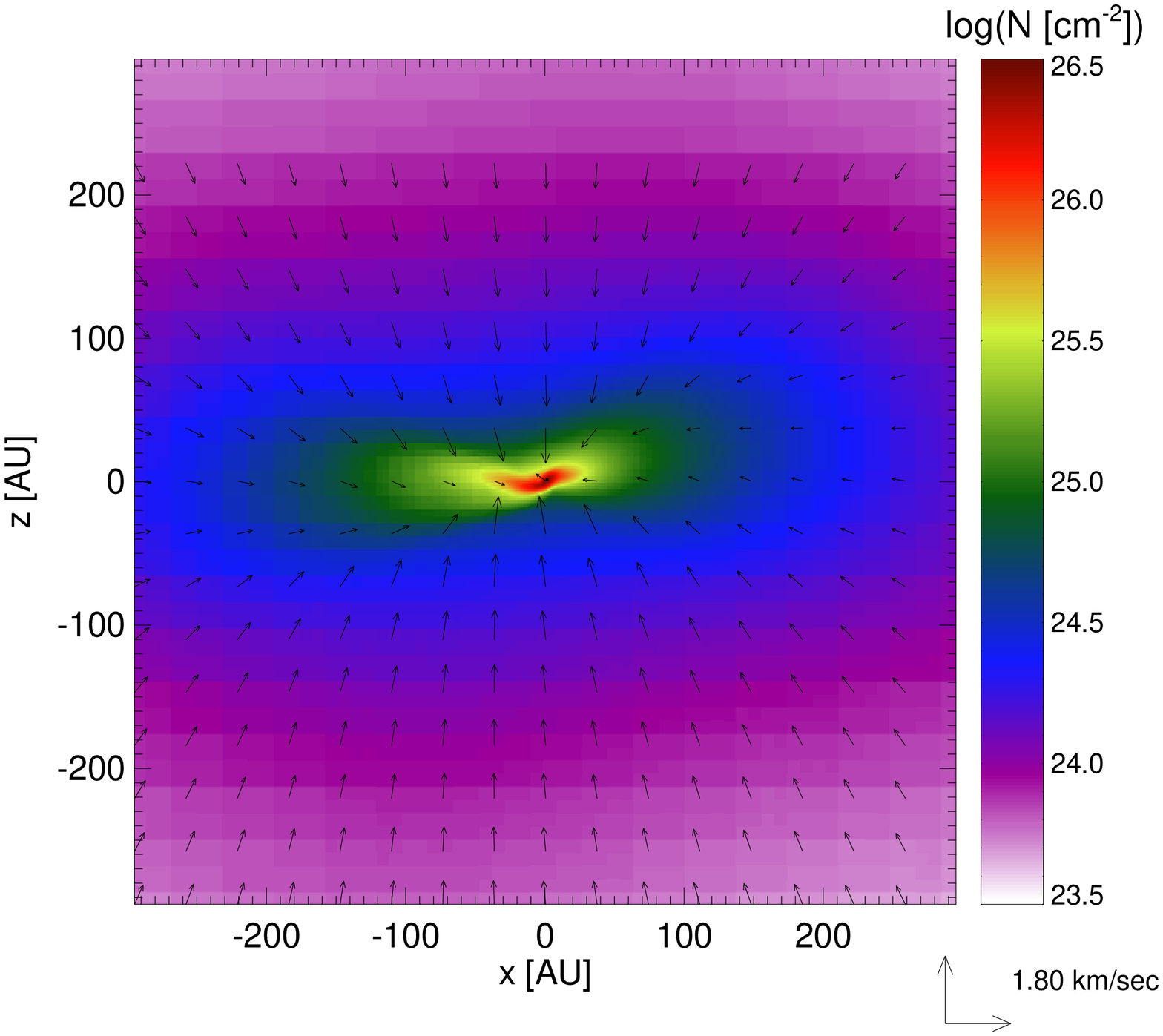}
\captionsetup{font=small}
\caption{Column densities along the z- (left) and y-axis (right) of the simulation box after the formation of the first hydrostatic core at $t\approx60\,\kys \approx1.07\,t_{\rm ff}$ including
rotation and turbulence. The rotational energy forces the gas to accumulate in a circumstellar disc (in the xy-plane) around the first core.}
\label{fig:col_dens_run06}
\includegraphics[height=0.36\linewidth]{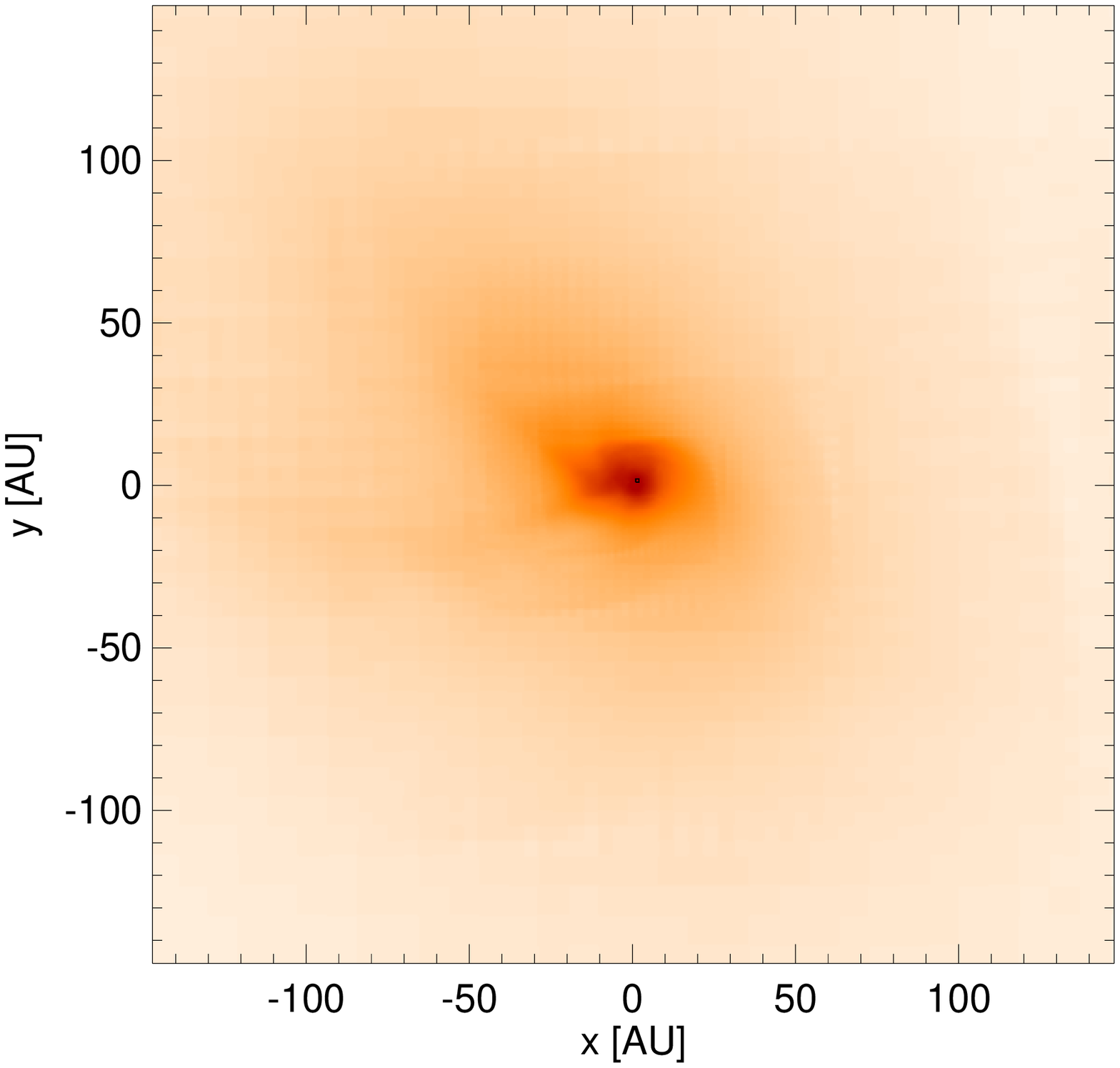}
\includegraphics[height=0.36\linewidth]{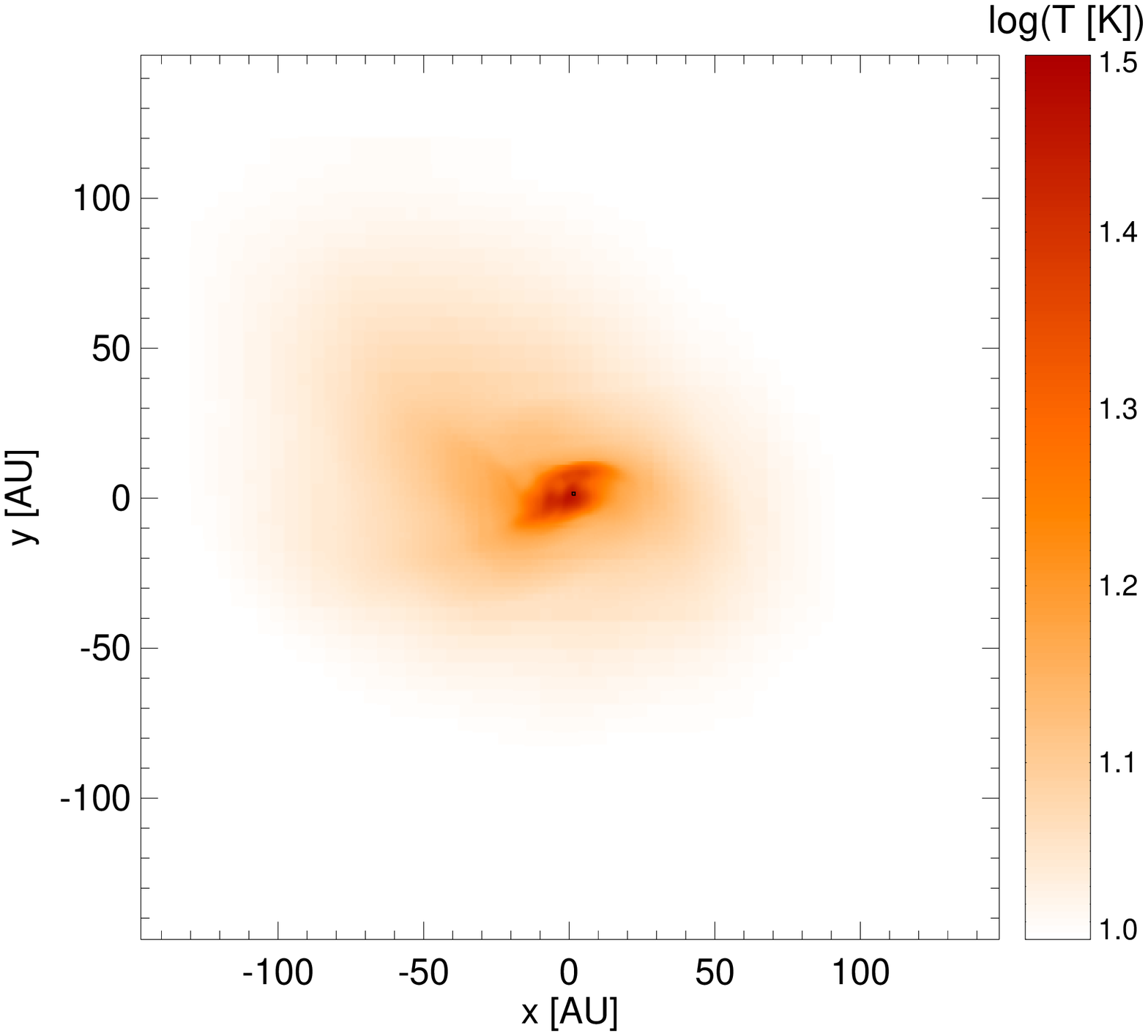}
\includegraphics[height=0.36\linewidth]{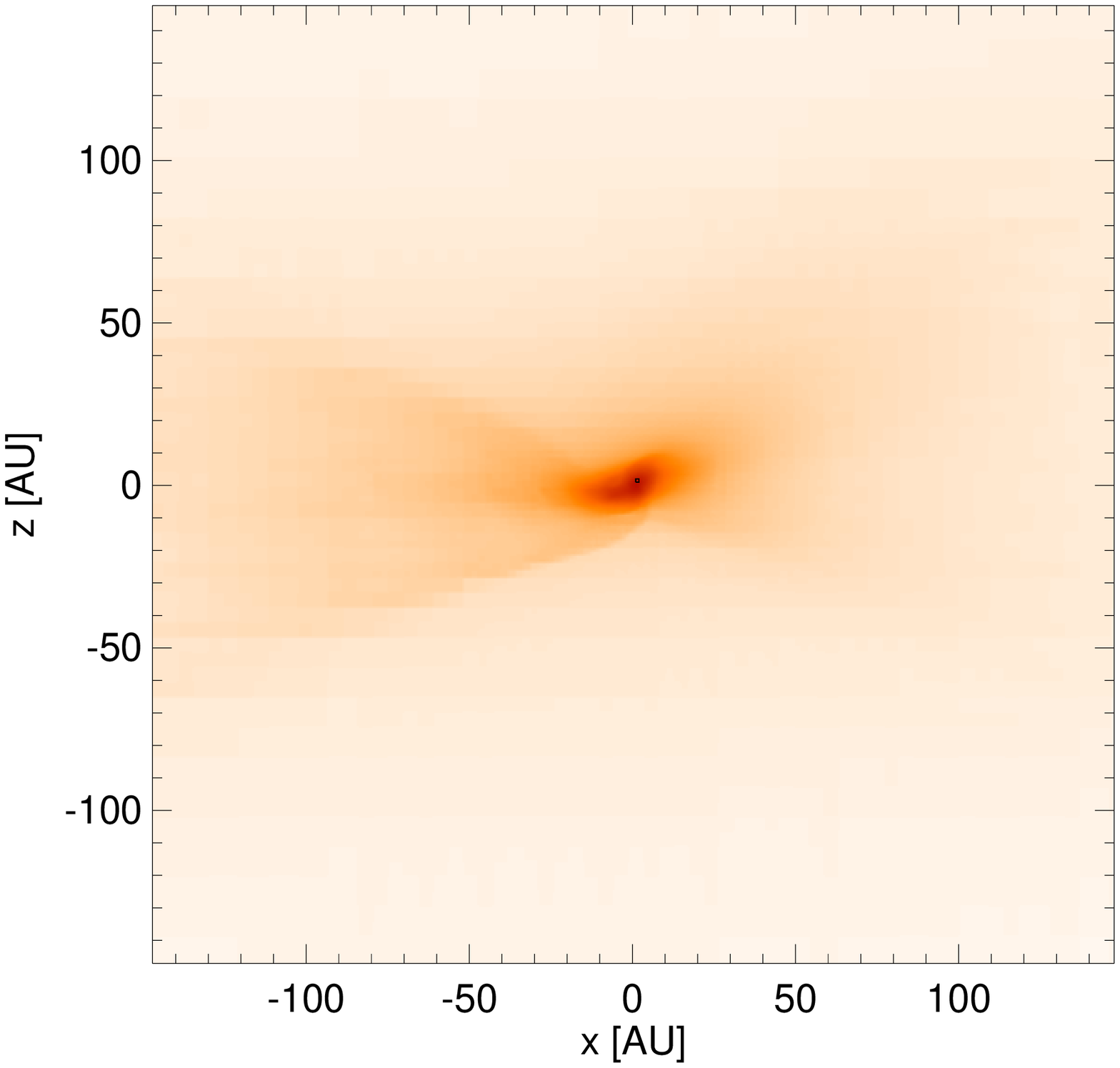}
\includegraphics[height=0.36\linewidth]{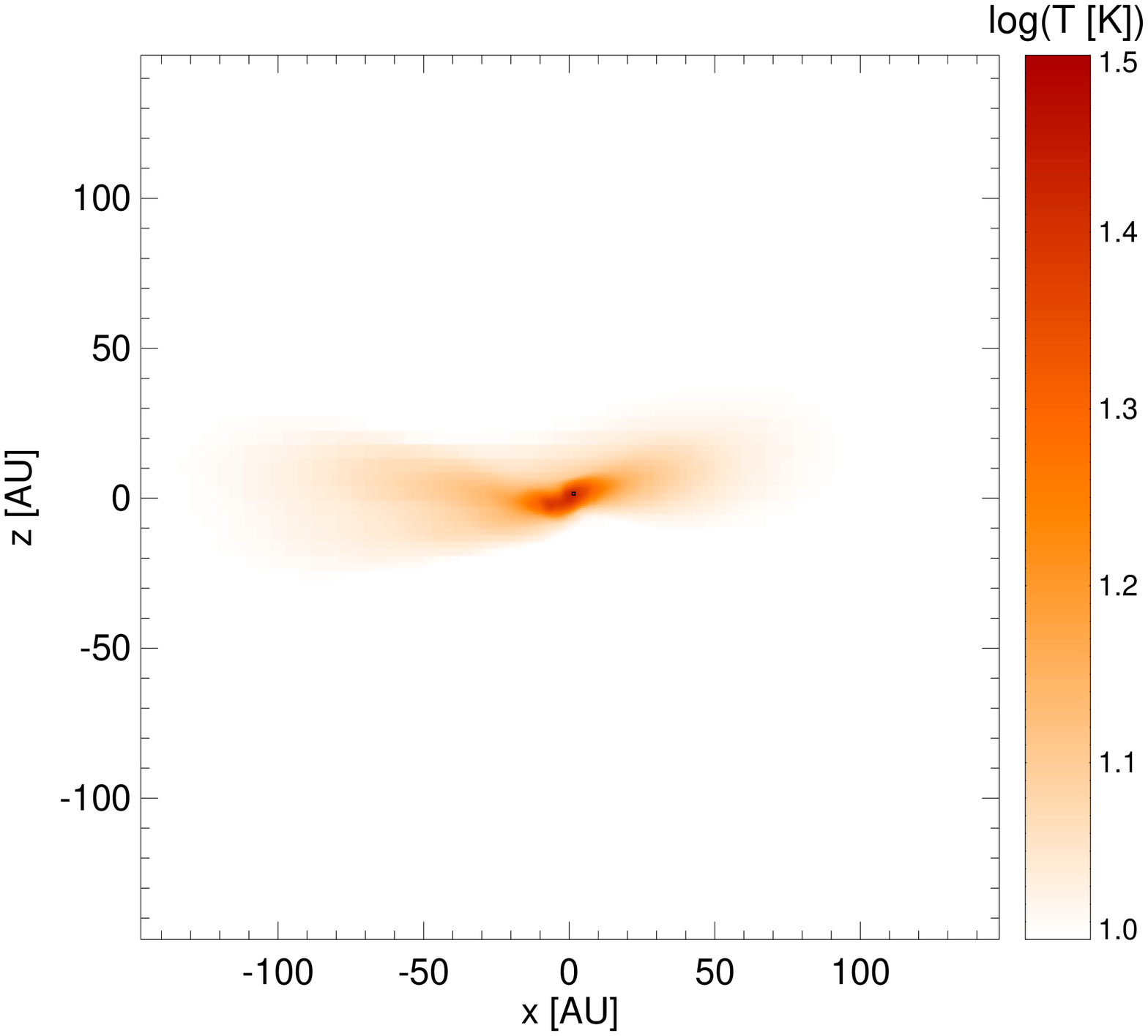}
\captionsetup{font=small}
\caption{The plots show density weighted temperature averages (e.g., $\int \rho\,T\,dz / \int \rho\,dz$) from a collapse
calculation including rotation and turbulence. 
{\it Left:} Results from the FLASH/RT calculations including radiative transfer. {\it Right:} Results from FLASH calculations using a barotropic EOS. The ambient gas temperature
in the FLASH/RT models is about 30\% higher.}
\label{fig:dens_temp_run06}
\end{figure*}
\bef
\centering
\includegraphics[width=0.98\linewidth]{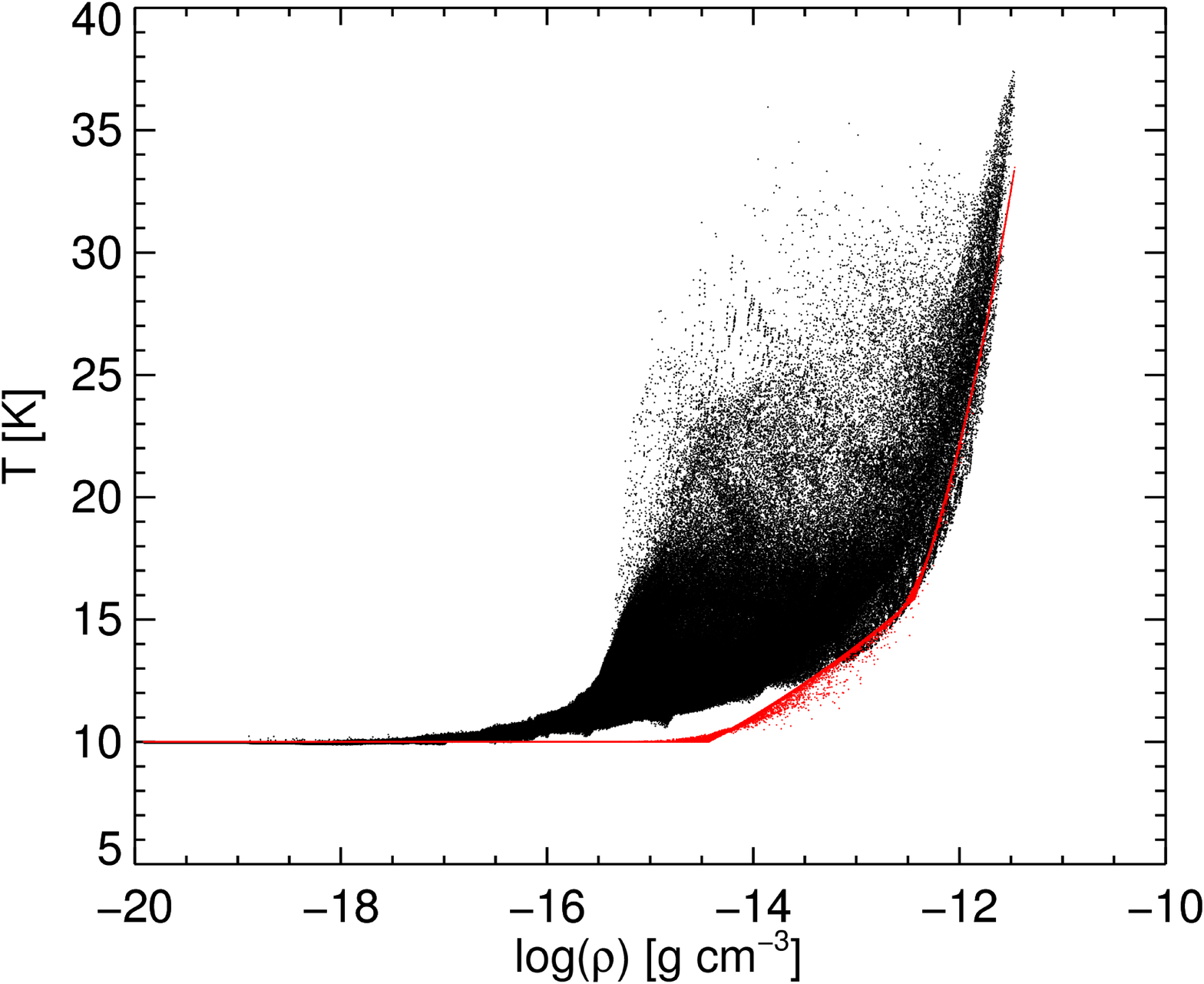}
\captionsetup{font=small}
\caption{Temperature distribution with respect to the gas density in the simulation box at the end of the collapse simulation including rotation and turbulence. 
Black dots show the temperature distribution from the FLASH/RT run, red dots resemble the temperature density dependence of the barotropic EOS.}
\label{fig:temp_dens_histo_run06}
\eef

\section{Performance}
\label{sec:performance}

The FLASH code shows excellent scaling behaviour on any computational infrastructure \citep[e.g.][]{FLASH00}.
For this work, the computations are clearly dominated by the solution of the RTE. Hence, the scaling behaviour
of the radiative transfer solver is crucial for the total performance of the FLASH/RT calculations. We investigate
the scaling performance of our radiation code using the disc benchmark setup (see Section \ref{sec:pascucci_benchmark}). 
We performed 50 formal solutions of the RTE using 192 directions on a spatial range covering 5 refinement levels. 
After the initial refinement, depending on the density structure and the radius, the computational domain consists of 
3648 valid subdomains (leaf blocks) each containing $8^3$ cells. The FLASH code distributes the blocks among
all available MPI ranks using a Morton space-filling curve\footnote[4]{\url{http://flash.uchicago.edu/site/flashcode/user\_support/flash4\_ug/}}.
The scaling tests were run at The North-German Supercomputing Alliance in Berlin on the Cray XC30 "Gottfried" using 12-core Xeon IvyBridge processors.
Figure \ref{fig:scaling} and Table \ref{tab:scaling} show the scaling results for the computation of the formal solution 
of the RTE averaged over 50 cycles. The scaling is normalized to the wall-clock time using 96 cores 
(e.g, 8 Xeon IvyBridge processors). "Gottfried" provides 2 Xeon processors with 24 cores in total per computing node, 
hence, adding 24 cores to the computation will increase the communication overhead. Figure \ref{fig:speedup} shows 
the speedup compared to a perfect scaling behaviour. The radiation solver scales reasonably well considering the 
communication of non-local information, which is necessary for the solution of the RTE. Figure \ref{fig:perf_block} shows 
that doubling the number of cores decreases the performance per block by approximately $10\%$, which we consider also as 
reasonable.\\
The cost of the radiative transfer solver from a 3D collapse simulation (Figure \ref{fig:perf_fracs}) is 
comparable to the cost for the computation of the self-gravitational potential which is done by a Poisson tree-solver.
However, the radiative transfer solver in this particular simulation uses a rather moderate angular resolution of 192 directions 
(using the HEALPix tessellation from \citet{Gorski05}). For runs including rotation and turbulence, the angular resolution 
probably needs a much higher resolution of at least 768 directions or higher. Since the cost of the radiative transfer 
solver scales linearly with the number of directions, it dominates the entire simulation run compared to the 
calculation of self-gravity. So far, we have tested the FLASH/RT code on our own 
computing cluster in Hamburg (32 nodes with 2x Intel Xeon Hexa-Core CPUs, 2.40 GHz) and at the North-German Supercomputing 
Alliance in Berlin on the Cray XC30.
\begin{figure*}
\centering
   \begin{subfigure}{0.49\textwidth}
     \caption{Wallclock Time}
     \includegraphics[width=\textwidth]{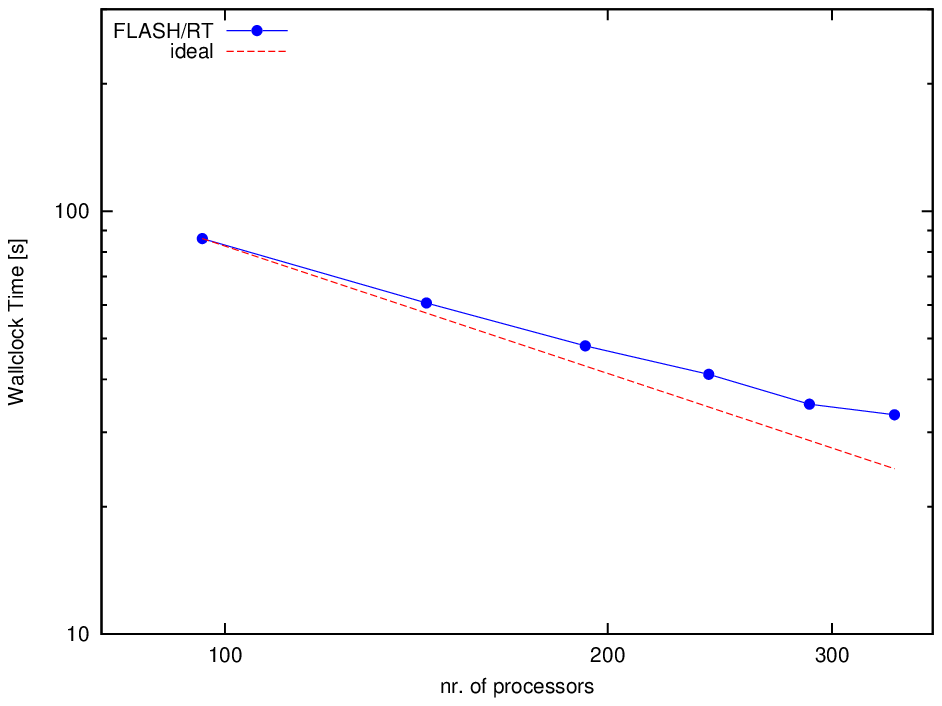}
     \label{fig:wc_time}
   \end{subfigure}
   \begin{subfigure}{0.49\textwidth}
     \caption{Speedup}
     \includegraphics[width=\textwidth]{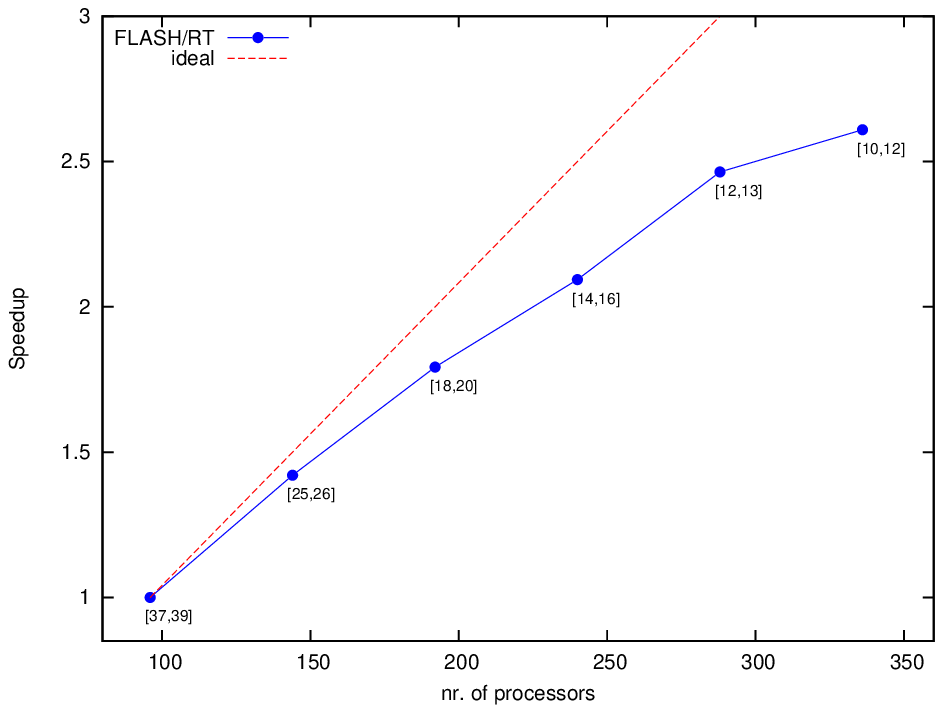}
     \label{fig:speedup} 
   \end{subfigure}
   \begin{subfigure}{0.49\textwidth}
     \caption{Performance per Block}
     \includegraphics[width=\textwidth]{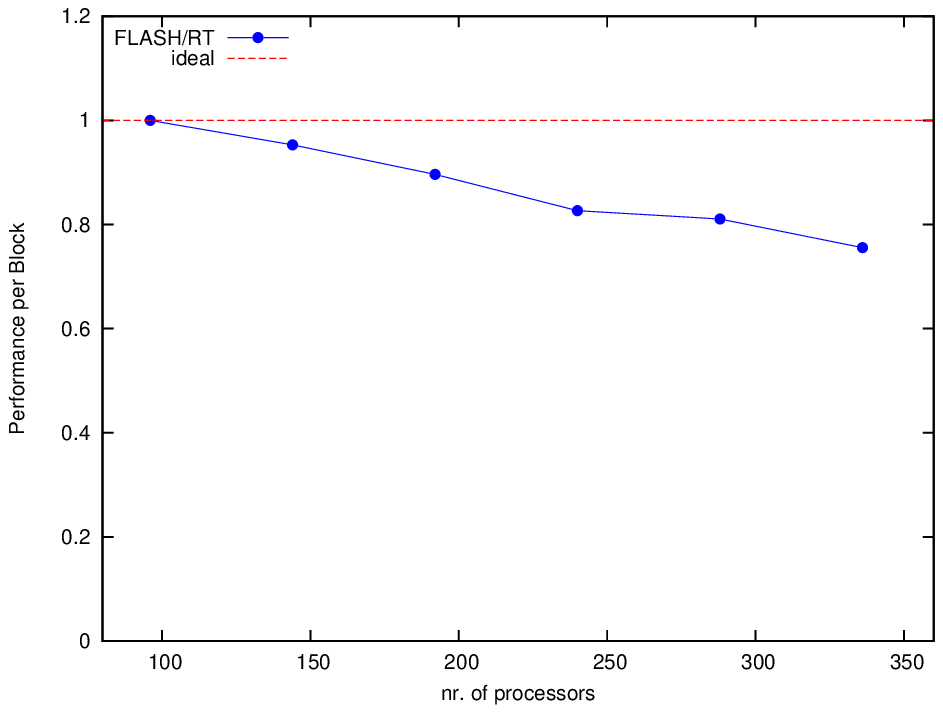}
     \label{fig:perf_block} 
   \end{subfigure}
   \begin{subfigure}{0.49\textwidth}
     \caption{fractional runtimes}
     \includegraphics[width=\textwidth]{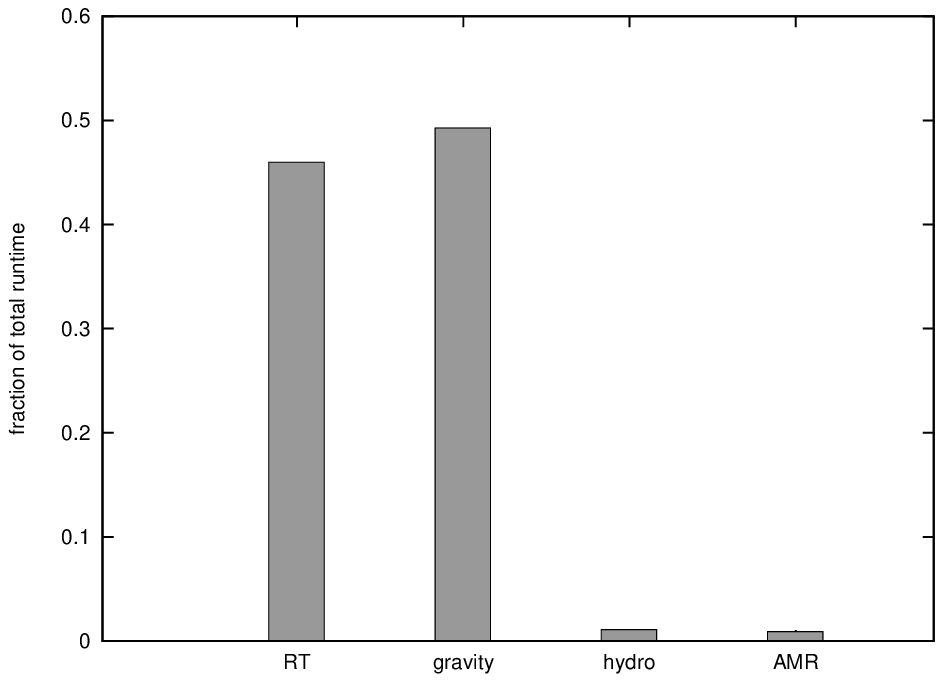}
     \label{fig:perf_fracs} 
   \end{subfigure}
\captionsetup{font=small}
\caption{Results from the parallel scaling test; In a) the total wall-clock time for the formal solution averaged over 50 iteration cycles from the Pascucci disc benchmark is shown. In b) the speedup normalized to the wall-clock time using 96 cores is shown. The number in brackets denote the minimum and maximum blocks per core, which is the result from the Morton space-filling curve. In c) the performance per block is shown, which decreases by roughly 10\% if the number of cores is doubled. In d) the fractional runtimes for the most costly steps of the the collapse simulation (Section \ref{sec:non_rotating_collapse}) is shown. All computations were performed using 192 directions.}
\label{fig:scaling}
\label{grid_step}
\end{figure*}
\begin{table*}
    \centering
    \begin{tabular}{  l | l | l | l | l }
    \hline 
    \hline
    nr. of cores & Time [s] & Speedup  & Blocks per cpu & Performace per Block [\%] \\ \hline
    96  & 86.06 & 1.0  & 37-39 & 100.0  \\ 
    144 & 60.60 & 1.42 & 25-26 & 95.2  \\ 
    192 & 48.01 & 1.79 & 18-20 & 89.6  \\ 
    240 & 41.11 & 2.09 & 14-16 & 82.6  \\ 
    288 & 34.93 & 2.46 & 12-13 & 81.0  \\ 
    336 & 32.98 & 2.60 & 10-12 & 75.5  \\ \hline 
    \end{tabular}
    \captionsetup{font=small}
    \caption{Results from the scaling test normalized to a run with 96 cpus; because of the increased communication overhead, each cpu should handle as many
    block as possible in terms of memory requirements.}
\label{tab:scaling}
\end{table*}

\section{Summary}
\label{sec:summary}
We have implemented a new radiation transfer solver based on the method of hybrid characteristics. The solver successfully reproduces standard
radiative transfer problems, including NLTE, thermal radiative transfer and the diffusion limit. We proved the feasibility of the method for 3D
collapse simulations where radiative transfer is the dominant cooling process during the formation of the first protostellar core. In contrast to the
FLD approximation, our method preserves the anisotropy of the radiation field which becomes crucial in the transition from optically thin to
optically thick regions (e.g, a protostellar disc). The radiation solver is implemented in the framework of the MHD code FLASH which allows
for a straight forward coupling of both codes (e.g., the collapse simulations.\\
However, the explicit energy coupling, as described in Section \ref{sec:timestep}, puts rather strong limitations on the time step. 
A possible improvement can be achieved by combining the raytracer with the solution of the moment equation for the radiative energy. 
In contrast to the FLD approach, one can compute an angle dependent diffusion coefficient in the form of the variable Eddington tensor (VET) \citep{Jiang12}
which can be achieved using our raytracer. The advantage is that the evolution of the radiative energy can be handled implicitly by solving the
linearized moment equation for the radiation temperature, like in \citet{Commercon11}, which resolves the problem of the time step restriction.
The framework for this has already been implemented in FLASH by \citet{Klassen14} and can be combined with our raytracer to implement the
VET approach.\\
Our method is a generalized and enhanced implementation of the hybrid characteristics method by \citet{Rijkhorst06} and \citet{Peters10a}. 
The original implementation was restricted to direct irradiation from point sources and the integration of optical depths respectively. The FLASH
code in combination with our radiative transfer framework allows for the solution of a much wider range of problems and can also very easily be
extended to handle a more complex form of the radiative transfer equation. Our implementation fits very well into the parallel design of the FLASH
code which is based on AMR with domain decomposition. Our method works within the AMR design of FLASH and is able to solve the 3D RTE on a
wide range of scales which is indispensable for star formation simulations.\\

\section*{Acknowledgments} 
L.B. acknowledges financial support by the Deutsche Forschungsgemeinschaft 
(DFG) mainly via the Emmy Noether grant BA 3706/1-1 {\it Theory of Massive Star Formation} and partly 
through the Graduiertenkolleg 1351 {\it Extrasolar Planets and their Host Stars} and the Priority 
Program 1573 {\it Physics of the Interstellar Medium}. 
T.P. acknowledges financial support through a Forschungskredit of the University of Z\"{u}rich, grant no. FK-13-112, 
and from the DFG Priority Program 1573 {\it Physics of the Interstellar Medium}. 
This work benefited from helpful discussions 
with Peter Hauschildt (Universität Hamburg) and Stefan Dreizler (Universität Göttingen). 
Most of the collapse simulations were carried out at The North-German Supercomputing 
Alliance in Berlin ({\it Gottfried}, HLRN).

\begin{appendix}
\section{Accelerated Lambda Iteration}
\label{sec:ALI}
The lambda operator $\Lambda$ describes the task to compute the radiation field from the source function. It is usually written as 
\begin{equation}
\label{lambda_step}
J = \Lambda[S].
\end{equation}
Formally, we can solve this by inverting the Lambda operator. When we arrange the cells of a 3D domain successively in a 1D vector, we can write the operator as a matrix. 
But the complete operator for {\it one cell} in the computational domain contains all radiative contributions from each other cell. Hence, the Lambda matrix
is far from being sparse. The explicit construction and storage of the Lambda matrix would easily reach computational limits in terms of memory requirements. 
Furthermore, the inversion of the Lambda operator is far too costly to be used in 3D radiative transfer.
Instead, the formal solution (\ref{equ:fs}) is used. Since the source function may depend on the mean intensity, this task requires iteration over Equations (\ref{equ:time_independent_rte}) to (\ref{equ:fs}).
This is called {\it Lambda iteration} but it usually fails in optically thick regimes. This happens because photons can be trapped and scattered many times, 
if a single cell of the computational domain is optically thick. The ordinary Lambda iteration is not able to account for these processes on scales smaller than the spatial resolution.\\
The idea behind the accelerated Lambda iteration (ALI), is to extract these sub-cell scattering contributions from the Lambda operator (and hence from the iteration), 
because we are not able to resolve them anyway. The extracted part of the ordinary Lambda operator is then put into a new {\it approximated} Lambda operator, which is solved quasi-analytically. 
Since the approximated Lambda operator usually only contains a small part of the whole Lambda operator (the subgrid part so to say), it is easy to compute, store and solve.
Mathematically, the Lambda operator becomes split
\begin{equation}
\Lambda = (\Lambda-\Lambda^*) + \Lambda^*
\end{equation}
where $\Lambda^*$ denotes the approximated Lambda operator. Inserting this into equation \ref{lambda_step} and using the source function for isotropic scattering (Equation \ref{equ:isotropic_source}), we get
\begin{equation}
S = \epsilon B + (1-\epsilon) (\Lambda-\Lambda^*)S + (1-\epsilon)\Lambda^*S
\end{equation}
Since the $\Lambda^*$-operator consists of only a small part of the whole Lambda-operator, it is sparse and easy to solve. We bring it to the left-hand side:
\begin{equation}
[1-(1-\epsilon)\Lambda^*]S = \epsilon B + (1-\epsilon) (\Lambda-\Lambda^*)S. 
\end{equation}
We introduce the iteration scheme, because there is still a contribution of the source function on the right-hand side. This remaining contribution can be regarded as the non-local contribution of the radiation field, 
which is solved by iteration. Inverting the approximated Lambda-operator then yields
\begin{equation}
\label{equ:ALI_scheme}
S^{n+1} = [1-(1-\epsilon)\Lambda^*]^{-1}(\epsilon B + (1-\epsilon) (\Lambda-\Lambda^*) S^n).
\end{equation}
The scheme in Equation \ref{equ:ALI_scheme} is a combination of iteration and analytic solution. The non-local contributions (in the Lambda matrix ($\Lambda-\Lambda^*$)) are accounted for by iteration while
the local subgrid scattering is handled by an inversion of the approximated Lambda operator ($\Lambda^*$). The computational cost of the inversion of the $\Lambda^*$-operator depends on its bandwidth, which
determines the range on which we solve analytically. 
Obviously, a diagonal $\Lambda^*$ is trivial to invert. But since the diagonal part of the Lambda operator describes only the local scattering in a 
single cell, it is not the best choice in terms of iterative performance. Usually, a tri-diagonal operator yields the best compromise between fast convergence and computational cost. 
But this requires the solution of a coupled set of linear equations, which is complex to implement. For now, we stay with a diagonal local $\Lambda^*$-operator, 
since it is the easiest one to implement and still has a tremendous effect on the convergence rate.
\section{The Angular Discretization using HEALPix}
\label{sec:healpix}
The choice of the solid angle grid is equivalent to the problem of discretizing the surface of a unit sphere. 
The method of characteristics requires the solution of the parameterized RTE along a large number of directions $\mathbf{n}$ depending 
on the anisotropy of the specific intensity $I(\mathbf{x},\mathbf{n})$. In general, this requires a homogeneous discretization 
of the solid angle $\Omega$ on the $4\pi$ unit sphere. For this purpose, we use the HEALPix\footnote[5]{Hierarchical Equal Area isoLatitude Pixelization} 
scheme introduced by \citet{Gorski05}. HEALPix ensures an optimal discretization of the unit sphere 
(also called {\it pixelation} or {\it tessellation}) into a number of finite solid angles $\Delta\Omega$. 
HEALPix in general addresses problems in which a function on domains of spherical topology has to be analyzed.
The pixelation scheme was originally developed to handle large datasets generated by cosmic microwave background experiments (e.g., WMAP, Planck)
and provides a software library\footnote[6]{\url{http://healpix.jpl.nasa.gov/}} with numerous subroutines for spherical discretization and 
numerical analysis of functions or datasets on the sphere.\\
The HEALPix pixelation has a base resolution of 12 pixels in three rings around the poles and the equator of the unit sphere each 
covering the same area. Based on these base pixels, the resolution is refined by dividing each base pixel into 
$N_{\text{side}}^2$ subpixels, where $N_{\rm{side}}$ has to be a power of 2 ($N_{\rm{side}}=1,2,4,8,...$). 
The total number of pixels (assuming an isotropic refinement) is then $N_{\rm{pix}}=12N_{\rm{side}}^2)$.
\begin{figure}
\centering
\includegraphics[width=0.34\textwidth]{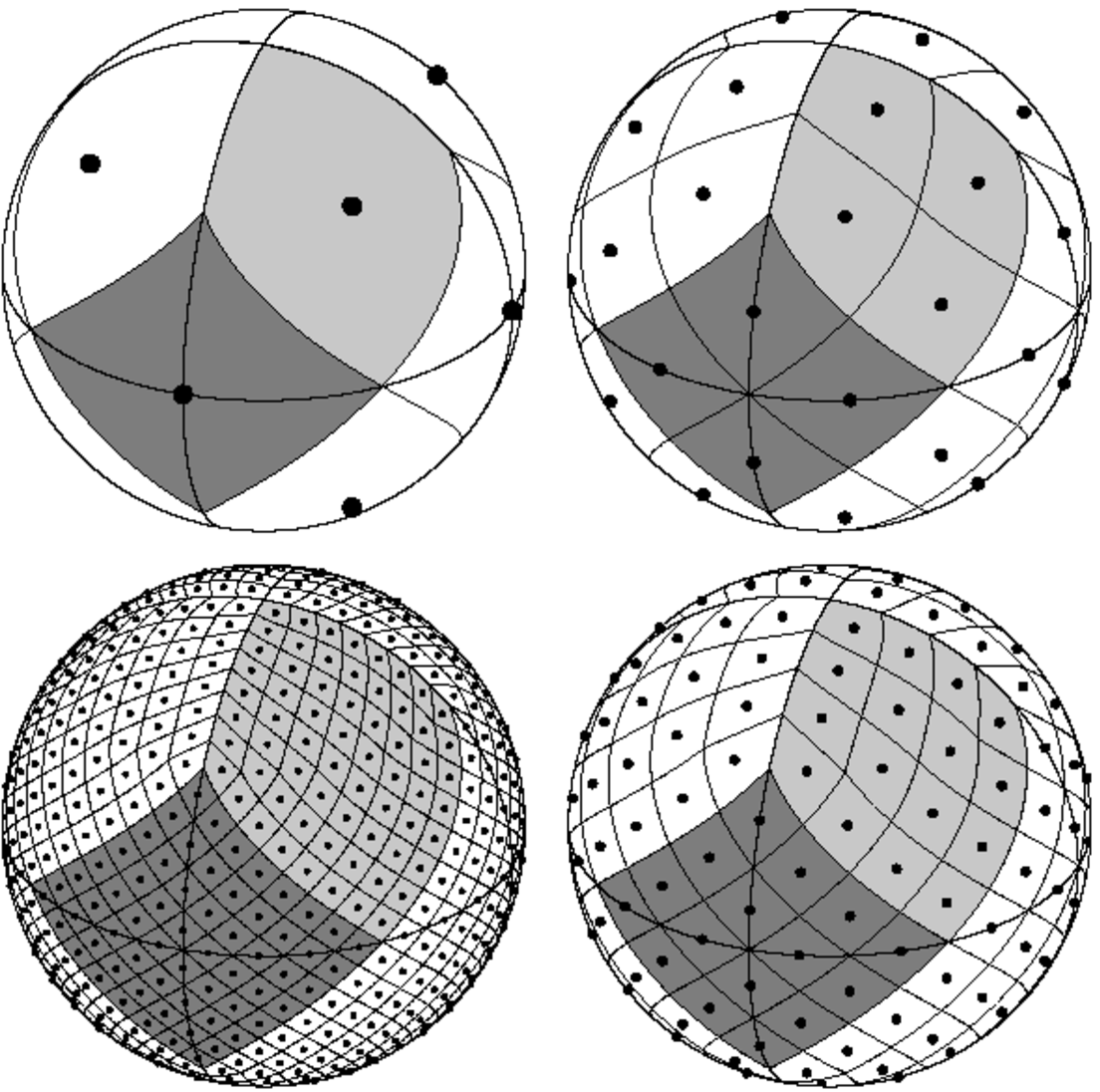}
\captionsetup{font=small}
\caption{The HEALPix tessellation scheme, from \citet{Gorski05}. 
The Area in light grey shows one of the eight (four north, and four south) polar base pixels and the dark grey area
shows one of the four equatorial base pixels. Moving clockwise from the upper left panel the base pixels are hierarchically 
subdivided with the grid resolution parameter equal to $N_{\rm side} = 1, 2, 4, 8$ and the total number of pixels
$N_{\rm pix}=12, 48, 192, 768$.}
\end{figure}

\end{appendix}
\newpage
\bibliographystyle{elsarticle-harv}
\bibliography{astro}
\end{document}